%% file: Pati-Memoir2.tex
\documentclass[12pt]{article}
\usepackage{graphicx}
\usepackage{url}
\usepackage{geometry}
\usepackage{enumitem}

\newcommand\pubnumber{SLAC--PUB--16937}
\newcommand\pubdate{\today}

\def\SLAC{SLAC,
    Stanford University, Menlo Park, California 94025 USA}
\def\doeack{\footnote{E-mail address:~pati@slac.stanford.edu.}\textsuperscript{,}\footnote{ Work supported in part by the US Department of Energy,
                     contract DE--AC02--76SF00515.}}

\def\Title#1{\begin{center} {\Large #1 } \end{center}}
\def\Author#1{\begin{center}{ \sc #1} \end{center}}
\def\Address#1{\begin{center}{ \it #1} \end{center}}

\newcommand\pubblock{\rightline{\begin{tabular}{l} \pubnumber\\
         \pubdate \end{tabular}}}
\newenvironment{Abstract}{\begin{quotation} \begin{center}
                       ABSTRACT
     \end{center}\bigskip  }{\end{quotation}}

\input mymacros.tex

\def\lesssim{\lsim}
\def\gtrsim{\gsim}

\begin{document}
\begin{titlepage}
\pubblock

\vfill
\Title{ {\bf {\LARGE Advantages of Unity With SU(4)-Color:}\\
Reflections Through Neutrino Oscillations,\\
Baryogenesis and Proton Decay\footnote{Presented at the Abdus Salam 90th birthday
Memorial Meeting,
IAS, Singapore, January 25--28, 2016\cite{WS-PROC}}}}
\vfill
\Author{Jogesh C. Pati\doeack}
\Address{\SLAC}
\vfill
\begin{Abstract}
{\small   As a tribute to Abdus Salam, I recall the initiation in 1972-73 of the idea of grand unification based on
 the view that lepton number is the fourth color. Motivated by aesthetic demands,
these attempts led to the suggestion that the existing $SU(2)\times U(1)$ symmetry be extended minimally to the quark-lepton and left-right symmetric
non-Abelian gauge structure $G(2,2,4) = SU(2)_L\times SU(2)_R\times SU(4)$-color.  This served to unify members of a family within a single L-R self-conjugate multiplet. It also explained: the
 quantization of electric charge,  the co-existence of quarks and leptons, and that of their three basic forces -- weak, electromagnetic, and strong -- while providing  the appealing possibility  that nature is fundamentally left-right symmetric (parity-conserving).The minimal extension of the symmetry
$G(2,2,4)$ to a simple group is given by the attractive symmetry $SO(10)$ that came a year later.  The advantages of the core symmetry $G(2,2,4)$, including those listed
above (which are of course retained by $SO(10)$ as well), are noted. These include the introductions of: (i){ \it the right-handed neutrino as a compelling member of each family}, (ii) (B-L) as a local symmetry, and (iii) the mass relation
$ m(\nu^\tau)_{Dirac} = m_{top}(M_{GUT})$.
 These three features, all arising due to $SU(4)$-color, as well as the gauge coupling unification scale
(identified with the (B-L)- breaking scale), are crucially needed to understand the tiny mass-scales of the neutrino oscillations within the seesaw mechanism, and to implement successfully the
 mechanism of baryogenesis via leptogenesis.
Implications of a well-motivated class of models based on supersymmetric $SO(10)$ or a string-unified $G(2,2,4)$  symmetry in 4D for (a) gauge coupling unification, (b) fermion masses and mixings, (c) neutrino osillations, (d) baryogenesis via leptogenesis,
and last but not least (e) proton decay are presented. Recent works on the latter providing upper limits on proton lifetimes suggest that the potential for discovery of proton decay in the next-generation detectors would
be high.}
\end{Abstract}
\vfill
\newpage
\tableofcontents
\end{titlepage}
\setcounter{footnote}{0}
%


\noindent {\bf Precis}

By way of paying tribute to Abdus Salam, I first recall the ideas of higher unification
that  the two of us introduced in 1972--73 to remove certain shortcomings in the status
of particle physics prevailing then, and then present their current role in theory as
well as experiments. These attempts initiated the {\it idea} of grand unification and
provided the core symmetry-structure 
$ G(2,2,4)= SU(2)_L \times  SU(2)_R \times SU(4)-color$
 towards
 such a unification. Embodied  with quark-lepton unification and left-right symmetry,
 the symmetry $G(2,2,4)$ is uniquely chosen as being the minimal one
that permits members of a family to belong to a {\it single multiplet}. The minimal
extension of $G(2,2,4)$ to a simple group is given by the attractive SO(10)-symmetry that
was suggested a year later. The new concepts, and the many advantages introduced by this
core symmetry (which are, of~course, retained by SO(10) as well) are noted. These include
explanations of the observed: (i) (rather weird) electroweak and color quantum numbers of
the members of a family; (ii) quantization of electric charge; (iii) electron-proton
charge-ratio being $-1$; (iv) the co-existence of quarks and leptons; (v) likewise that
of the three basic forces~--- the weak, electromagnetic and strong; (vi) the non-trivial
cancelation of the triangle anomalies within each family on symmetry grounds; {\it and} opening the door for
(vii) the appealing concept of parity being an exact symmetry of nature at the
fundamental level.

In addition, as a {distinguishing feature}, both because of SU(4)-color and independently
because of SU(2)$_{\rm R}$ as well, the symmetry $G(2,2,4)$ introduced, to my knowledge,
for the first time in the literature: (viii) {\it a new kind of matter}~--- the
right-handed (RH) neutrino ($\nu_R$)~--- {\it as a compelling member of each family}, and
together with it; (ix)~\hbox{(B-L)} as a local symmetry. The RH neutrinos~--- contrary to
prejudices held in the 1970's against neutrinos being massive and thereby against the
existence of $\nu_R$'s as well~--- have in fact turned out to be an asset. They are
needed to (a) understand naturally the tiny mass-scales observed in neutrino oscillations
by combining the seesaw mechanism {\it together with} the unification ideas based on the
symmetry SU(4)-color, and also (b)~to implement the attractive mechanism of baryogenesis
via leptogenesis. The quantitative success of the attempts as regards understanding both
(a) and (b) are discussed in Sec.~6. These provide a clear support simultaneously for the
following three features: (i) the seesaw mechanism, (ii) the SU(4)-color route to higher
unification based on a symmetry like SO(10) or a string-derived $G(2,2,4)$ symmetry in
4D, as opposed to alternative symmetries like SU(5) or even [SU(3)]$^3$, and (iii) the
(B-L)~- breaking scale being close to the gauge coupling unification scale ${\sim}2\times 10^{16}$\,GeV.

The observed dramatic meeting of the three gauge couplings in the context of low-energy
supersymmetry, at a scale ${\rm M}_{\rm U} \sim 2 \times 10^{16}$\,GeV, providing strong
evidence in favor of the ideas of both grand unification and supersymmetry, is discussed
in Sec.~3. The implications of such a meeting in the context of string-unification are
briefly discussed. Weighing the possibility of a stringy origin of gauge coupling
unification versus the familiar problem of doublet-triplet splitting in supersymmetric
SO(10) (or SU(5)), I discuss the common advantages as well as relative merits and
demerits of an effective SO(10) versus a string-derived $G(2,2,4)$ symmetry in 4D. Some desirable features of five or six-dimensional orbifold GUT
models are noted in here as well, 

In Sec.~7, I discuss the hallmark prediction of grand unification, viz. proton decay,
which is a generic feature of most models of grand unification. I present results of
works carried out in collaboration with Babu and Wilczek and most recently with Babu and
Tavartkiladze on expectations for decay modes and lifetimes for proton decay, including
upper limits for such lifetimes, in the context of a well-motivated class of
supersymmetric SO(10)-models. In view of such expectations, I~stress the pressing need
for having the next-generation large underground detectors~--- like DUNE and
HyperKamiokande~--- coupled to long-baseline neutrino beams to search simultaneously with
high sensitivity for (a) proton decay, (b) neutrino oscillations and (c) supernova
neutrinos. It is remarked that the potential for major discoveries through these searches
would be high.

Some concluding remarks on the invaluable roles of neutrinos and especially
of proton decay in probing physics at the highest energy scales are made in the last
section. The remarkable success of a class of supersymmetric grand unification models
(discussed here) in explaining a large set of distinct phenomena is summarized. Noticing
such a success and yet its limitations in addressing some fundamental issues within its
premises, such as an understanding of the origin of the three families and that of the observed dark energy ( cosmological constant),
some wishes are expressed on the possible emergence of a suitable grand-unified theory in 4D, together with a resolution of the issues as above
in the context of a well-understood quantum theory of of gravity, such as what one might expect from a better understanding of string/M-theory.

\section{Introduction}

\subsection{ Salam in Perspective}

Abdus Salam was a rare phenomenon: a great scientist, a humanitarian and a strong
promoter of his message that science is the common heritage of all mankind.  He will
surely be remembered for his many seminal contributions to physics, some of which have
proven to be of lasting value. These include his pioneering work on electroweak
unification for which he shared the Nobel Prize in physics in 1979 with Sheldon Glashow
and Steven Weinberg.

But I believe his most valuable contribution to science and humanity, one that is perhaps
unparalleled in the world, is the sacrifice he has made of his time, energy and personal
comfort, including his family-life, in promoting the cause of science in the third world.
His lifelong efforts in this direction have led to the creation of some outstanding
research centres, including especially the International Centre for Theoretical Physics
(ICTP) at Trieste, Italy,\footnote{Now named the Abdus Salam International Centre for
Theoretical Physics.} an International Centre for Genetic Engineering and Biotechnology
with components in Trieste and Delhi, and an International Centre for Science and High
Technology in Trieste. All these centres focus on serving scientists from the third
world. Salam dreamed of creating twenty international centres like the ICTP, spread
throughout the world, emphasizing different areas of science and technology. Approaching
developed as well as developing nations, for funding of such institutions, Salam often
used the phrase:~``science is not cheap, but expenditures on it will repay
tenfold''\cite{one}. Salam's dreams in this regard could not come to fruition during his
lifetime due to his illness in his later years. Fortunately, they have begun to be
realized in part recently through the creation of ICTP partner institutions in Brazil,
Mexico, Rwanda and China, thanks to the initiative of the present director of ICTP,
Fernando Quevedo, and others.

\subsection{Aspects of My Collaboration with Salam}

My close collaboration with Salam started spontaneously through a tea conversation in the
summer of 1972, during my short visit to ICTP, Trieste, and remained strong for over a
decade.\footnote{A brief account of how our collaboration started in May 1972 leading to
the ideas of higher unification and to the origin of the notion of ``Lepton Number as the Fourth Color'', in this context, is given in my
article in the Proceedings of the Salamfestschrift\cite{two} which was held at ICTP in 1993
(that is probably the last scientific meeting that Salam attended), and a shorter version
is given in the articles written in his honor after he passed away\cite{three,four}.The first
two sections of this talk are based in part on these three articles.} Of this period, I
treasure most the memory of many moments which were marked by the struggle and the joy of
research that we both shared. While I have mentioned this in my previous
writings,\cite{two,three,four} befitting the present occasion, let me mention again one aspect of
Salam's personality. During the ten year period of our collaboration, there have been
many letters, faxes, arguments over the phone and in person and even heated exchanges,
about tastes and judgements in physics, but always in a good natured spirit. In our
discussions, Salam had some favorite phrases.  For example, he would sometimes come up
with an idea and get excited.  If I expressed that I did not like it for such and such
reason, he would get impatient and say to me: ``My dear sir, what do you want: Blood?'' I
would sometimes reply by saying: ``No Professor Salam, I would like something better''.
Whether I was right or wrong, he never took it ill.  It is this attitude on his part that
led to a healthy collaboration and a strong bond between us.  Most important for me, by
providing strong encouragement from the beginning, yet often arguing, he could bring out
the best in a collaborator.  For this I will remain grateful to him.

\subsection{A Preview}

As a preview of the topics to be covered in the following sections, I first recall in
Sec.~2 the status of particle physics existing in mid-1972, together with its
shortcomings, and then present the ideas of higher unification which Salam and
I~introduced in 1972--73\  \cite{five,six,seven,eight} to remove some of these shortcomings. Current
status of these ideas in the context of subsequent developments in theory and experiments
are discussed in Secs.~2--6. To begin with, this would include a discussion in Sec.~2 of
the chronological evolution of the ideas on unification during 1972--74 starting from:
(1)~those of the standard model symmetry $G(2,1,3)=SU(2)_{L}\times U(1)_{Y}\times
SU(3)$-color; to (2)~the minimal quark-lepton and left-right  symmetric non-Abelian
symmetry ${\rm G}(2,2,4)={\rm SU}(2)_{L}\times {\rm SU} (2)_{R}\times {\rm
SU}(4)$-color~\cite{seven,eight} that brought a host of attractive features by removing some of
the major shortcomings of the standard model and introducing certain new ingredients,
like the right-handed (RH) neutrinos ($\nu_{{\rm R}}$'s), that turned out to be an asset
(See Secs.~2, 5 and~6); to (3) the smallest left-right asymmetric simple group
SU(5)\cite{9}, containing the standard model symmetry G(2,1,3), that had the virtue of
demonstrating explicitly the idea of grand unification; to (4) the minimal extension of
G(2,2,4) to the attractive simple group SO(10)\cite{10} that possesses all the benefits
of G(2,2,4), and in addition offers gauge coupling unification. SO(10) even retains the
left-right self-conjugate 16-component family-structure of G(2,2,4), as opposed to the
15-component family, composed of ($\bar{5}+10$) of SU(5). The single extra member in the
16-plet of G(2,2,4) or SO(10) is the RH neutrino.

The special advantages of the SU(4)-color route to higher unification offered by the
symmetry G(2,2,4), and therefore SO(10) as well, are noted in Secs.~2, 4, 5 and~6. These
include an understanding of the tiny mass-scales observed in neutrino
oscillations\cite{11,12}, in the context of the seesaw mechanism\cite{13}, as well as
implementing the promising mechanism of baryogenesis via leptogenis\cite{14,14a}. These
desirable features are not available, however, within the alternative routes of SU(5) or
even [SU(3)]$^3$\cite{15}.

Implications of the precision measurements of the three gauge couplings at LEP, revealing
their unification\  \cite{17,18} in the context of low-energy
supersymmetry \cite{19,20}, are discussed in Sec.~3. The changes in theoretical
perspective pertaining to gauge coupling unification and proton decay\cite{21} brought
about by the ideas of supersymmetry and superstrings~\cite{22} are discussed briefly in
here, as well.

In Sec.~7, I discuss works carried out in collaboration with Babu and
Wilczek~\cite{23,24} and more recently with Babu and Tavartkiladze\cite{25,25a} on
expectations for decay modes and lifetimes of proton decay, including upper limits, in
the context of a well-motivated class of supersymmetric SO(10) models. It is stressed
that these expectations are within a striking range of the next-generation detectors
being planned at DUNE and HyperKamiokande. Some concluding remarks are made   in
Sec.~8.


\section{Status of Particle Physics in 1972: The Growth of New Ideas}

\subsection{From $SU(2) \times U(1)$ to $G(2,2,4)$ and Beyond}

As noted above, my collaboration with Salam started during my two-months visit to ICTP,
Trieste in the summer of 1972. To put the growth of ideas during that summer in
perspective, I will first provide a historical background of the status of particle
physics existing in May 1972 and then provide motivations for the idea of higher
unification, which developed over the next two years. This was a time when the
electroweak ${\rm SU(2)} \times {\rm U(1)}$ model\cite{26} based on the Higgs mechanism
for symmetry breaking~\cite{27} existed. And the renormalizability of such theories had
been proven\cite{28} creating much excitement in the field.

But there was no clear idea of the origin of the fundamental strong interaction. The
latter was thought to be generated, for example, by the vector bosons ($\rho,\omega, {\rm
K}^{*}$ and $\phi$) along the lines suggested\footnote{Although the idea of generating
strong interactions by the gauge principle is attractive, it might have been argued that
the flavor-SU(3) gauge symmetry is not a suitable choice to generate the fundamental
strong interaction because weak interaction (viewed perturbatively) was ``known'' to use
part of the same symmetry, as in the ${\rm SU(2)} \times {\rm U(1)}$ model. This would
suggest that ($\rho,\omega,{\rm K}^{*}$) are not fundamental gauge bosons. \label{GaugeFootnote}} by
Sakurai,\cite{29} inspired by the beautiful Yang-Mills idea\cite{30}, or even by the
spin-o mesons ($\pi, {\rm K}, \eta, \acute{\eta},\sigma$) assumed to be elementary, or a
neutral U(1) vector gluon coupled universally to all the quarks\cite{31}.

By this time, based on the need to satisfy Pauli principle for the baryons treated as
three-quark composites, the idea of SU(3)-color as a global symmetry had been introduced
implicitly with quarks satisfying parastatistics of rank 3 in \cite{32} and
explicitly though quarks obeying familiar Fermi-Dirac statistics in \cite{33}. In this
context, the suggestion of generating a ``superstrong'' force by gauging the SU(3)-color
symmetry had also been made by Han and Nambu as early as in 1965 \cite{33}, though in a
variant form compared to its present usage (see remarks later). But the existence of the
SU(3)-color degree of freedom even as a global symmetry was not commonly accepted in 1972
because many thought that this would require an undue proliferation of elementary
entities. And, of course, asymptotic freedom had not been discovered yet. Thus the
standard model including the SU(3)-color symmetry had not been born,  

In the context of such a background, inspired by 't Hooft's proof of renormalizability of
spontaneously broken gauge theories, there was a number of papers appearing almost daily
at the ICTP preprint library which tried to build variants of the SU(2) $\times$ U(1)
model. For example, there were even attempts\cite{34} to get rid of the weak neutral
current weak interactions because experiments at that time (May 1972) hinted at their
absence.

As I was trying to catch up with these papers, it appeared to me that the heart of the
matter laid not in trying to find variants of the ${\rm SU(2)} \times {\rm U(1)}$ model,
but in removing its major shortcomings, first in its gauge sector. These included: (i) in
particular the arbitrary choice of the five scattered multiplets for each family
consisting of quarks and leptons with rather weird assignment of their quantum numbers
including the weak hypercharge which were put in by hand without a guiding principle;
(ii) the lack of a reason based on symmetry arguments for the co-existence of quarks and
leptons, and likewise (iii) that of the three forces-weak, electromagnetic and strong;
and (iv) the absence of a compelling reason for the quantization of the electric charge
and that for the observed charge-relation: ${\rm Q}_{\rm electron}=-{\rm Q}_{\rm
proton}$. (v)~In addition, I was bothered by the disparity with which the ${\rm
SU(2)}\times {\rm U(1)}$ model treated the left and the right chiral fermions (see
Eq.~(1)). This amounted to putting in non-conservation of parity by hand. I thought (in
Pauli's words) that God can't be weakly left-handed and at a deeper level the underlying
theory ought to treat left and right on par, conserving parity.

I mentioned these concerns of mine, based on aesthetic grounds, about the ${\rm SU(2)}
\times {\rm U(1)}$ model to Salam at a tea-gathering at ICTP\cite{two}. I also expressed
that in order to remove these shortcomings one would need to put quarks and leptons in
one common multiplet of a higher symmetry group (so that one may understand the
co-existence of quarks and leptons and explain why ${\rm Q}_{\rm e^{-}}=-{\rm Q}_{\rm
p}$) and gauge such a symmetry group to generate simultaneously the weak, electromagnetic
and strong interactions in a unified manner.

Now, the idea of putting quarks and leptons in the same multiplet was rather
unconventional at that time. Rather than expressing any reservation about it, as some
others did, Salam responded immediately by saying ``That seems like an excellent idea!
Let us develop it together''. It is this sort of spontaneous appreciation and
encouragement from Salam that helped to enrich our collaboration at every step. Thus
started our collaboration from that tea-conversation.

Searching for a higher symmetry to incorporate the features noted above, it became clear
within about two weeks$^{\rm 2}$ 
that quarks and leptons can be united in an elegant
manner by assuming that quarks do in fact possess the SU(3)-color degree of
freedom,\footnote{For reasons alluded to in footnote \ref{GaugeFootnote}, one could argue that the
SU(3)-color degree of freedom of quarks in the explicit sense\cite{33} is {\it essential}
not only to achieve a higher unification but just to realize a pure gauge-origin of the
three forces-weak, electromagnetic and strong (see below). I thank O.W. Greenberg for a
discussion on the need of Fermi-Dirac rather than parastatistics especially in the
context of grand unification.} obeying the familiar Fermi-Dirac statistics\cite{33}
rather than parastatistics \cite{32}, like the electrons do, and extending SU(3)-color to
the gauge symmetry SU(4)-color that treats lepton number as the fourth color. {\it Within
this picture, the neutrino and the electron emerged as just the up and down ``quarks'' of
lepton color}.

With SU(4)-color, the whole spectrum of quarks and leptons (then consisting of only two
families) fitted beautifully into a $4 \times 4$ structure of a global symmetry group
${\rm SU(4)}^{\rm flavor}\times {\rm SU(4)}^{\rm color}$ operating on four flavors
(u,d,c,s) and four colors (r,y,b,l).\footnote{Effectively such a $4 \times 4$-structure
is to be viewed as a merger of two families each being a $2 \times 4$. In reality, one
must of course gauge either the chiral ${\rm SU(4)}_{\rm L}^{\rm f}\times {\rm
SU(4)}^{\rm f}_{\rm R}$ (by assuming mirror fermions to avoid anomalies), or a suitable
anomaly-free subgroup of it. This is what we did (see below). But for the purposes of
classification and assignment of electric charge, which is vectorial, it sufficed to use
the non-Abelian vectorial ${\rm SU(4)}^{\rm f}\times {\rm SU(4)}^{\rm c}$.} Such a
structure accounted naturally for the vanishing of the sum of quark and lepton charges
and that of the combination (${\rm Q}_{\rm e^{-}}+ {\rm Q}_{\rm p}$), as desired. The
spontaneous breaking of SU(4)-color at high energies to ${\rm SU(3)}^{\rm c} \times {\rm
U(1)_{B-L}}$ was then suggested to explain the observed distinction between quarks and
leptons at low energies, as regards their response to strong interactions; such a
distinction must then disappear at sufficiently high energies.


Uniting quarks and leptons by the SU(4)-color gauge symmetry thus naturally implied the
idea that {\it the fundamental strong interaction of quarks arises {\it entirely} through
the octet of gluons generated by its subgroup of the SU(3)-color gauge symmetry, which is
exact in the lagrangian}\cite{35,36,37}.

In the course of our attempt at a higher unification \cite{five,six}, it thus followed that
the effective gauge symmetry describing electroweak and strong interactions at low
energies (below a TeV) must minimally be given by the {\it combined gauge
symmetry}\footnote{I should comment here on a common impression that exists in the
literature, including especially popular writings, as regards the origin of the standard
model versus that of the idea of grand unification. It is often stated that the successes
of the standard model naturally suggested that it be extended to include grand
unification. Historically, the story is, however, quite different. In the beginning of
May 1972, neither the standard model in its entirity, including SU(3)-color, nor any of
its empirical successes, such as the discovery of weak neutral current interactions,
existed. As indicated in the text, the combined gauge symmetry of the standard model
emerged\cite{five,six} at this juncture {\it simultaneously} with the attempts at higher
unification, based (in our case) on SU(4)-color. See further remarks in
Ref.~\cite{38}.} ${\rm G}(2,1,3)= {\rm SU}(2)_{\rm L}\times {\rm U}(1)_{\rm Y}\times
{\rm SU}(3)^{\rm c}$. This became known eventually as the standard model symmetry (SM).
It, of course, contains the electroweak symmetry  ${\rm SU}(2)_{\rm L}\times {\rm
U}(1)_{\rm Y}$\cite{26}.

It is instructive to note the family-multiplet structure with respect to the SM symmetry.
The 15 two-component members of the electron family belong to five disconnected
multiplets under the symmetry ${\rm G}(2,1,3)$ as shown below:
\begin{eqnarray}
\left(\begin{array}{ccc}{u_{r}}\,\,\,\,{u_{y}}\,\,\,\,{u_{b}}\\{d_{r}}\,\,\,\,{d_{y}}\,\,\,\,{d_{b}}\end{array}\right)^{\frac{1}{3}}_{L}\,;\,\,
\left(\begin{array}{ccc}{u_{r}}\,\,\,\,{u_{y}}\,\,\,\,{u_{b}}\end{array}\right)^{\frac{4}{3}}_{R}\,;\,\,
\left(\begin{array}{ccc}{d_{r}}\,\,\,\,{d_{y}}\,\,\,\,{d_{b}}\end{array}\right)^{-\frac{2}{3}}_{R}\,;\,\,
\left(\begin{array}{c}{\nu_{e}}\\{e^{-}}\end{array}\right)^{-\,1}_{L}\,;\,\,
\left(e^{-}\right)^{-\,2}_{R}\,. \label{e1}
\end{eqnarray}
Likewise for the muon and the tau families. Here the superscripts denote the respective
weak hypercharges $Y_{W}$ (where $Q_{em}=I_{3L}+Y_{W}/2$), which are chosen by hand
simply to fit the ``observed'' electric charges. The subscripts L and R denote the
chiralities of the respective fields. The symmetry SU(3)-color acts horizontally treating
quarks of three different colors of either chirality in each row as a triplet, while
SU(2)$_{\rm L}$ acts vertically on each column treating all LH fields as doublets, but
all RH ones only as singlets. Note the sharp distinction between the ways the SM treats
the left and the right chiral fermions. This is of course needed to conform with
observations (at presently available energies). I will discuss shortly how these five
disconnected multiplets become parts of a single multiplet under certain higher
unification symmetries, which would also treat the left and the right symmetrically.

We wrote up this aspect of our thinking in a short draft, which we submitted to J.D.
Bjorken for presentation at the 1972 Batavia conference \cite{five}, and then in a paper
which appeared in \cite{six}. In here, we suggested the concept of quark-lepton unification
through SU(4)-color. {\it In addition, unknown to many, we also initiated in the same paper ( in the third para)
the idea of a gauge-unification of the three forces in terms of a single coupling
constant, without exhibiting explicitly a symmetry to implement this~idea}.
We~\hbox{conjectured}\footnote{The fulfillment of this conjecture is, of course, a
prerequisite for the idea of grand unification to work. See remarks in Sec.~1 of
Ref.~\cite{six}.} that the differing renormalization effects on the three gauge
couplings following spontaneous breaking of the unifying symmetry, may cause the observed
differences between these three couplings at low energies. Fortunately, this conjecture
(hope) was borne out precisely by the discovery of asymptotic freedom about four months
later\cite{39}. I will return to a discussion of the success of the idea of gauge
coupling unification in Sec.~3.

\subsection{Advantages of the Standard Model}

Before continuing on the idea of higher unification, certain advantages of the standard
model, viewed as an effective low-energy theory, are worthnoting. They include the
following:

(i) The triangle anomalies~\cite{ABJ} coming from the quarks for this gauge-structure, with the
$4\times4$-spectrum of quarks and leptons mentioned above, are found to cancel
beautifully against those coming from the leptons\cite{41} provided the quarks possess
three colors. Such a cancelation, which is crucial to the renormalizability of the
theory, thus provided a strong {\it independent evidence in favor of the SU(3)-color
degree of freedom of the quarks}. The cancelation, which is non-trivial, can not however
be an accident. As we will see, a deeper reason for the automatic cancelation of the
anomalies for each family would in fact arise on symmetry grounds within certain higher
unification symmetries, including some based on   ${\rm SU}(4)$-color.

(ii) With the presence of four flavors (u,d,c,s), the standard model naturally
incorporates the Glashow-Iliopoulos-Maiani (GIM) protection mechanism\cite{42} in the
presence of cabibbo mixing\cite{43} so as to avoid excessive flavor-changing neutral
current processes including ${\rm K}^{\circ}-\bar{\rm K}^{\circ}$ mixing (see below for
results including radiactive corrections). Such a protection is essentially unaffected
when the standard model is extended to incorporate the third family in the context of
the Cabibbo-Kobayashi-Masakawa (CKM) mixing, including one CP-violating phase\cite{44}.

(iii) With strong interactions generated by SU(3)-color and the electroweak interactions
by the commuting ${\rm SU}(2)_{\rm L}\times {\rm U}(1)_{\rm Y}$- symmetry, it was
shown\cite{45} that, despite radiative corrections,` violations of parity and of stangeness
by one unit are of order ${\rm G}_{\rm F} {\rm m}^{2}$ (rather than of order $\alpha$)
and the $|\Delta S|=2 $ transitions (${\rm K}^{\circ}-\bar{\rm K}^{\circ}$) are of order
$({\rm G}_{{\rm F}}{\rm m}^{2})^{2}$ as desired, where m is a typical hadronic mass.

(iv) Importantly, with strong interactions generated entirely by the non-Abelian
SU(3)-color gauge force, the short-distance processes involving the hadrons are governed
by the property of asymptotic freedom\cite{39}. This served to explain the Bjorken
scaling\cite{46} in deep inelastic electron-nucleon scattering observed at SLAC,
providing justification for the approximate validity of the parton model\cite{47}, and
subsequently explaining remarkably well a host of short-distance processes including the small logarithmic
deviations from scaling in them. In turn, this gave a strong boost to the
purely SU(3)-color gauge-origin of the strong force. This defined the theory of quantum
chromodynamics (QCD), which is an integral part of the standard model. In addition, the
perturbative growth of the QCD coupling at long distances, together with the infrared
divergences of the non-Abelian QCD, made the idea of confinement of quarks and massless
gluons at least plausible\cite{48}, which has been put on a firmer footing by QCD-lattice
calculations\cite{49}.{\it In short, QCD with its non-Abelian self-interactions of gluons
served to account simultaneously for the two rather mysterious properties of quarks and
gluons- i.e. asymptotic freedom at short distances, and yet confinement at long
distances}.

(v) Last but not least, the standard model provides a renormalizable self-consistent
quantum field theory of the three basic forces with enormous predictive power involving a
rich variety of phenomena, well beyond those of QED. And, its predictions are brilliantly
successful. Even a brief mention of the empirical successes of the standard model, both
in its electroweak and the QCD sectors, culminating with the discovery of the Higgs boson in
2012, will take me outside the theme of my talk. At the same time, reflecting on the latter, this discovery seems to 
reaffirm the view that nature, peculiarly enough, utilizes renormalizability as one of the principles in formulating her laws.
{\it In other words, she cares that the laws should possess not only beauty but also reliability in the associated predictions} !. Let me then return to the growth of
ideas on higher unification.

\subsection{Why Choose The Symmetry $G(2,2,4)$?}

In searching for a desirable symmetry to meet certain aesthetic demands there appeared to
be two different ways which lead to the same answer as regards the choice of such a
symmetry. First, if one asks the question: which is the {\it minimal gauge symmetry} that
contains quark-lepton unification through ${\rm SU}(4)$-color, together with the
electroweak symmetry ${\rm SU}(2)_{\rm L}\times {\rm U}(1)_{\rm Y}$, and simultaneously
provides a rationale for the quantization of electric charge, the answer is clear and
simple. One must minimally gauge the quark-lepton and left-right symmetric gauge
structure \cite{seven,eight}:
\begin{equation}
G(2,2,4) \equiv SU(2)_L \times SU(2)_R \times SU(4)^c.
\end{equation}
Note the need for the non-Abelian left-right symmetric flavor gauge-structure ${\rm
SU}(2)_{\rm L}\times {\rm SU}(2)_{\rm R}$ (rather than ${\rm SU}(2)_{\rm L}\times {\rm
I}_{\rm 3R}$), accompanying SU(4)-color, that arises due to the requirement of
quantization of electric charge, together with that of minimality. Here SU(2)$_{\rm L}$
and SU(2)$_{\rm R}$ are the exact left-right analogs of each other. While SU(2)$_{\rm L}$
groups the LH fermions of a family into doublets (see Eq.~(1)), SU(2)$_{\rm R}$ does the
same for the corresponding RH fermions, thereby providing the basis for left-right
symmetry in the gauge interactions. The deeper implications of left-right symmetry will
be noted shortly following the presentation of the family-multiplet structure in Eq.~(3), with respect to the symmetry G(2,2,4).

Before moving on, it is interesting to note that the need for choosing ${\rm G}(2,2,4)$ as
the minimal symmetry arises by starting from a completely different angle. {\it Without
assuming SU(4)-color or the L-R symmetric gauge structure to begin with, if one just asks
the question: which is the minimal gauge symmetry that would group the five disconnected
multiplets of the SM belonging to a family (see Eq.~(1)) into a single multiplet, then
the answer is simple and unique}. The {\it minimal extension} of the SM symmetry ${\rm
G}(2,1,3)$ that is needed to serve the purpose is once again given by the symmetry ${\rm
G}(2,2,4)$, possessing the three features: (i) quark-lepton unification through
SU(4)-color, (ii)~the L-R symmetric gauge structure ${\rm SU}(2)_{\rm L}\times {\rm
SU}(2)_{\rm R}$, {\it and} (iii) the rationale for quantization of electric charge. {\it
In short, the three aesthetically desired features (i)--(iii) emerge simultaneously as
necessary features to provide a unique answer to the single question posed above, without
being assumed}. Aesthetically, this particular aspect appeared to be quite appealing to
us  and suggested, starting in 1972--73, that the SU(4)-color route to higher
unification, embodied in the symmetry ${\rm G}(2,2,4)$ , may well be
used by nature at some level as being part of an ultimate picture. Fortunately, as we
will see, such a route which of course includes extension of ${\rm G}(2,2,4)$ into a
simple group, the smallest one being SO(10), can clearly be distinguished empirically from
alternative ones such as those based on SU(5) devoid of SU(4)-color, through phenomena
such as neutrino oscillations, leptogenesis, fermion masses and mixings, and even proton
decay.

As an added desirable feature, since the symmetries SU(2)$_{\rm L,R}$ and the traceless vectorial
SU(4)-color are individually free from the triangle anomalies, {\it the gauge interactions of ${\rm G}(2,2,4)$,
subject to L-R discrete symmetry, can be chiral (as desired), yet anomaly-free, and
importantly parity conserving (see discussion below)}\cite{Witten-anom}. As alluded to before, this in turn
accounts on symmetry-grounds for the {\it non-trivial anomaly-cancelation} within each fifteen-member 
family of the SM symmetry, since the extra RH neutrino of G(2,2,4) is a singlet of the SM. 

Before discussing some of the additional major advantages of the symmetry ${\rm
G}(2,2,4)$, let me mention one that is perhaps the most striking.

\subsection{The Unwanted Right-Handed Neutrino}


Either one of the symmetries ${SU}(4)$-color or ${SU}(2)_{R}$ implies, however, that
\emph{there must exist the right-handed counterpart ($\nu_{R}$) of the left-handed
neutrino ($\nu_{L}$).} This is because the RH neutrino ($\nu_{R}$) is the fourth color
partner of the RH up quark; and it is also the ${SU}(2)_{R}$ doublet partner of the RH
electron. Thus, given the symmetry ${G}(2,2,4)$, one necessarily had to postulate the
existence of an \emph{unobserved new member in each family}~--- the right-handed
neutrino.

This in turn meant, especially within the SU(4)-color symmetry, that the neutrinos must
be massive, like the quarks, posing the dilemma as to why they are so light. A natural
resolution of this dilemma did emerge within the same ${\rm G} (2,2,4)$-framework within
four years through the realization of the seesaw mechanism\cite{13} and I will discuss
the same shortly. Meanwhile, there was a strong prejudice, however, in the 1970's, and
even through the 1990's till neutrino oscillations were discovered, against neutrinos
being massive and therefore against the existence of the RH neutrinos as well. Given that
the upper limits on neutrino masses were known to be so small: $(m(\nu_{\rm e})/m_{\rm e}
\lesssim 10^{-6}$, and after the discoveries of $\nu_\tau$ and the top quark, $m(\nu_\tau)/m_{{\rm top}}
< 10^{-9})$, many, perhaps most in the community, believed that~they must be exactly
massless.\footnote{The extent to which this belief was ingrained among many leading
physicists even in the 1990's may be assessed by an interesting remark by C.N. Yang at
the 2002 Stony Brook conference on neutrinos \cite{50}. See also remarks in Ref.~\cite{51}.} This
is in fact what the two-component theory of the neutrino\cite{52} as well as the standard
electroweak model of particle physics\cite{27} and the~SU(5) grand unification
model\cite{9}, possessing only LH neutrinos ($\nu_{{\rm L}}$'s), would naturally
suggest.\footnote{This is barring, of course, possible contributions\cite{53} to the
Majorana mass of $\nu_L$ from lepton-number violating quantum gravity effects \(\sim
(v^{2}_{\mathrm{EW}}/M_{Pl}) \sim (250 \mbox{\ GeV})^2/10^{19} \mbox{\ GeV} \sim 10^{-5}
\mbox{\ eV}\), which are tiny, compared to the presently observed mass scales of
atmospheric and even solar neutrino oscillations. As expressed elsewhere \cite{54}, this
smallness of (possible) quantum gravity effects prompts one to regard the atmospheric and
solar neutrino oscillations as clear signals for physics beyond what one may expect
within the standard model combined with quantum gravity.} In this sense, {\it the RH neutrino
was regarded perhaps by most as an unwanted child (the ugly duckling) in the 1970's}, and
I faced much resistance from the community in my seminars as regards the unavoidable need
for such an unwanted object.

 My only defense at that time (1972-74) was, what appeared in my view, the grand beauty of the 
 $[ (2 \times 4)_L + (2 \times 4)_ R ]$ quark-lepton and L-R symmetric family-structure, which is naturally suggested by the ${\rm G}(2,2,4)$ symmetry, explaining neatly the quantum numbers of all its members (see Eq.(3)).
It is interesting that subsequent developments in theory and experiments, which were crucial, upheld this inner conviction. With the realization of the
seesaw mechanism\cite{13}, and importantly the discovery of neutrino oscillations in
1998\cite{11}, {\it the RH neutrino has turned out to be an asset (a beautiful swan) to
understand neutrino oscillations as well as the observed baryon asymmetry of the
universe}. I will discuss these in Sec.~6. Let me now return to presenting the
other key features of the symmetry ${\rm G}(2,2,4)$.

The introduction of the RH neutrino requires that {\it there be sixteen two-component fermions
in each family}, as opposed to fifteen for the standard model (SM) or the SU(5) symmetry.
{\it Subject to left-right discrete symmetry (L $\leftrightarrow$ R) which is natural to
${G} (2,2,4)$, all 16 members of the electron family now became parts of a whole~--- a
single left-right self-conjugate multiplet ${F} =\{{F}_{L} \oplus {F}_{R}\}$, where}
\begin{equation}
{F}_{{L,R}}^{e}=\left[\begin{array}{c@{\quad}c@{\quad}c@{\quad}c}
u_r & u_y & u_b & \nu_e \\
d_r & d_y & d_b & e^{-}
\end{array} \right]_{{\rm L,R}}.
\end{equation}
The multiplets ${F}_{{L}}^{e}$ and ${F}_{{R}}^{e}$ are left-right conjugates of each
other transforming respectively as (2,1,4) and (1,2,4) of ${{\rm G}}(2,2,4)$; likewise
for the muon and the tau families. The symmetry ${SU}(2)_{{L,R}}$ treat each column of
${F}_{{L,R}}^{e}$ as a doublet; while the symmetry ${SU}(4)$-color unifies quarks and
leptons by treating each row of $F_{{L}}^{e}$ \emph{and} $F_{{R}}^{e}$ as a quartet;
\emph{thus lepton number is treated as the fourth color.} A very special feature of the
symmetry ${\rm G}(2,2,4)$ is now worth noting.

\subsection{Left-Right Symmetry in the Fundamental Laws:}

Because of the parallelism between the actions of SU(2)$_{\rm L}$ and SU(2)$_{\rm R}$ on
the left and the right chiral fermions respectively, and because SU(4)-color is
vectorial, the gauge symmetry ${\rm G}(2,2,4)$, with the choice $g^{(0)}_{\rm L}=g_{\rm
R}^{(0)}$, naturally opened the door for the {\it novel and attractive concept} that the
laws of nature possess left $\leftrightarrow$ right discrete symmetry~--- i.e. parity
invariance~--- at a fundamental level \cite{55}\footnote{In the discussion to follow, the L-R discrete symmetry should thus be understood
to accompany the symmetry ${\rm G}(2,2,4)$, with or without being specified as such.\label {LR}}, that interchanges ${\rm F}^{\rm
e}_{\rm L}\leftrightarrow {\rm F}^{\rm e}_{\rm R}$ together with ${\rm W}_{\rm
L}\leftrightarrow {\rm W}_{\rm R}$. Subject to suitable restrictions on the Higgs
system, the observed parity violation could then be interpreted as being a low-energy
phenomenon arising entirely through a spontaneous breaking of the ${\rm L}
\leftrightarrow {\rm R}$ discrete symmetry,
which should disappear at appropriately high energies. Thus within this picture, (using
Pauli's words) God is no longer ``weakly left-handed''; {\it left and right are treated
on par at the fundamental level}.

Briefly, I may mention that if one did not insist on quark-lepton unification through
SU(4)-color, parity invariance at a fundamental level can still be realized through the
so-called ``left-right symmetric (LRS)'' model \cite{eight,55} based on the symmetry ${\rm
G}(2,2,1,3)={\rm SU}(2)_{\rm L}\times {\rm SU}(2)_{\rm R}\times {\rm U}(1)_{\rm
B\hbox{-}L}\times {\rm SU}(3)^{\rm c}$, a sub-group of ${\rm G}(2,2,4)$. For a possible motivation for this sub-group, which might have been relevant in the pre-neutrino-oscillation era, 
see footnote 11 of Ref.~\cite{eight}.  Much
work\cite{57a}, especially in connection with having light W$_{\rm R}$'s which could
possibly be observed at the LHC, has recently been carried out in the context of this
minimal left-right symmetric (LRS) model.

I will return shortly to the relevance of having the RH neutrinos for an understanding of
the neutrino masses. First, it is worth noting a few additional features of the symmetry
${G}(2,2,4)$ and its relationship to still higher symmetries.

1) {\it The Charge Formula}: The symmetry ${G}(2,2,4)$ introduces an elegant charge
formula:
\begin{equation}
{Q}_{{em}} = {I}_{{3L}} + {I}_{{3R}} + ({B-L})/2,
\end{equation}
that applies to all forms of matter (including quarks and leptons of all six flavors,
Higgs and gauge bosons).\footnote{The 15th diagonal generator of $SU$(4)-color entering
into the electric charge formula of Ref.~\cite{eight} is naturally proportional
(as per Eq.~(11)) to the charge (${\rm B}_{{\rm q}}-3{\rm L}$) where
${B}_{{q}}=\mbox{quark number}$. It was pointed out in Ref.~\cite{56} that this can be
expressed in terms of the more familiar charge ($B$-$L$), since ${B}_{{q}}=3{\rm B}$,
resulting in Eq.~(4).} Note that the quantum numbers of all members of a family,
including the weak hypercharge ${Y}_{W} = {I}_{{3R}} + ({{B}}-{{L}})/2$, are now
completely determined by the symmetry group ${G}(2,2,4)$ and the tranformation-property
of $(F_{L}\oplus F_{R})$. This is in contrast to the case of the SM for which the
15~members of a family belong to five disconnected multiplets, with unrelated quantum
numbers. Quite clearly the charges $I_{{3L}}$, $I_{{3R}}$, and ${B-L}$ being generators
of ${SU}(2)_{L}$, ${SU}(2)_{R}$, and ${SU}(4)^c$ respectively are quantized; so also then
is the electric   charge $Q_{em}$.

2) {\it An Intimate Link Between SU(4)-color and the L-R Symmetry}: At this point, an
intimate link between ${SU}(4)^c$ and ${{SU}}(2)_{{L}} \times {SU} (2)_{{R}}$ is worth
noting. As remarked before, assuming that ${SU}(4)^c$ is gauged and requiring an
explanation for the quantization of electric charge as above leaves one with no other
choice but to gauge minimally the commuting symmetry ${{SU}}(2)_{{L}} \times {{SU}}(2)_{{R}}$ (rather than
${{SU}}(2)_{{L}} \times {{U} }(1)_{{{I} }_{3R}}$). Likewise, assuming that ${{SU}}(2)_{{L}}
\times {SU}(2)_{{R}}$ is gauged and again asking for a compelling reason for the quantization of
electric charge dictate that one must minimally gauge the symmetry ${SU}(4)^c$ (rather than
${{SU}}(3)^c \times {{U} }(1)_{{{B}}-{{L}}}$). The resulting minimal gauge symmetry is
then ${G}(224) = {SU}(2)_{L} \times {SU}(2)_{R} \times {SU}(4)^c$ that simultaneously
achieves quantization of electric charge, quark-lepton unification and left-right
symmetry. \emph{In short, the concepts of ${SU}(4)$-color and left-right gauge symmetry
in its minimal form (symbolized by ${SU}(2)_{L} \times {SU}(2)_{R}$) become inseparable
from each other, if one demands that there be an underlying reason for the quantization
of electric charge.} Assuming one automatically implies the other.

3) {\it Universality of Weak Interactions}: It is furthermore worth noting that the
extension of the standard model symmetry to the level of the symmetry ${\rm G}(2,2,4)$
provides a compelling reason for why weak interactions are universal with respect to
quarks {\it and} leptons, though strong interactions are not. This is because ${\rm
SU}(2)_{\rm L}$, generating weak interactions, commutes with SU(4)-color (as it must for
the sake of renormalizability) and thus treats all four colors representing quarks and
leptons universally. Non-universality of strong interactions can be attributed, as
mentioned before, to spontaneous breaking SU(4)-color to SU(3)-color $\times$ U(1)$_{\rm
B-L}$ at high energies. These features, which were known from the 1940's and were
essentially put in by hand to satisfy observations, thus found a rationale through   
quark-lepton unification as in ${\rm G}(2,2,4)$.

4) {\it The Two Simple Mass Relations}: The symmetry SU(4)-color leads to two simple
relations between the masses of quarks and leptons at the unification-scale ${\rm
M}_{\rm U}$: 
\begin{eqnarray}
m_{b}(M_U)&\approx& m_{\tau}\\[3pt]
m({\nu^{\tau}_{Dirac}})&\approx& m_{top}(M_{U})
\end{eqnarray}
These two relations arise from the SU(4)-color-preserving leading entries in the fermion mass
matrices which contribute to the masses of the third family [see Ref.~\cite{24} for a
detailed discussion]. The sub-leading corrections that arise from SU(4)-color breaking in
the (B-L) direction turn out to be important for the masses and mixings of only the first
two families \cite{24},and that, of course, goes well with observations.This is discussed in Sec.~5. Now, of the two
relations given above, the first is successful empirically. As we will see in Sec.~4, the
second is crucial to the success of the seesaw formula for $m(\nu^\tau_{\rm L})$ and
thereby for the observed $\delta m^2(\nu)_{23}$.

5) {\it B-L as a Local Symmetry}: The symmetry SU(4)-color contains B-L as a generator,
which provides some essential benefits. First, with B-L remaining intact at least upto
the unification scale ${\rm M}_{\rm U}\approx 2\times 10^{16}$\,GeV, it serves to protect
the RH neutrinos from acquiring a Planck or string-scale (${\sim}10^{18}$\,Gev) Majorana
mass through quantum gravity effects.\footnote{Such an ultraheavy Majorana mass
(${\sim}10^{18}$\,GeV) for the RH neutrino would be unacceptable because it would lead to
too tiny a mass (${<}10^{-4}$\,eV) for even the heaviest Dirac mass ${\sim}200$\,GeV for
any neutrino species through the seeaw formula.} Following limits from the
E\"{o}tvos-type experiments, however, one can argue that B-L must be violated at some
scale ${\rm M}_{\rm B-L}$ (considerations based on MSSM gauge coupling unification, that
predicts the weak angle \hbox{successfully,} would in fact suggest that ${\rm M}_{{\rm
B-L}}\approx {\rm M}_{\rm U}\approx 2\times 10^{16}$\,GeV, see Secs.~5 and 6). Now ${\rm
M} _{\rm B-L}$ sets the scale for the (superheavy) Majorana masses of the RH neutrinos,
and thereby plays a crucial role in determining the masses of the light LH neutrinos via
the seesaw formula (see Sec.~6). Furthermore, spontaneous breaking of B-L allows one to
implement the mechanism of baryogenesis via leptogenesis in the presence of the
elctroweak sphaleron effects which wipe out any (B-L)-conserving matter-antimatter
asymmetry (see Sec.~6).

In brief, viewed against the background of particle physics in 1972--73, the symmetry
${G}(2,2,4)$ brought some attractive features for the first time. These
include:\makeatletter \renewcommand\theenumi{\roman{enumi}}\makeatother
\begin{enumerate}[label=(\roman*)]
\item Unification of all 16 members of a family within \emph{one} left-right
self-conjugate multiplet, with a neat explanation of all their quantum numbers;
\item Quantization of electric charge, with ${\rm Q}_{{\rm e^{-}}}+{\rm Q}_{{\rm p}}=0$;
\item Quark-lepton unification through ${SU}(4)$-color;
\item A rationale for the co-existence of the weak, electromagnetic and strong forces, which, together with the features listed above, set the stage for their
unity \cite{six,9} possessing ${\rm SU}(4)$-color, within a symmetry like $SO(10)$ \cite{10}.
\item A compelling reason for the universality of the weak interactions with respect to
quarks {\it and} leptons;
\item Conservation of parity at a fundamental level \cite{55};
\item The RH neutrino as a compelling member of each family; 
\item B-L as a local symmetry;
\item A rationale for the now-successful mass-relations 5) and 6); and
\item The right set-up, due to (vii) and (viii), for implementing baryogenesis via
leptogenesis.
\end{enumerate}

These ten features, together with the two intriguing predictions mentioned below, constitute the hallmark of the symmetry $G(2,2,4)$, fulfilling major
aesthetic ((i)--(vi)) as well as practical ((vii)--(x)) needs. As mentioned before, the
three distinguishing features of $G(2,2,4)$~--- i.e. the existence of the RH neutrinos
and B-L as a local symmetry, together with the mass-relation 6), now seem to be
essential to understand naturally the tiny mass-scales observed in neutrino oscillations, in the context of the seesaw mechanism,
and to implement the mechanism of baryogenesis via leptogenesis. This will be discussed
in more detail in Secs. 5 and 6.
 
\begin{enumerate}[label=(\roman*)]
\setcounter{enumi}{10}
\item {\bf Existence of Magnetic Monopoles}
\end{enumerate}

In addition to the features listed above, a deep and crucial consequence of the idea that electromagnetism has its origin within a spontaneously broken
non-Abelian gauge symmetry ( semi-simple or simple), so that electric charge would be quantized, is the existence of topological 
't Hooft-Polyakov magnetic
 monopoles\cite{hooftpolyakov}, with masses ${\sim} {\rm M}_{\rm X}/\alpha_{\rm X}$, where ${\rm M}_{\rm X}$ is the mass of the gauge boson(s) associated with the relevant symmetry breaking and $\alpha_{\rm X}$ is the corresponding gauge coupling.
{\it The symmetry $ G(2,2,4)$, being the first realistic example of such a symmetry, thus necessarily predicts the existence of  magnetic monopoles}. As a distinction, the  $G(2,2,4)$- monopole would carry two units of Dirac magnetic charge ${\rm g}_{\rm D} = 2\pi/{\rm e}$, in contrast to one unit for the  
SU(5)-monopole \cite{Shafi-twoDirac}.

An interesting point here is the following. Dirac observed that the existence of even a single magnetic monopole (in his case, point-like) would imply quantization of electric charge \cite{Dirac}. A satisfactory quantum theory of such Dirac monopoles is, however, lacking. The argument is now reversed in the context of 
higher unification symmetries like $G(2,2,4)$  or its extensions to simple groups. These theories necessarily quantize electric charge and thereby predict the existence of topological monopoles which are fully consistent with relativistic quantum theory. Based on the simplest 
interpretations of the observed gauge coupling unification and the mass-scales of neutrino oscillations,
discussed in Secs.3 and 6, a grand-unified symmetry like SUSY $SO(10)$ or a string-unified $G(2,2,4)$  symmetry
( see Sec.~3) is expected to break in {\it one step} at the unification-scale ${\rm M}_{\rm U} \sim 2\times 10^{16}$ GeV to the standard model. In this case, the associated magnetic monopoles (to be called GUT-monopoles) would be superheavy with masses $\sim few\times10^{17} GeV$. These monopoles are thus too heavy to be produced by accelerators in the conceivable future.

 At the same time,
considering that the only known consistent quantum theory of a magnetic monopole is the topological one, as noted above, the discovery of a single magnetic monopole ( either superheavy or medium-heavy (see below)) would provide a conclusive evidence ( stronger than any other) for the existence of a GUT or GUT-like symmetry at high energies. {\it In other words, magnetic monopole is a very precious 
property of GUT}.

Despite their heaviness, in the context of a spontaneously broken GUT or GUT-like symmetry, one expects that these beautiful objects must have been produced in abundance in the very early universe.
Unfortunately for their detectability now, though "fortunately" for our own existence, it seems most likely that, soon after being created, they were super-diluted by the inflationary expansion of the universe \cite{Guth}, for which the theoretical motivations and empirical evidence are now strong.
The severe dilution of GUT- monopoles, expected within the inflationary picture, with the number of e-folds being greater than about 50 or so, is of course in accord with the stringent upper limits on the relic cosmic monopole flux which have been set by a variety of sensitive searches for a wide range of the monopole mass ( see the two papers in Ref.~\cite{physicstoday} for a review of the current status and references there in). Thus, within the inflationary picture, one would expect that the superheavy GUT monopoles have been inflated away beyond the level of observability.

There is, however, the possibility that relic monopoles of intermediate mass-scales $\sim 10^{14} GeV$ (say), subject to a reduced number of e-foldings ($\sim 30$ or so) may still be around so as to be detectable through improved searches. Such a possibility would arise if a symmetry like $SO(10)$ breaks in {\it two steps} to the SM symmetry
( instead of one) via the symmetry $G(2,2,4)$, with the latter breaking at an intermediate scale of $\sim 10^{13} GeV$
 (say) to the SM. In this case, the superheavy monopoles associated with $SO(10)$- breaking would be inflated away, but those associated with the subsequent step of $G(2,2,4)$-breaking, subject to a milder
 e-folding (of $\sim 30$, say), can survive in accord with all cosmological constraints and current empirical limits 
 so as to be detectable through improved searches. Consistency of such a scenario with the desired cosmological 
 parameters (including the spectral index $n_s$ and the tensor to scalar ratio r), and also with the limits from the monopole searches \cite{physicstoday}, has been
 shown within a non-supersymmetric inflationary $SO(10)$ model \cite{senoguz}.
 
 Now, as alluded to above, based on the simplest interpretation of the observed gauge coupling unification that leads to a successful prediction for the weak angle (Sec.~3) and an understanding of the mass-scales of the neutrino oscillations (Sec.~6), it is the one-step breaking of supersymmetric $SO(10)$ to the SM that is favored over the two-step breaking scenario. Nevertheless, given the importance of the discovery of a magnetic monopole as noted above,
 allowance should be made for the lack of certainty in our understanding of the pattern of GUT symmetry-breaking.
 In matter such as this, we must let experiment alone be the guide. Thus searches for magnetic monopoles covering a wide range of masses need to be pursued so as to push the current limits as close to the threshold of observability as possible, unless monopoles are discovered in the process.
 
 Let me now turn to one of the most novel features that emerged as a consequence of attempts at higher unification. The symmetry
$G(2,2,4)$ needing a spontaneous violation of B-L introduced a {\it new line of thinking}
leading to a questioning of the conservations of baryon and lepton numbers\cite{seven}\footnote{The case of spontaneous violation arises because B-L is gauged in $G(2,2,4)$. Now, a
massless gauge particle coupled to any linear combination of B and L (which is
gauged) must acquire a mass through SSB in order to conform with the limits from the
E\"{o}tvos type experiments.  The corresponding charge (B and/or L) must then be violated
spontaneously. \label{foonote_12}}, which
were otherwise held sacred owing to the extraordinary stability of the proton. Such a
questioning in turn evolved in the context of unification-ideas going beyond that of the
symmetry $G(2,2,4)$ leading to the hallmark prediction of grand unification~--- i.e.
proton decay\cite{seven,9}. This is briefly noted below, together with other fundamental
processes violating B and/or L. Proton decay will be discussed in more detail in Sec.~7.

\subsection{Non-Conservations of B and L: Proton Decay as a Generic Feature}

As mentioned before, B-L being a local symmetry in $G(2,2,4)$, the combination
\hbox{(B-L)} and therefore B and/or L must be violated spontaneously\textsuperscript{\ref{foonote_12}}. It was
\hbox{recognized} that this feature is a reflection of a more general phenomenon
involving non-conservations of baryon and lepton numbers which are most likely to occur
in unified gauge theories\cite{seven,9}, including those going beyond the symmetry
$G(2,2,4)$, that unify quarks and leptons as well as their three basic forces. Depending
upon the nature of the gauge symmetry and the multiplet structure, the violations of B
and/or L could be spontaneous, as is the case for non-conservation of B-L in SU(4)-color, and those of B
and L in the maximal one-family symmetry SU(16) which gauges both B and L
Ref.~\cite{58}. Alternatively, the violations could be explicit; that is what
happens for the subgroups of SU(16), like SU(5)\cite{9} and SO(10)\cite{10}, for which
the gauge boson interactions violate B and L. One way or another, baryon and lepton
number conservations cannot be absolute in the context of such higher unification.

The simplest and most dramatic consequence of such non-conservation is proton decay
($\Delta {B} \ne 0,\Delta L\ne 0$), which as will see provides an indispensable tool to
probe physics at truly high energies ${\sim}10^{16}$ GeV, for proton decaying via the
canonical ${\rm d} = 6 (\Delta (B-L)=0)$ decay modes, such as: $p\to e^+\pi^\circ$ and
$p\to \overline{\nu}K^+$. 

The other fundamental ``processes'' are the lepton-number violating Majorana masses of
the RH and LH neutrinos ($\Delta {B} = 0, |\Delta L|=2$), which are relevant to an
understanding of the neutrino masses via the seesaw mechanism, and in a related context
the nutrinoless double beta decay ($nn \to pp\ e^{-} e^{-}$)\cite{61,62}, both of which
arise naturally within $G(2,2,4)$ and $SO(10)$ symmetries. Based on our current
understanding of masses and mixings of all fermions including neutrinos in the context of
such unification symmetries, the neutrinoless double beta decay should occur at some
level and could quite plausibly lie within the realm
of observation in several ongoing or next-generation experiments\cite{62}. 

The third
equally fundamental process violating only baryon number is ${n}-\bar{n}$ oscillation\cite{63,64} ($|\Delta
{B}|=2,\Delta L=0$) which probes into physics typically at energy-scales of
${\sim}10^{5}$\,GeV (or possibly lower). Several promising models have been constructed
within symmetries like $G(2,2,4)$, SO(10) and $G(2,2,1,3)$ (see e.g. Ref.~\cite{64})
which suggest that ${n}-\bar{{n}}$ oscillation could well be discovered if the current
sensitivity can be improved by one to two orders of magnitude, reaching free
${n}-\bar{{n}}$ oscillation time ${\sim}10^{9}-10^{10}$\,s. 

Now any of these three
processes, if seen, would be a breakthrough in particle physics, shedding light on
fundamental physics at a deeper level. Thus the need for an improved sensitive study of
\textit{each of these three processes} cannot be overemphasized.

In this talk, I will focus primarily on three aspects: (i) gauge coupling unification
(Sec.~3), (ii) an understanding of the mass-scale(s) of neutrino oscillations (Secs.~5
and 6), and (iii) proton decay (Sec.~7). These turn out to be intimately linked in the
context of a promising class of grand unification models. As we will see, proton decay
arises as a \textit{generic feature} of most grand unification models if one seeks for
the simplest realization of gauge coupling unification that permits a successful
prediction of the weak angle (see Sec.~3), as well as an understanding of the mass-scale
of the observed atmospheric neutrino oscillation (Sec.~6 ). One can in fact argue,
within a class of well-motivated models of grand unification that proton decay should
occur at accessible rates, with a lifetime bounded above by about $10^{35}$ years, for
proton decaying into ${\rm e}^{+}\pi^{0}$, and a lifetime of less than about
$(1{-}8)\times 10^{34}$ years for proton decaying into $\bar{\nu}{\rm K}^{+}$. These
lifetimes are within factors of 5--10 above the current limits on proton lifetimes coming
from Super--Kamiokande\cite{65,65a}, raising hopes that proton decay should most likely
be discovered in the planned DUNE and/or HyperKamiokande experiments.

I next consider the idea of extending the symmetry $G(2,2,4)$ to simple groups which
serve to unify the three forces.

\subsection{Going Beyond ${\rm G}(2,2,4)$: $SO(10)$ and $SU(5)$}

To realize the idea of a single gauge coupling governing the three forces,\cite{six,9} one
must embed the standard model symmetry, or ${\rm G}(2,2,4)$, in a simple or effectively
simple group (like SU(N) $\times$ SU(N)). Several examples of such groups have been
proposed. Howard Georgi and Sheldon Glashow proposed the first such group SU(5) \cite{9}
which contains the standard model symmetry, but not ${\rm G}(2,2,4)$.  Following the
discovery of asymptotic freedom of non-Abelian gauge theories\cite{39} and the suggestion
of SU(5), Georgi, Helen Quinn and Weinberg showed how renormalization effects, following
spontaneous breaking of the unification symmetry, can account (to a good approximation)
for the observed disparity between the three gauge couplings at low energies\cite{17}.
Subsequently, such a unification of the gauge couplings was shown to hold to a much
better accuracy in the context of low-energy supersymmetry\cite{18} and more precise
measurements of the gauge couplings at LEP. Each of these contributions played a crucial role
in strengthening the ideas of higher unification.

Now one can retain all the aesthetic and practical advantages ((i)--(x), listed in Sec.~2.5) of
the symmetry group ${\rm G}(2,2,4)$ and in addition achieve gauge coupling unification,
if one extends the symmetry ${\rm G}(2,2,4)$ (which is isomorphic to \(SO(4) \times
SO(6)\)) minimally into the simple group $SO(10)$\cite{10}. As a historical note, it is
worth noting, however, that leaving aside gauge coupling unification\cite{six,9}, \emph{all the attractive features of $SO(10)$, arise entirely at the level of the symmetry ${\rm G}(2,2,4)$, subject to the L-R discrete 
symmetry\cite{55}, and were introduced as such\cite{seven,eight}, well before $SO(10)$ was proposed\cite{10}}.
These include in particular: members of a family belonging to a single multiplet, quantization of electric charge, the existence of the right-handed neutrino, $B-L$ as a local symmetry, {\it and} quark-lepton unification through $SU(4)$-color
 ( see the list in Sec.~2.5 ). The advantages of the last three features in understanding neutrino masses and implementing baryogenesis via leptogenesis will be discussed in Sec.~6.

The
\hbox{symmetry} $SO(10)$ of course fully preserves these \hbox{features}\footnote{In the
context of $SO(10)$, the L-R discrete symmetry associated with ${\rm G}(2,2,4)$ is
replaced by an equivalent generalized charge conjugation symmetry, which is a gauge symmetry in $SO(10)$\cite{66}.} because it contains
${\rm G}(2,2,4)$ as its maximal subgroup, with the latter being non-Abelian. It is furthermore remarkable that $SO(10)$
\hbox{preserves} even the left-right self-conjugate sixteen-component family-structure of ${\rm
G}(2,2,4)$ by using the set \(F = (F_L \oplus (F_R)^c)\) as its spinorial representation. 
Interestingly, following the discovery of neutrino oscillations \cite{11,12}, {\it this sixteen-component family-structure
has emerged as being not too small ( unlike the \(\bar{5}+10\) of $SU(5)$) and not too big ( unlike the 27 of $E_6$\cite{67}
with several unobserved members), but just right, possessing the extra RH neutrino}. The latter, added to the
\(\bar{5}+10\) of $SU(5)$, makes a family of sixteen, which is special to both ${\rm G}(2,2,4)$ and $SO(10)$.

In contrast to the extension of ${\rm G}(2,2,4)$ to $SO(10)$ or $E_6$, if one wished to
extend only the SM symmetry ${\rm G}(2,1,3)$ to a simple group, the minimal such
extension would be $SU(5)$\cite{9}. In the 1970s, long before the discovery of neutrino
oscillations, the symmetry $SU(5)$, being the smallest simple group possessing the SM
symmetry, served the important purpose of demonstrating the ideas of grand unification
simply. It, however, does not contain ${\rm G}(2,2,4)$ as a subgroup. As such, except for the features of
quantization of electric charge (feature (ii)), universality of weak interactions
((iv)), and b-$\tau$ mass-equality of Eq.~(5),  $SU(5)$ does not possess the other advantages of
of ${\rm G} (2,2,4)$ listed in Sec.~2.5, including: (a) the
RH neutrino , (b) $B-L$ as a local symmetry, and (c) the
mass-relation of Eq.~(6), based on $SU(4)$-color.\footnote{Like $SO(10)$, $SU(5)$ does
possess, however, fermion to anti-fermion gauge transformations which are absent in ${\rm
G}(2,2,4)$.} As I will discuss in Sec.~6, the last three features play crucial roles in
providing an understanding of neutrino masses and in implementing baryogenesis via
leptogenesis. Furthermore $SU(5)$ splits members of a family (not including $\nu_R$ or
$(\nu_R)^c$) into two multiplets: \(\bar{5}+10\), and it violates parity, like the SM,
manifestly.

Comparing ${\rm G}(2,2,4)$ with $SO(10)$, as mentioned above, $SO(10)$ possesses
 {\it all} the
features (i) to (x) of ${\rm G}(2,2,4)$ listed in Sec.~2.5,  but in addition it offers gauge coupling
unification. I should, however, mention at this point that the perspective on coupling
unification and proton decay changes considerably in the context of supersymmetry and
superstrings.  In balance, a string-unified ${\rm G}(2,2,4)$ symmetry offers some advantages over
a string-derived $SO(10)$ symmetry in 4D, while the reverse is true as well.  Thus,
 it seems that a definite choice of one
over the other is hard to make at this point.  This will be discussed briefly in the next
section.


\section{Gauge Coupling Unification}

It has been recognized from the 1970's that the concept of higher unification~--- now
commonly called Grand Unification (abbreviated as GUT)~--- has three dramatic
consequences:
\begin{enumerate}[label=(\roman*)]
\item a meeting of the three gauge couplings at a high scale;

\item the prediction of nonzero but superlight neutrino masses and thereby of
neutrino oscillations, which is a {\it special feature} of a sub-class of grand
unification symmetries, in particular those containing SU(4)-color, like SO(10) or E$_6$
or a string-derived ${\rm G}(2,2,4)$ symmetry (see below), in the context of the seesaw
mechanism; and last but not least;

\item proton decay.
\end{enumerate}

I will discuss the empirical verification of the first two features and the consequent
implications in this and the next section. The third feature~--- proton decay~--- which
now constitutes the missing piece of evidence for grand unification is discussed in
Sec.~7.

\begin{figure}
\centering
\includegraphics[width=0.6\hsize]{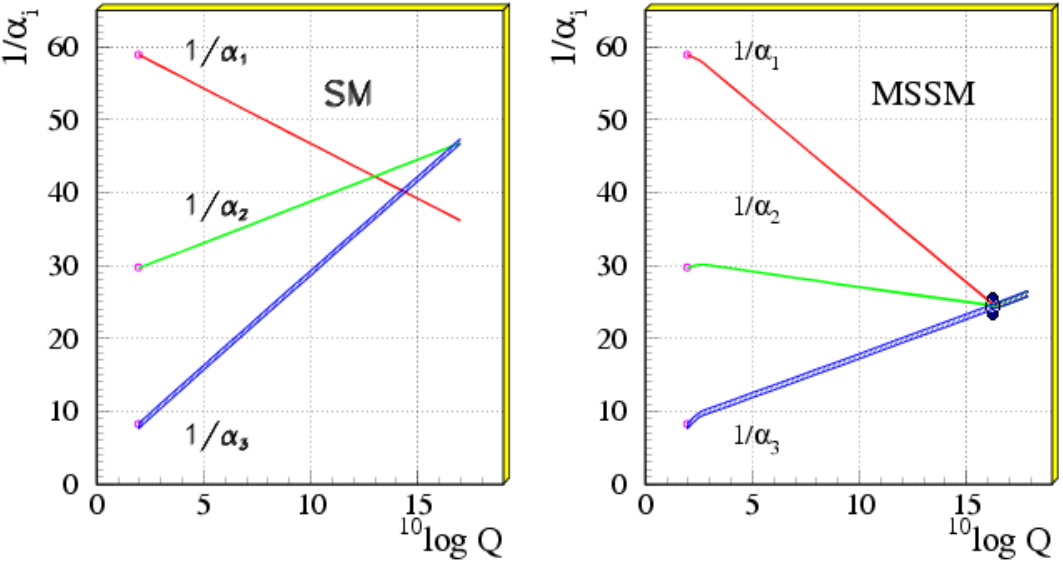}
\caption{Evolution of the three
gauge couplings $\alpha_i$ with momentum $Q$:  Standard Model (left panel) and Minimal
Supersymmetric Standard Model (right panel) }\label{4fig1}
\end{figure}

It has been known for some time that the precision measurements of the standard model
coupling constants (in particular ${\rm sin^2 \theta_W)}$ at LEP put severe constraints
on the idea of grand unification. Owing to these constraints, the non-supersymmetric
minimal ${SU}(5)$, and for similar reasons, the {\it one-step breaking} minimal
non-supersymmetric ${SO}(10)$-model as well, are now excluded\cite{68}. For example,
minimal non-supersymmetric ${SU}(5)$ predicts:  $\sin^2 \theta_W(m_Z)) \mid_{\bar{MS}} =
0.214 \pm 0.004$, where as current experimental data show:  $\sin^2
\theta_W{(m_Z)_{expt}}^{LEP} = 0.23153 \pm 0.00016$\cite{69}. The disagreement with
respect to $\sin^2 \theta_W$ is reflected most clearly by the fact that the three gauge
couplings $(g_1,g_2$ and $g_3)$, extrapolated from below, fail to meet by a fairly wide
margin in the context of minimal \underline{non-supersymmetric} $SU(5)$ (see left panel
of Fig.~\ref{4fig1}).

But the situation changes radically if one assumes that the standard model is replaced by
the minimal supersymmetric standard model (MSSM), above a threshold of about 1\,TeV,
which is motivated independently by requiring naturalness in the understanding of the
Higgs mass\cite{20}. Furthermore, subject to the assumption of R-parity or matter-parity conservation,
 which serves to avoid dangerous d=4 proton-decay operators, low-energy supersymmetry provides the
 lightest supersymmetric particle (LSP) as a natural candidate for the cold dark matter. With low-energy supersymmetry, 
 the three gauge couplings are found to meet \cite{18}
to a very good approximation (See Fig.~\ref{4fig1}, right panel) at a scale given by
\begin{eqnarray}
M_{U}\approx 2\times10^{16}\,\mbox{GeV\,\,\,\,(MSSM or SUSY\,\,SU(5))} \label{e5}
\end{eqnarray}

This dramatic meeting of the three gauge couplings, or equivalently the excellent
agreement of the MSSM-based prediction of
$\sin^{2}\theta_{W}(m_{Z})_{\mbox{Th}}=0.2315\pm0.003$\cite{18,68} with the observed
value given above, provides a strong support for the ideas of both grand unification and
low-energy supersymmetry, as being relevant to physics at short distances.

The simplest interpretation of the observed meeting of the three gauge couplings at the
scale $M_{U}$, that preserves the successful prediction of the weakangle
$\sin^2\theta_W$, is that a supersymmetric grand unification symmetry (often called GUT
symmetry), like SU(5) or SO(10), breaks spontaneously {\it in one step}, (as opposed to
possible multiple steps, relevant to SO(10)) at the scale $M_{U}$ into the standard
model symmetry $G(2,1,3)$.

Before discussing possible alternative interpretation of the observed meeting of the
gauge couplings in the context of string/M theory, it is worthnoting that, unlike SU(5),
a symmetry like SO(10) can in general break spontaneously in multiple steps (two or more)
to the SM via a symmetry like $G(2,2,4)$ (see e.g. Refs.~\cite{70} and \cite{71},
and two recent works\cite{72,73}) involving intermediate scale(s) of symmetry breaking.
In these cases, even without supersymmetry, gauge coupling unification can be made to
work by choosing appropriately the intermediate scale(s) of symmetry breaking. {\it But
in such cases, the weak angle is no longer a prediction; it needs to be used as an input
to fix the intermediate scale(s)}.\footnote{The case of Ref.~\cite{72} is somewhat of
an exception in this regard because the single intermediate scale associated with the
breaking of the $G(2,2,4)$ symmetry to the SM is identified with the Peccei-Quinn
symmetry breaking scale. But, even in this case, intermediate-scale breaking of B-L does
not go well with attempts to obtain a natural understanding of the mass-scale of the
atmospheric neutrino oscillation (see Sec~6).} Thus, in these cases, the success of the simplest
interpretation of the observed meeting of the gauge couplings in predicting the weak
angle in excellent agreement with experiments, which is realized in the context of an one-step breaking of supersymmetric SO(10), will have to be regarded only as an
accident. Remarkably enough, an identical conclusion emerges in attempting to have a
natural understanding of the mass-scale of the atmospheric neutrino oscillation (see
Sec.~6). In short, {\it two independent phenomena}~--- gauge coupling unification and the
mass-scale of the atmospheric neutrino oscillation~--- {\it seem to lead to one and the
same conclusion favoring the simplest picture of the one-step breaking of a SUSY GUT
Symmetry} (like SO(10)) or a string-unified $G(2,2,4)$ symmetry (see below)) to the SM
symmetry at a high scale $M_U\approx 2 \times 10^{16}\,{\rm GeV}$.

\subsection{SUSY SO(10) Versus a String-Unified G(2,2,4) Symmetry in 4D}

Now, the simplest interpretation, mentioned above, would require a SUSY GUT-symmetry,
like SO(10), to be effective at and above the unification-scale $M_U$ in 4D. An
alternative interpretation that would permit a supersymmetric non-GUT symmetry like
$G(2,2,4)$ breaking in one step into the SM symmetry at the unification-scale $M_U$ is,
however, possible in the context of string or M theory,
 which seems to be needed  to unify all the
forces of nature including gravity and also to obtain a good quantum theory of gravity.
This is because, even if the effective symmetry in four dimensions emerging from a higher
dimensional string theory is non-simple, like $G(2,2,4)$ or $G(2,1,4)$, or even the SM
symmetry $G(2,1,3)$, string theory can still ensure familiar unification of the gauge
couplings at the string scale. In this context, one could, therefore, first ask: can the
effective symmetry below the string-scale in 4D be as small as the SM symmetry
$G(2,1,3)$? I would argue (see Sec.~5 and 6) that attempts to understand naturally the
neutrino mass-scales utilizing the seesaw mechanism, and to implement the mechanism of
baryogenesis via leptogenesis successfully, provide an answer to the contrary. Both
suggest clearly that the effective symmetry in 4D, below the string scale, should {\it
minimally} contain either the $G(2,2,4)$ or its close relative $G(2,1,4)=SU(2)_L\times
I_{3R}\times SU(4)^c$ symmetry. This is because, as mentioned before, B-L should survive
at least up to the unification scale $M_U$ so as to protect the RH neutrinos from
acquiring Majorana masses of the string or Planck scale.\footnote{Such an ultraheavy
Majorana mass (${\sim}10^{18}$\,GeV) for the RH neutrino would be unacceptable because it
would lead to too tiny a mass (${<}10^{-4}$\,eV) for even the
heaviest LH neutrino through the seeaw formula.} Furthermore, the
full SU(4)-color symmetry, that contains B-L, is needed to ensure the two Dirac
mass-relations given in Eqs.~(5) and (6). The first is empirically successful, while
the second is needed crucially for the success of the seesaw mechanism (see Sec.~6).

While keeping $G(2,1,4)$ in the picture, to simplify discussion, I will thus proceed to
consider the two alternatives of a (hopefully) realistic string/M-theory solution
possessing either an effective $G(2,2,4)$ symmetry or the GUT-symmetry SO(10), both with
supersymmetry, emerging at the string-scale in 4D. Needless to say, among the vast
landscape of string/M-theory solutions, which are essentially perturbative in nature,
there is no criterion of selectivity at present, and identifying a completely realistic
solution, even with only the SM symmetry in 4D, remains a challenge. Thus, at present, combining a top-down with a bottom-up 
approach, hoping that the two would meet at just the desired structure, seems to be the best way. Within the vast landscape of string solutions,
 there do exist, however, 
promising semi-realistic solutions containing an effective $G(2,2,4)$ symmetry in
4D \cite{74,leonta-dbrane,cvetic-Fth}, with three generations, and some of them with the right set of Higgs-like
multiplets to do the desired symmetry breaking in 4D ( see e.g.~B. Assel et al.~in \cite{74} and Refs.~\cite{leonta-dbrane} and
\cite{cvetic-Fth}). Semi-realistic string-derived
SO(10) solutions also exist\cite{75} (though these have not been studied as much as those for the case of $G(2,2,4)$). While the emergence of such solutions from 
within a fundamental theory is encouraging, much work still needs to be done to realize the essential ingredients in 4D within one such solution. {\it Thus at present, the existence of a completely ( or close to)
realistic string-$G(2,2,4)$ or string-SO(10) solution in 4D, with the desired spectrum and couplings, can only be regarded as an
assumption}. In view of the strong empirical motivations for either one of these two
symmetries being effective in 4D, and the theoretical motivations for the string/M theory being the mother
theory, consequences of such an assumption seems worth pursuing, and I will do so in the following.

Let me now return to the question of a consistency that arises for a non-GUT
$G(2,2,4)$-like string solution in 4D. While string theory will ensure gauge
\hbox{coupling} unification (i.e. $g_{\rm 2L}=g_{\rm 2R}=g_4$) at the string-scale, one
needs to account for the mismatch by about a factor of 20 between the MSSM unification
scale $M_{U}$ (given~above), and the
 string-unification scale, given by $M_{st}\approx g_{st}\times 5.2\times10^{17}\,{\rm GeV}\,\approx 3.6\times10^{17}$\,GeV (Here we have put
$\alpha_{st}=\alpha_{GUT}(\mbox{MSSM})\approx 0.04$) \cite{76}. Possible resolutions of
this mismatch have been proposed. These include:

(i) {\bf utilizing the idea of string-duality}\cite{77} which allows a lowering of
$M_{st}$ compared to the value shown above, or alternatively

(ii) {\bf allowing the possibility of highly anisotropic string compactification} in which one or two of the compact radii are extremely 
large compared to the others\cite{hebecker}, 

(iii) {\bf using the idea of a semi-perturbative unification} that assumes the existence of two
vector-like families, transforming as $(16+\overline{16})$, at the TeV-scale. The latter
raises $\alpha_{GUT}$ to about 0.25--0.3 and simultaneously $M_{U}$, in two loop, to
about $(1/2-2)\times10^{17}$\,GeV \cite{78}.

(iv) {\bf Assuming a Few Extra Multiplets at GUT-scale}: A fourth
possibility\cite{79}, which has not appeared in a published form yet, would arise if a supersymmetric $G(2,2,4)$-solution emerges from string theory
in 4D, possessing {a few extra Higgs-like multiplets} (as mentioned below) having
GUT-scale masses; most of these will be needed to break the symmetry G(2,2,4) at the GUT-scale and/or 
give appropriate masses and mixings to the fermions any way. These extra multiplets are in addition, of course, to the familiar
low-energy spectrum of MSSM possessing: (a) the three generations of quarks and leptons,
(b)~one light (massless) multiplet $(2,2,1)_{\rm H}$ of $G(2,2,4)$ which contains H$_{\rm u}$ and H$_{\rm d}$ of
MSSM and serves to break the EW symmetry, and (c) their superpartners. The assumed GUT-scale
multiplets, that would serve the desired purpose of  GUT-scale symmetry-breaking, while preserving L-R discrete symmetry {\it and} accounting for the mismatch
between ${\rm M}_{\rm U}$ and ${\rm M}_{\rm st}$ (mentioned above), are given by the set:
\beqa
{\rm H}_{\rm GUT}&=&{\rm H}_{\rm SSB}+{\rm H}_{\rm extra}^\prime  \CR
&=& \left[(1,1,15)_{\rm H} + \{ (2,1,4)_{\rm H} +
(1,2,\overline{4})_{\rm H}\}_{\rm A}+\,\{(2,1,\overline{4})_{\rm H} + (1,2,4)_{\rm H}\}_{\rm B}\right]_{\rm SSB}\CR
&&{}+\left[(1,1,15)^\prime_{\rm H}+(2,2,1)^\prime_{\rm H}\right]_{\rm extra}\label{eq:representation}
\eeqan
Note that, of these extra multiplets, the $(1,1,15)_{\rm H}$ is the analog of 45$_{\rm
H}$ of SO(10) and the sub-sets A and B correspond precisely to 16$_{\rm H}$ and
$\overline{16}_{\rm H}$ of SO(10) respectively, while $(2,2,1^\prime)_{\rm H} \subset
10^\prime_{\rm H}$ of SO(10). The VEVs of $\{(1,2,4)_{\rm H}+(1,2,\overline{4})_{\rm H}\}$ 
of GUT-scale are in fact needed to break the gauge symmetry $G(2,2,4)$ to
the SM, while preserving supersymmetry; and the corresponding L-R conjugate multiplets
$\{(2,1,4)_{\rm H} + (2,1,\overline{4})_{\rm H}\}$ should be present, if L-R discrete symmetry survives compactification, which we assume. This is the exact analog of the roles of
$\{16_{\rm H}+\overline{16}_{\rm H}\}$ for breaking SO(10) (see Sec.~4). Also one of the
two $(1,1,15)_{\rm H}$ (${\subset}45_{\rm H}$ of SO(10)) having a GUT-scale VEV along the
(B-L)-direction is needed to introduce the observed (B-L)-dependence in fermion masses
and mixings, through the use of higher dimensional operators (see Sec.~5). 

In this sense, all the
multiplets listed in the first square bracket on the RHS of Eq.(8)  
 are in fact needed, minimally, to implement necessary symmetry breaking and/or to give desired masses and mixings to the fermions, while preserving supersymmetry and L-R symmetry.
{\it Only the two multiplets- i.e. the $(1,1,15)^\prime_{\rm H}$ and the
$(2,2,1)^\prime_{\rm H}$ -  in the second square bracket of Eq.(8) are really ``extra''}. For comparison, it may be noted that analogous Higgs multiplets
like 16$_{\rm H}$, $\overline{16}_{\rm H}$ and 45$_{\rm H}$ are needed also for the case of an intact SUSY SO(10) in 4D
 to serve identical purposes. Furthermore, for the latter, a few additional multiplets like a pair of ($16^\prime_{\rm H}$ and $\overline{16}^\prime_{\rm H}$) and a $10_{\rm H}^\prime$ with specified
couplings would be needed to implement natural and stable doublet-triplet splitting (see secs.~5, 7 and Ref.~\cite{24}).

 It is worth noting that the spectrum of the type exhibited in Eq.~(\ref{eq:representation}), including the adjoint of SU(4)-Color $(1,1,15)$ (in fact two of them) and a second $(2,2,1)_{\rm H}$ which is light (massless), is quite feasible within the D-Brane and F-theory constructions, as in Refs.~\cite{leonta-dbrane} and \cite{cvetic-Fth} respectively\footnote{I thank Mirjam Cvetic, Alon Faraggi, Arthur Hebecker, George Leontaris and Stuart Raby for discussions on these points.}. It, of course, still needs to be checked whether precisely the spectrum of Eq.~(\ref{eq:representation}), together with the three-generation MSSM spectrum and desired gross pattern of Yukawa couplings can be realized within such constructions or variants thereof. Note also that the spectrum of Eq.~(\ref{eq:representation}) is devoid of (1,1,6) and thereby of possibly dangerous color triplets which might induce rapid $d=5$ proton decay depending upon their Yukawa couplings. Such a spectrum in fact possesses some specially desirable features with no obvious problems, as noted below:

\emph{(a) Removal of Mismatch:}~The spectrum of Eq.~(\ref{eq:representation}) helps remove the mismatch between $M_{st}$ and $M_U$ as follows. The gauge couplings $g_{2L}, g_{2R}$ and $g_4$ of the $G(2,2,4)$ symmetry, unified
at the string scale $M_{\rm st}\approx 4\times 10^{17}\,$GeV, would run down to lower
energies with their respective $\beta$-functions. It may be verified that, in the
presence of the multiplets given by ${\rm H}_{\rm extra}^\prime$ and the MSSM-GUT spectrum, the
one-loop beta functions for $g_{2L}, g_{2R}$ and $g_4$ (including the gauge, Higgs and
matter contributions) are equal, with $b_{2L}=b_{2R}=b_4=6$. Since, for MSSM, the gauge
coupling $\alpha_U$ at the unification scale $M_U$ is small ${\approx}0.04$, we expect
that two-loop corrections will not alter the running between $M_{st}$ to $M_U$
significantly. Thus, the three gauge couplings of $G(2,2,4)$, unified at the
\hbox{string-scale,} will still remain unified while running down upto the conventional
MSSM GUT-scale $M_U \approx 2\times 10^{16}$\,GeV, where the symmetry $G(2,2,4)$ will
break to the SM symmetry by utilizing the VEVs of  $\{(1,2,4)_{\rm H}+(1,2,\overline{4})_{\rm H}\}$ and of $(1,1,15)_{\rm H}$
listed in the subset ${\rm H}_{\rm SSB}$. This resolves the mismatch between the MSSM-unification
at $M_U$ and the string-unification of $G(2,2,4)$ at $M_{\rm st}\approx 4\times
10^{17}$\,GeV.

\emph{(b) Preserving the Ratio $M_U/M_{st}$:}~I should add that of the four possible resolutions of the mismatch between $M_{st}$ and
$M_U$ for a string-unified $G(2,2,4)$ symmetry, mentioned above, the fourth one, needing just two extra
multiplets, beyond those needed for SSB ( see the second subset on the RHS of Eq.(8)), has a practical advantage. 
This is because, for the first three, one way or another, the ratio
$(M_U/M_{st})$ is near unity, {\it where as for the fourth one the conventional values of}
$M_U\approx 2\times 10^{16}$\,{\it GeV and} $M_{st}\approx 4\times 10^{17}$\,{\it GeV are
preserved}, and thereby their ratio of $M_U/M_{st}\approx 1/20$ as well. This latter
value of the ratio of $M_U/M_{st}$ clearly goes well with the hierarchical entries that
are needed for an understanding of the masses and mixings of the charged fermions
( see Sec.~5) and of the neutrinos ( see Sec.~6), that are realized through higher dimensional operators.

\emph{(c) Desired $\alpha_3(m_Z)$:}~There is an additional benefit of the extra multiplets listed above. It is well known
that with a ``typical'' SUSY spectrum at the electroweak scale, and without any GUT-scale
threshold corrections, the MSSM gauge couplings unify, with 2-loop running from the weak
to the GUT-scale, for $\alpha_3(m_Z)\approx 0.127$. This is somewhat higher than the
observed PDG average value of $\alpha_3(m_Z)_{\rm expt}=0.1176 \pm 0.002$ (see second
reference in Ref.~\cite{69}). Now, the one-loop GUT-scale threshold corrections, {\it
including contributions from the extra multiplets} shown in Eq. (8), can be calculated in terms of two
parameters: (i) the ratio of the two GUT-scale VEVs $\langle (1,2,\overline{4})_{\rm
H}\rangle/\langle (1,1,15)_{\rm H}\rangle$ and (ii) the ratio $M/M_U$, where $M$ denotes
an ``average'' mass of the superheavy GUT-scale particles, assumed to have a common mass,
for simplicity. One finds\cite{79} that, allowing for a reasonably wide range of
variation of these two ratios, the GUT-scale threshold correction to $\alpha_3(m_Z)$
ranges between~--- ($0.006-0.012$), which gives $\alpha_3(m_Z)$ within one standard
deviation of the observed value. This demonstrates that with the addition of just two
extra multiplets (i.e. one $(1,1,15)^\prime_{\rm H}$ and one $(2,2,1)^\prime_{\rm H}$) of an assumed
string-origin, given by ${\rm H}_{\rm extra}^\prime$, together with one $(1,1,15)_{\rm H}$ and the sub-sets A and B listed in
${\rm H}_{\rm SSB}$ that are needed to implement symmetry breaking, a string-unified
$G(2,2,4)$ symmetry can be fully consistent with the observed GUT-based MSSM gauge
coupling unification at $M_U\approx 2\times 10^{16}$\,GeV, with a clear benefit as
regards the predicted value of $\alpha_3(m_Z)$.

\emph{(d) Achieving Doublet-Triplet Splitting and Realizing the MSSM Spectrum below the GUT-Scale:}~ As noted above the spectrum
of Eq.(8) is devoid of the multiplet $(1,1,6)$, while it possesses two $(1,1,15)$'s. Without the $(1,1,6)$, the four dimensional theory is free from
the color-triplets and thus does not face the familiar doublet-triplet splitting problem. The latter could pose the danger of extra rapid d=5 proton decay,
even if the color-triplets have masses ${\sim} {\rm ( one~ to~ a~few)}\times {\rm M}_{\rm GUT}$, unless the Yukawa couplings of the
$(1,1,6)$ are adequately suppressed compared to those of the light $(2,2,1)_{\rm H}$. Now, many of the string solutions ( see e.g. the works of
T. Kobayashi et al. and B. Assel et al. in \cite{74} and those in Refs.~\cite{leonta-dbrane} and\cite{cvetic-Fth} ) do in fact possess the $(1,1,6)$-multiplet
in 4D and assign heavy GUT-scale mass to it, which is allowed. However, there can also exist solutions ( see e.g. the Appendix B.2 of Ref.\cite {leonta-dbrane}, and also
Ref.\cite {cvetic-Fth}, with the remarks there in), which are devoid of the (1,1,6) of $G(2,2,4)$ and thus of the color triplets in 4D.

I should add that the multiplet $(1,1,6)$ has in fact been utilized in these works to give GUT-scale masses to the $d^c$ and $\overline{d^c}$-like 
components lying within the multiplets $S_H\sim (1,2,4)_H$ and $\overline{S}_H\sim (1,2,\bar{4})_H$, which are 
used to break the symmetry $G(2,2,4)$ to the Standard Model. This is done by using a superpotential coupling of the form 
$$W\supset \left[S_H^2 . (1,1,6) + \overline{S}_H^2 . (1,1,6)\right]$$ 
and utilizing the GUT-scale VEVs of the RH sneutrino-like fields 
in $S_H$, and likewise in $\overline{S}_H$. This in turn leads to just the MSSM-like spectrum below the GUT-scale, as desired.
The same result can, however, be achieved by utilizing the multiplet $(1,1,15)$ of Eq. (8), instead of $(1,1,6)$, through a 
superpotential coupling of the form 
$$
W \supset S_H.(1,1,15).\overline{S}_H,
$$
and using the same VEVs as above. In short, the spectrum of Eq. (8), being present in 4D, would naturally 
provide a resolution of the doublet-triplet splitting problem through string-theoretic 
compactification in higher dimensions,.Simultaneously it would lead to to just the MSSM spectrum below the GUT-scale, 
which goes well with the observed gauge coupling unification.

In this sense, the spectrum of Eq. (8) accompanying three chiral families and a light bi-doublet $(2,2,1)_H$, appears to have definite 
advantages. It would thus be worth checking if precisely such a spectrum, and (ambitiously) with the desired gross pattern of Yukawa 
couplings, and supersymmetry breaking, can be derived as an allowed solution within a suitable string-theoretic construction.

 Returning to the task of a comparison between a non-GUT string-derived  symmetry versus a GUT-symmetry in 4D, 
 we see that the mismatch between  $M_{st}$ and $M_U$ can thus quite plausibly be removed ( as in cases (i)-(iii)), or accounted for (as in case (iv)), for a non-GUT string-derived
symmetry like $G(2,2,4)$. At the same time, a SUSY GUT symmetry like SU(5) or SO(10) in 4D would have an advantage in
this regard because it is guaranteed to keep the gauge couplings together between
$M_{st}$ and $M_{U}$ (even if $M_{U}\sim M_{st}/20$), and thus not even encounter the
problem of a mismatch between the two scales. A supersymmetric GUT-solution (like SU(5)
or SO(10)), however, has a possible disadvantage as well, because it needs certain color
triplets to become superheavy by the so-called double-triplet splitting mechanism (see
Ref.~\cite{25} and references therein and discussion in Sec.~7), in order to avoid the
problem of rapid proton decay. However, no such mechanism has emerged yet, in string
theory, for the GUT-like solutions \cite{75}. Four-dimensional SUSY SO(10) models possessing technically natural
and stable doublet-triplet splitting have been constructed \cite{25}, and I will discuss their consequences 
in Sec.~7. I may add, however, that they are not so simple, and one may wonder if such a mechanism can arise  from an underlying theory, like string theory.

Non-GUT string solutions, based on symmetries like $G(2,2,4)$ or $G(2,1,4)$ for example, can
have a distinct advantage in this regard, in that the dangerous color triplets, which
would induce rapid proton decay, can be naturally projected out for such solutions
through string compactification , see e.g. \cite{80,81}. This is also feasible within the D-brane ( see Appendix B.2 of Ref.\cite{leonta-dbrane}) and very likely within F-theory
constructions (see remarks in Ref.\cite{cvetic-Fth}. Furthermore, the non-GUT solutions
invariably possess new ``flavor'' gauge symmetries, which distinguish between families.
These symmetries are immensely helpful in explaining qualitatively the observed fermion
mass-hierarchy (see e.g. Ref.~\cite{81}) and resolving the so-called naturalness
problems of supersymmetry such as those pertaining to the issues of
squark-degeneracy \cite{82}, and quantum gravity-induced rapid proton decay \cite{83}.

Weighing the advantages and possible disadvantages of both, it seems hard at present to
make a priori a clear choice between a presumed SUSY GUT (like SO(10)) versus a non-GUT (like
$G(2,2,4)$) string-solution. As expressed elsewhere \cite{four}, it therefore seems prudent
to keep both options open and pursue their phenomenological consequences. Now, the
advantages of an effective $G(2,2,4)$ or SO(10) symmetry in understanding masses and
mixings of all fermions including neutrinos turn out to be essentially identical,
especially if one uses low-dimensional Higgs multiplets (as opposed to large-dimensional
ones) to break SO(10) or $G(2,2,4)$ to the SM in 4D (See discussion in Secs.~4--6 and
Ref.~\cite{24}). I will thus proceed by assuming that either a suitable
$G(2,2,4)$-solution with a mechanism to resolve the mismatch between $M_U$ and $M_{\rm
string}$ of the sort mentioned above (most preferably the fourth one), or a realistic
SO(10)-solution with the needed doublet-triplet mechanism, will emerge from string
theory. As we will see in Sec.~7, a study of proton decay can help distinguish between
these two alternatives. 

Before discussing in more detail the consequences of an intact SUSY $SO(10)$ or a string-unified $G(2,2,4)$-symmetry in 4D, 
I should mention briefly an interesting line of attempt which proposes to achieve unification through higher-dimensional 
(D=5 or 6) orbifold GUT-models based on SUSY SU(5)~\cite{Kawamura} or SO(10)~\cite{Asaka}.

\noindent{\bf D=5 or 6 Orbifold GUT-Models:} 
These models are motivated, on the one hand, by the successes of SUSY grand unification 
in 4D (as listed, for example, in (i)-(x) in Sec. 2 for the $G(2,2,4)/SO(10)$ case), as well as by its shortcomings in the Higgs-sector, 
such as the problem of doublet-triplet splitting, which is relevant to the case of an intact SUSY $SU(5)$ or $SO(10)$ in 4D.
On the other hand, they have also been inspired by the promising structure of orbifold compactification~\cite{Dixon} of the 
ten-dimensional heterotic string-theory~\cite{Gross}, with the inclusion of background fields, such as the Wilson lines~\cite{Wilson}.

Combining some of the virtues of both sets of ideas, the higher dimensional orbifold GUT-models~\cite{Kawamura,Asaka} assume that 
a SUSY GUT symmetry, like $SU(5)$ or $SO(10)$, is effective as a point-particle field theory in a higher-dimensional space-time 
(D=5 for the $SU(5)$-case~\cite{Kawamura} and D=5 or 6 for the $SO(10)$-case~\cite{Asaka}). Such a symmetry is presumed to break down 
either to the SM or to an $U(1)_X$-extended SM, with broken supersymmetry, operating in 4D (the latter  
would be applicable if the parent symmetry in 6D is $SO(10)$), through compactification on an orbifold of the extra spatial dimension(s), 
subject to a choice(s) of suitable GUT-symmetry breaking boundary conditions. Assuming that such a choice(s) would be allowed by an underlying 
theory, these models have the merit that they can achieve: (a) gauge coupling unification at least in the leading order, (b) desired GUT-symmetry 
breaking, as well as (c) doublet-triplet splitting through compactification, without involving the Higgs mechanism of 4D GUTs.

Predictions on proton decay in these orbifold GUT-models, which have been studied better only for the $SU(5)$-case, 
vary widely as regards the lifetime and branching ratios of different decay 
modes, depending upon the choice of location of the multiplets and of their orbifold parities. For the $SU(5)$-case, the choice includes the possibility of 
separate locations of the $\bar{5}$ and $10$ of a single $SU(5)$-family. These can be either on the GUT-symmetry-breaking or symmetry -preserving branes, or 
in the bulk (which is symmetry-preserving). As a generic feature, however, all sources of $d=4$ and $d=5$ proton decay operators (including the 
color-triplet Higgsino-exchange contribution) are forbidden in these models.

Now, depending upon the choice of locations 
and boundary conditions (including parities) as mentioned above, there are several possibilities for contributions from the effective 
$(X,Y)$ gauge-boson mediated $d=6$ proton decay operators, especially for the D=5 $SU(5)$-case, which is better studied. These operators can lead to either (a) suppressed or even forbidden $d=6$ 
proton decay (see e.g. Altarelli and Feruglio in~\cite{Kawamura}); or (b) to $d=6$ proton decays with estimated lifetimes $\sim 10^{34}$ years (subject to large uncertainties), 
with a rather distinctive flavor structure that the branching ratios for proton decaying via $e^+ \pi^0$, $\mu^+\pi^0$, 
$e^+ K^0$, $\mu^+ K^0$, $\bar\nu \pi^+$ and $\bar\nu K^+$ can all be comparable (see e.g. the third paper by Hall and Nomura in~\cite{Kawamura}); 
or (c) to $d=6$ proton decays with an estimated lifetime $\sim 10^{35}$ years (for a compactification scale $M_c\sim 10^{14}$ GeV) with the 
novel feature that the $\mu^+ K^0$ and $\bar{\nu}_\mu K^+$ decay modes of the proton are dominant (see e.g. the second paper by 
Hebecker and March-Russell in~\cite{Kawamura}). Clearly the flavor structures of (b) and (c) are very distinct from those of the 
predictions of SUSY $SU(5)$ and $SO(10)$ models in D=4 (see Sec. 7).

Given the merits of these higher-dimensional orbifold GUT-models as noted above, and their possible 
distinctive signatures in proton decay (especially for the $SU(5)$-case), they seem to provide a viable 
path to higher unification which is worth pursuing. Noting that, as point-particle field theories in 5D or 6D, they are non-renormalizable, 
they should  be viewed as effective theories needing an ultraviolet completion. This poses the question as to whether such an effective 
D=5 or D=6 orbifold GUT-model can be derived, along with the desired boundary conditions, from an underlying theory, in particular the 
superstring theory. The task here is similar in spirit to that of deriving, for example, a string-unified $G(2,2,4)$ symmetry, with the desired Higgs 
multiplets in 4D to achieve symmetry-breaking, but without the problem of doublet-triplet splitting in 4D. As discussed in the preceding, there are promising 
semi-realistic string-solutions of the latter type (see e.g.~\cite{leonta-dbrane,cvetic-Fth} and remarks there in). Deriving such a string solution corresponding to a 
suitable 5D or 6D orbifold GUT-model with the desired boundary conditions remains a challenge ( A promising attempt in this regard has in fact been made by T. Kobayashi {\it et al.},
referenced in~\cite{74}).

Before returning to the main aspects of my talk, I should mention that, despite their merits as mentioned above, the D=5 orbifold $SU(5)$ models \cite{Kawamura}, 
 and even the D=6 orbifold $SO(10)$ models, that rely entirely on orbifold boundary conditions to break $SO(10)$ to an $U(1)_X$ - extended SM symmetry in 4D (see e.g. the first two papers of Ref.~\cite{Asaka}), suffer from a drawback in that they do not seem to provide a natural understanding of the mass-scales of the observed neutrino oscillations. ( By contrast, this is readily possible within a SUSY $SO(10)$ or a suitable string-unified $G(2,2,4)$ symmetry of the type mentioned above, in 4D ( see Sec.~6)) 
 To be specific, for the D=5 $SU(5)$-case, the difficulty in this regard is similar to that of $SU(5)$ in D=4. That is: (i) the three RH neutrinos have to be added by hand as singlets; but (ii) without the (B-L) local symmetry in $SU(5)$, there is no good control over their Majorana masses; and (iii) without 
 the $SU(4)$-color-relation between the Dirac mass of $\nu_{\tau}$ and the mass of the top quark (see Eq.~(6)), the former can vary widely. 
 As a result, the seesaw-mass for the corresponding light neutrino, depending crucially on the last two features, becomes uncertain by more than five  orders of magnitude ( see discussion in Sec.~6). Now, one might have expected that the D=6 orbifold $SO(10)$ models to do better in this regard. But, as noted in the second paper of Ref.~\cite{Asaka},
generating suitable Majorana masses for the RH neutrinos does not seem to go well, for this case,with an understanding of the weak angle, together with the removal of anomalies of the 6D bulk.

There exists an alternative line of attempt based on D=5 orbifold $SO(10)$ models ( see the last three papers in Ref.~\cite{Asaka}), which combine orbifold boundary conditions to achieve a partial breaking of $SO(10)$, for example to the symmetry $G(2,2,4)$, together with the Higgs mechanism on the brane to 
break $G(2,2,4)$ to the SM symmetry in 4D. This class of orbifold $SO(10)$ models fare better by providing at least a gross understanding of the mass-
scales of neutrino oscillations, to within factors of 10 to 100, depending upon the compactification scale ${\rm M}_{\rm c}$ (see e.g. the fourth and fifth papers of 
Ref.~\cite{Asaka}).\footnote{I thank Radovan Dermisek for a communication in this regard.}
Thus the idea of combining the orbifold boundary conditions with the Higgs mechanism to break $SO(10)$ in 5D to the SM symmetry in 4D, via the intermediary of 
 the $G(2,2,4)$ symmetry, seems  to be promising.

For comparison, it is worth noting that this alternative approach to symmetry-breaking within 5D $SO(10)$ models, that combines the orbiflod boundary conditions with the Higgs mechanism to complete the task,
 is in fact quite analogous in its spirit to the attempts within string theory that aim to derive a string-unified $G(2,2,4)$ symmetry in 4D through compactification of the ten-dimensional string theory and break the $G(2,2,4)$ symmetry through the Higgs mechanism in 4D to the SM at the GUT-scale. As discussed in the beginning of this sub-section, 
semi-realistic string-unified $G(2,2,4)$ -solutions do exist, with the desired Higgs system to break the $G(2,2,4)$ symmetry in 4D to the SM, some of which can be free from the doublet-triplet spltting problem ( like the orbifold GUT models in 5D or 6D). Such a string-unified 
 $G(2,2,4)$-symmetry, as also a SUSY $SO(10)$ model, both in 4D, have a clear advantage of providing a natural understanding of the mass-scales of neutrino oscillations 
 ( see Sec.~6). Furthermore, subject to the assumption of a suitable flavor symmetry ( which may have a string-origin), they also provide a predictive framework for understanding at least the gross pattern and some intriguing features of the masses and mixings of all fermions including neutrinos ( see discussion in Sec.~5).

Returning now to the main task of my talk, I next discuss the spontaneous breaking of SO(10) or $G(2,2,4)$
to the SM, which will be relevant to our understanding of the masses and mixings of all
fermions including neutrinos (Secs.~5 and 6).

\section{Breaking SO(10) or $G(2,2,4)$ to the SM}

For compactness of notation, I will present the Higgs system that would be responsible
for breaking SO(10) to the SM. A completely analogous set of Higgs multiplets can be used
to break $G(2,2,4)$ to the SM. Now two distinct sets of multiplets have been used to
break SO(10) to the SM, each having some advantages over the other.

One of these, which I will follow in the rest of my talk, utilizes a minimal set of
low-dimensional Higgs multiplets (LOH)\cite{84,85,86,87,88,89,90} to break a
supersymmetric SO(10) symmetry to ${\rm SU}(3)^{\rm c} \times {\rm U}(1)_{\rm cm}$. The
same set is also used to generate a predictive framework for understanding masses and
mixings of all fermions including neutrinos ( see e.g. \cite{24,93,94}). Minimally, it consists of the
multiplets:
\begin{eqnarray}
{\rm H}_{\rm LOH} = \{10_{\rm H}, 16_{\rm H}, \overline{16}_{\rm H}, 45_{\rm H}, S_{\rm
H}\}
\end{eqnarray}
The 16$_{\rm H}$ is paired with $\overline{16}_{\rm H}$ to preserve supersymmetry at the
GUT-scale. Of these, the VEV of $\langle 45_{\rm H}\rangle \sim {\rm M}_{\rm U}$ in the
(B-L)-direction breaks SO(10) to $G(2,2,1,3)$ and those of $\langle 16_{\rm
H}\rangle=\langle \overline{16}_{\rm H}\rangle \sim {\rm M}_{\rm U}$ along the $\langle
\tilde{\overline{\nu}}_{\rm RH}\rangle$ and $\langle\tilde{\nu}_{\rm
RH}\rangle$-components break $G(2,2,1,3)$ to the SM $G(2,1,3)$ at the unification-scale
${\rm M}_{\rm U}$. Now $G(2,1,3)$ breaks at the EW scale by the VEV of $\langle 10_{\rm
H}\rangle$ and that of the EW doublet of the down-type in $\langle 16_{\rm H}^d\rangle$
to ${\rm SU}(3)^{\rm c} \times {\rm U}(1)_{\rm em}$. The singlet $S$, with a VEV $\sim$
M$_{\rm U}$, is added to serve as a flavon field carrying, in the simplest case, a gauged
${\rm U}(1)$-flavor charge \cite{91}. It serves to generate a hierarchical pattern of
fermion masses and mixings of the type envisaged in Ref.~\cite{24}. This will be
discussed briefly in the following section.

Before entering into the role of the low-dimensional Higgs system in describing masses
and mixings of fermions, I should mention a notable alternative, which has been used
widely in the literature (for an incomplete list of references see
Ref.~\cite{92}). It
utilizes a large-dimensional Higgs system (LGH) consisting of SO(10)-tensorial multiplets
like $\{126_{\rm H}, \overline{126}_{\rm H}, 210_{\rm H}$ and possibly $120_{\rm H}$
and/or 54$_{\rm H}\}$. It is useful to note the decomposition of the SO(10)-multiplet
$126_{\rm H} = \{\Delta_{\rm L}=(3_{\rm L}, 1_{\rm R}, 10^{\rm c})+
\overline{\Delta}_{\rm R}=(1_{\rm L}, 3_{\rm R}, \overline{10}^{\rm c}) + (2_{\rm L},
2_{\rm R}, 15^{\rm c})+(1,1,\overline{6}^{\rm c})\}$ under the subgroup $G(2,2,4)$.

One advantage of $\overline{126}_{\rm H}$ is that it has a renormalizable Yukawa coupling
of the form $h_{\it ij}\ 16_i\ 16_j\ \overline{126}_{\rm H}$ (and likewise also 120$_{\rm
H}$ for $i\neq j$), where 16$_i$ (with $i=1,2,3$) denotes the three families. The VEV of
$\langle \Delta_R\rangle \sim {\rm M}_{\rm U}$ (with $\langle\Delta_{\rm L}\rangle
\approx {\rm ``0}$'') breaks B-L by two units and provides superheavy Majorana masses to
the RH neutrinos, which are needed to implement the Type~I seesaw mechanism. Such a
breaking of B-L automatically preserves the familiar ${\rm R\hbox{-}parity} =
(-1)^{3({\rm B-L})+2{\rm S}}$ of SUSY theories, and thereby avoids the dangerous $d=4$
proton-decay operators and also yields a stable LSP to serve as cold dark matter (CDM).
By contrast, 16$_{\rm H}$ and $\overline{16}_{\rm H}$ of the low-dimensional Higgs system
break B-L by one unit and thereby break the familiar R-parity. This difference is,
however, not significant in practice, because for LOH one can still define consistently a
matter-parity (under which 16$_i$ is odd, but all other fields exhibited in Eq.~(9) are
even), which fully serves the desired purpose by allowing all the desired interactions
but forbidding the dangerous $d=4$ proton decay operators and yielding a stable LSP to
serve   as CDM.

Now both the LGH \cite{92} and the LOH systems (see e.g. Refs.~\cite{24}, \cite{91},
\cite{93} and \cite{94}), the latter often with the assumption of a suitable flavor
symmetry, have been developed to yield predictive and successful frameworks for
describing masses and mixings of all fermions including neutrinos. An example of this
kind (Ref.~\cite{24}) for the case of LOH system will be presented briefly in the
following section.

While both the LGH and the LOH systems seem to have some merits and are worth pursuing, I
will base my discussion in the following on masses and mixings of all fermions including
neutrinos (Secs.~5 and 6) and on proton decay (Sec.~7) by confining to the LOH system
only. Briefly, the reasons are as follows: (i) There exists, for the case of SUSY SO(10),
a simple mechanism for a natural\cite{84,85} and stable\cite{25} doublet-triplet
splitting in the case of LOH, which thus avoids rapid $d=5$ proton decay without any fine
tuning. It is not clear whether such a situation can emerge for LGH. (ii) The
large-dimensional Higgs multiplets (like $126_{\rm H}, \overline{126}_{\rm H}, 210_{\rm
H}$ and possibly 120$_{\rm H}$) tend to give rather large and
differing GUT-scale threshold corrections (exceeding even 30 to
50\%) to the three gauge couplings from the split \hbox{sub-multiplets.} Such corrections
are more controlled for the case of LOH. \hbox{Incorporating} stable doublet-triplet
splitting, the latter tends to provide sharper predictions for proton lifetimes including
upper limits for a given SUSY spectrum\cite{25} (see Sec.~7). (iii) The models based on
LOH (see e.g. \cite{24}, \cite{91}) predict that the heaviest of the three light
neutrinos has a mass that is naturally of order 1/10\,eV, their masses being hierarchical
(see Sec.~6). This is in good agreement with the atmospheric neutrino oscillation data.
(iv) For the LGH system with multiplets like $126_{\rm H}, \overline{126}_{\rm H}$ and
others, the unified gauge coupling blows up rapidly just above the GUT-scale $M_U$,
making the physics between string/Planck-scale and GUT scale unclear. Finally, (v) while
our understanding of the non-perturbative aspects of string theory is still lacking, the
weakly interacting heterotic string theory solutions do not seem to yield the
large-dimensional tensorial multiplets like 126$_{\rm H}$ and 120$_{\rm H}$, but they do
yield the low-dimensional ones as listed in Eq.~(9) \cite{95}.

With these distinctions in mind, I present in the next section an attempt to understand
the masses and mixings of all fermions including neutrinos within the
SO(10)/$G(2,2,4)$-framework by using the low-dimensional Higgs (LOH) system as in
Eq.~(6).


\section{Masses and Mixings of Fermions in SO(10) or $G(2,2,4)$}

There are some distinct advantages of describing the masses and mixings of the three
families, including the neutrinos, within a symmetry like SO(10). Essentially the same
holds for the case of a string-unified $G(2,2,4)$-symmetry as well,
\hbox{especially} if one
confines to the low-dimensional Higgs system (LOH), as in Eq.~(9). In either case, owing
to quark-lepton unification through SU(4)-color and up-down correlation through ${\rm
SU}(2)_{\rm L}\times {\rm SU}(2)_{\rm R}$, the four mass-matrices ${\rm M}_{\rm U}, {\rm
M}_{\rm D}, {\rm M}_l$ and ${\rm M}_{\nu}^{\rm D}$ (the last one denotes the Dirac mass-matrix of
the neutrinos) get interrelated (see e.g. Ref.~\cite{24}). This reduces the number of
independent parameters significantly and thereby increases predictivity. An example of
such a correlation is given by Eqs.~(5) and (6), of which the first one is known to go well with
observations, while the second one plays an
important role in ensuring the success of the seesaw (see Sec.~6). Some of these
relations, like Eq.~(5), hold for ${\rm SU}(5)$ as well, but not Eq.~ (6).

In what follows (for reasons noted in the previous section), I will limit my discussion
to using only the minimal low-dimensional Higgs system (LOH) for generating the masses
and mixings of all fermions including neutrinos (Dirac as well as Majorana) for the case
of a supersymmetric SO(10) model. The Higgs multiplets in this case are given by the set
(see Eq.~(9)):\setcounter{equation}{8}
\begin{eqnarray}
{\rm H}_{\rm LOH} = \{10_{\rm H}, 16_{\rm H}, \overline{16}_{\rm H}, 45_{\rm H}, {\rm
S}_{\rm H}\}
\end{eqnarray}

\noindent An analogous Higgs system consisting of the multiplets [$(2,2,1)_{\rm H}
\{(2,1,4)_{\rm H}+(1,2,\overline{4})_{\rm H}\}, \{(2,1,\overline{4})_{\rm H}+(1,2,4)_{\rm
H}\}, (1,1,15)_{\rm H}, {\rm S}_{\rm H}$] can be used for the case of
the supersymmetric
$G(2,2,4)$ symmetry both to break $G(2,2,4)$ to ${\rm SU}(3)^{\rm c}\times {\rm
U}(1)_{\rm cm}$ and to generate a desirable and identical pattern for the masses and
mixings of the fermions as in the case of ${\rm SO}(10)$.

To illustrate the advantages alluded to above, let me present briefly the supersymmetric
SO(10)-framework developed in Ref.~\cite{24} (elaborated further in \cite{91}) for treating
the masses and mixings of the fermions within the minimal low-dimensional Higgs system
 (Eq.~(6)).\footnote{Alternative SO(10)-models of fermion masses and mixings based on
 the low-dimensional Higgs system have been developed with some overlapping features and
some distinctions (see e.g. Refs.~\cite{93} and \cite{94}).} 
The question
that one faces in the very beginning is this: can this minimal Higgs system provide a
realistic pattern for fermion masses and mixings?  Now $\mathbf{10_{H}}$ (even several
$\mathbf{10}$'s) cannot provide certain desirable features~--- i.e. family-antisymmetry
and (B-L)-dependence in the mass matrices --- which are, however, needed to suppress
$V_{cb}$ while enhancing $\theta^{\nu}_{23}$ on the one hand, and accounting for features
such as \(m_{\mu}^{0} \neq m_{s}^{0}\) on the other hand. Furthermore, a single
$\mathbf{10_{H}}$ cannot generate CKM mixings.  At the same time, $\mathbf{10_{H}}$ is
the only multiplet among the ones in the minimal Higgs system (Eq.~(6)) which can have
cubic couplings with the matter fermions which are in the $\mathbf{16}$'s. This apparent
impasse disappears as soon as one allows for not only cubic but also effective
non-renormalizable quartic and even higher dimensional couplings of the minimal set of
Higgs fields with the fermions.  Such effective couplings can of course arise quite
naturally through exchanges of superheavy states (e.g. those in the string-tower or those
having GUT-scale masses $\sim$ a few ${\rm M}_{\rm GUT}$ (say)) involving renormalizable
couplings. And importantly, all such effective couplings that are allowed by the symmetry
relevant to the `lower' energy (including SO(10) and possible flavor symmetries, see for
example\cite{91}) are expected to arise at least through quantum gravity effects, which
would be scaled by ${\rm M}_{\rm string}$ or ${\rm M}_{\rm planck}$.

\vspace*{2pt} The $3\times3$ Dirac masses matrices for the four sectors ($u$, $d$, $l$,
$\nu$) proposed in Ref.~\cite{24} are motivated in part by the group theory of
$SO(10)/G(2,2,4)$, which severely restricts the effective cubic and quartic couplings
(and thus the associated mass-patterns), for the minimal Higgs system. They are also
motivated in part by the notion that flavor symmetries\cite{96} distinguishing between
the three families lead to a hierarchical pattern for the mass matrices (i.e. with the
element ``33'' $\gg$ ``23'' $\gg$ ``22'' $\gg$ ``12'' $\gg$ ``11'' etc.), \emph{so that
the lighter family gets its mass primarily through its mixing with the heavier ones}.
Such a hierarchical pattern can in fact be realized by introducing (for example) just a
single gauged U(1)-flavor symmetry and an SO(10)-singlet flavon field S, together with an
appropriate assignment of the flavor charges to all the fields \cite{91}. Subject to the
constraints as mentioned above, the effective Yukawa couplings turn out to be rather
unique. Confining to the relevant lowest dimensional terms, they are given
by\cite{24,91}{:}%
 
\vspace*{-14pt}{\small
\begin{eqnarray}
&&\lefteqn{\mathcal{L}_{\mathrm{Yuk}} = h_{33}{\bf 16}_3{\bf 16}_3{\bf 10}_H}
\nonumber \\
& &\quad +\left[ h_{23}{\bf 16}_2{\bf 16}_3{\bf 10}_H(S/M)+a_{23}{\bf 16}_2{\bf
16}_3{\bf 10}_H ({\bf 45}_H/M')(S/M)^p\right.\nonumber\\
\
&&\quad \left.+\,g_{23}{\bf 16}_2{\bf 16}_3{\bf 16}_H^d
({\bf 16}_H/M'')(S/M)^q\right] \nonumber \\
& &\quad +\left[h_{22}{\bf 16}_2{\bf 16}_2{\bf 10}_H(S/M)^2+g_{22}{\bf 16}_2{\bf 16}_2
{\bf 16}_H^d({\bf 16}_H/M'')(S/M)^{q+1} \right] \nonumber \\
& & \quad + \left[g_{12}{\bf 16}_1{\bf 16}_2 {\bf 16}_H^d({\bf 16}_H/M'')(S/M)^{q+2}+
a_{12}{\bf 16}_1{\bf 16}_2 {\bf 10}_H({\bf 45}_H/M')(S/M)^{p+2} \right]
\label{eq:Yuk}\nonumber\\
\end{eqnarray}
}%

\vspace*{-3pt}

Typically we expect $M^{\prime}$, $M^{\prime\prime}$ and $M$ to be of order $M_{\rm
string}$\cite{97}.The VEV's of $\langle{\bf 45}_H\rangle$ (along B-L), $\langle{\bf
16}_H\rangle=\langle{\bf\bar {16}}_H\rangle$ (along standard model singlet sneutrino-like
component) and of the SO(10)-singlet $\langle S \rangle$ are of the GUT-scale, while
those of ${\bf 10}_H$ and of the down type SU(2)$_L$-doublet component in ${\bf 16}_H$
(denoted by ${\bf 16}_H^d$) are of the electroweak scale \cite{24,98}. Depending upon
whether $M^{\prime}(M^{\prime\prime})\sim M_{\rm GUT}$ or $M_{\rm string}$ (see comment
in Ref.~\cite{97}), the exponent $p(q)$ is either one or zero \cite{99}.

With the effective Yukawa couplings given in Eq.~(10), the Dirac mass matrices of quarks
and leptons of the three families, at the unification scale, take the form
(Ref.~\cite{24}):\footnote{The zeros in ``11'', ``13'', ``31'', and ``22'' elements
signify that they are relatively small.  For instance, the ``22''-elements are set to
zero because (restricted by flavor symmetries, see below), they are meant to be less than
(``23'')(``32'')/``33'' $\sim 10^{-2}$, and thus unimportant for our purposes. Likewise,
for the other ``zeros.''}
\begin{eqnarray}
\label{eq:mat}
\begin{array}{cc}
M_u=\left[
\begin{array}{ccc}
0&\epsilon'&0\\-\epsilon'&0&\sigma+\epsilon\\0&\sigma-\epsilon&1
\end{array}\right]{\cal M}_u^0;&
M_d=\left[
\begin{array}{ccc}
0&\eta'+\epsilon'&0\\ \eta'-\epsilon'&0&\eta+\epsilon\\0& \eta-\epsilon&1
\end{array}\right]{\cal M}_d^0\\
&\\
M_\nu^D=\left[
\begin{array}{ccc}
0&-3\epsilon'&0\\-3\epsilon'&0&\sigma-3\epsilon\\
0&\sigma+3\epsilon&1\end{array}\right]{\cal M}_u^0;& M_l=\left[
\begin{array}{ccc}
0&\eta'-3\epsilon'&0\\ \eta'+3\epsilon'&0&\eta-3\epsilon\\0& \eta+3\epsilon&1
\end{array}\right]{\cal M}_d^0\\
\end{array}
\end{eqnarray}

\noindent These matrices are defined in the gauge basis and are multiplied by
$\bar\Psi_L$ on left and $\Psi_R$ on right. For instance, the row and column indices of
$M_u$ are given by $(\bar u_L, \bar c_L, \bar t_L)$ and $(u_R, c_R, t_R)$ respectively.
{\it Note the group-theoretic up-down and quark-lepton correlations}: the same $\sigma$
occurs in $M_u$ and $M_\nu^D$, and the same $\eta$ occurs in $M_d$ and $M_l$. It will
become clear that the $\epsilon$ and $\epsilon'$ entries, arising through the coupling of
($10_{\rm H}\cdot 45_{\rm H}$) in the $a_{23}$ and $a_{12}$-terms, are proportional to
B-L and are antisymmetric in the family space (as shown above). Thus, the same $\epsilon$
and $\epsilon'$ occur in both ($M_u$ and $M_d$) and also in ($M_\nu^D$ and $M_l$), but
$\epsilon\rightarrow -3\epsilon$ and $\epsilon'\rightarrow -3\epsilon'$ as $q\rightarrow
l$. Such correlations result in an enormous reduction of parameters and thus in increased
predictivity. Furthermore, it is precisely because of this family-antisymmetric
(B-L)-dependent contribution that one obtains a group theoretic understanding of why
$V_{\rm cb}$ is small while $\theta^{\nu}_{23}$ is large. Although the entries $\sigma$,
$\eta$, $\epsilon$, $\eta'$, and $\epsilon'$ will be treated as parameters, consistent
with assignment of flavor-symmetry charges (see below), we would expect them to be
hierarchical with \((\sigma, \eta, \epsilon) \sim 1/10\) and (\(\eta', \epsilon')\ \sim
10^{-3}-10^{-4}\) (say).

The entries 1 and $\sigma$ arise respectively from $h_{33}$ and $h_{23}$ couplings, while
$\hat\eta\equiv\eta-\sigma$ and $\eta'$ arise respectively from $g_{23}$ and
$g_{12}$-couplings. As mentioned above, the (B-L)-dependent antisymmetric entries
$\epsilon$ and $\epsilon'$ arise respectively from the $a_{23}$ and $a_{12}$ couplings.
This is because, with $\langle{\bf 45}_H\rangle\propto$ B-L, the product ${\bf
10}_H\times{\bf 45}_H$ contributes as a {\bf 120}, whose coupling is
family-antisymmetric.  Thus, for the minimal Higgs system (see Eq.~(6)), (B-L)-dependence
can enter only through family off-diagonal couplings of
\(\mathbf{10_{H}}\,{\cdot}\,\mathbf{45_{H}}\) as in $a_{23}$ and $a_{12}$-terms.
\emph{Thus, for such a system, the diagonal ``33'' entries are necessarily
(B-L)-independent (as shown in Eq.~(8)). This in turn makes the relations like
\(m_{b}(M_{X}) \approx m_{\tau}\) (barring corrections of order $\epsilon^{2}$, which
turn out to have the right sign and magnitude (see Ref.~\cite{24}) robust}. This
feature would, however, be absent if one had used
\(\overline{\mathbf{126}}_{\mathbf{H}}\), whose coupling is family-symmetric and can give
(B-L) dependent contributions to the   ``33''-elements.

As alluded to above, such a hierarchical form of the mass-matrices, with $h_{33}$-term
being dominant, is attributed in part to flavor gauge symmetry(ies) that distinguishes
between the three families,\cite{91} and in part to higher dimensional operators
involving for example $\langle{\bf 45}_H\rangle/M'$ or $\langle{\bf 16}_H\rangle/M''$,
which are supressed by $M_{\rm GUT}/M_{\rm string}\sim 1/10$, if $M'$ and/or $M''\sim
M_{\rm string}$.

To discuss the neutrino sector one must specify the Majorana mass-matrix of the RH
neutrinos as well. As in the case of effective Yukawa couplings giving Dirac masses, this would arise through effective couplings
of the form\cite{101}:\footnote{There is an impression, sometimes conveyed in the
literature, that a multiplet like $\overline{126}_{\rm H}$ of SO(10) (or equivalently
$(1,3_{\rm R},10)_{\rm H}$ of $G(2,2,4)$) is {\it needed} to give Majorana masses to the
RH neutrinos. While $\overline{126}_{\rm H}$ would contribute through renormalizable
Yukawa couplings, as explained before for the case of the low-dimensional Higgs multiplets exhibited in Eq.~(9), the effective
non-renormalizable interactions, allowed by all relevant symmetries, should be expected
to arise through exchange of heavy states in the string-tower (see Ref.~\cite{101}),
and at least through quantum gravity. Thus $\overline{126}_{\rm H}$ is not really needed
for the purpose. The virtue of Eq.~(9) in understanding the mass-scale of atmospheric
neutrino oscillation, owing to the suppression of the operator by a factor
${\sim}(M^2_{\rm GUT}/M_{\rm st})$ is discussed   in
Sec.~6.}
\begin{eqnarray}
\label{eq:LMaj} {\cal L}_{\rm Maj}=f_{ij}{\bf 16}_i{\bf
16}_j{\bf\bar{16}}_H{\bf\bar{16}}_H/M
\end{eqnarray}
where the $f_{ij}$'s include appropriate powers of $\langle S \rangle/M$, in accord with
flavor symmetry. For the $f_{33}$-term to be leading, being ${\sim}1$, we must assign the
flavor-charge of $\overline{16}_{\rm H}$ to be $-a$ (see Ref.~\cite{100}). This leads
to a hierarchical form for the Majorana mass-matrix\cite{24}:
\begin{eqnarray}
\label{eq:MajMM} M_R^\nu=\left[
\begin{array}{c@{\quad}c@{\quad}c}
x&0&z\\0&0&y\\z&y&1
\end{array}
\right]M_R
\end{eqnarray}

\noindent Following the flavor-charge assignments given in Ref.~\cite{100}, we expect
\(|y|\sim \langle S/M\rangle\sim 1/10\), \(|z|\sim (\langle S/M\rangle)^2\sim 10^{-2}(1
\mbox{ to } 1/2)\), \(|x|\sim (\langle S/M\rangle)^4\sim (10^{-4}$-$10^{-5})\) (say). The
``22'' element (not shown) is $\sim (\langle S/M\rangle)^2$ and its magnitude is taken to
be $< |y^2/3|$, while the ``12'' element (not shown) is $\sim (\langle S/M\rangle)^3$.

As it turns out, the mass-scale of the heaviest of the three light neutrinos ($m(\nu_3)$)
that would emerge from Eqs.~(11) and (13) and would correspond (because of the
hierarchical pattern of the masses) to the mass-scale of the atmospheric neutrino
oscillation, would have an important implication for physics at truly high energies. I
would therefore discuss a derivation of this mass-scale and its implications separately
in the next section. But for now, it would be interesting to note the prediction of the
simple patterns of the mass-matrices given in Eqs.~(11) and (13) for the gross features of
the masses and mixings of quarks and leptons including the neutrinos, by stating some
features of the neutrino-sector in advance   of Sec.~6.

To make predictions (postdictions), one needs to determine the parameters appearing in
Eqs.~(11) and (13). For the purposes of this talk, I will ignore possible phases in these
parameters and thus CP violations, as was done in Ref.~\cite{24}. The patterns of
mass-matrices suggested in Ref.~\cite{24} (i.e. in Eqs.~(11) and (13)) has, however,
been studied subsequently by allowing for phases in the parameters (which could arise
from complex effective couplings and/or VEVs) in Ref.~\cite{102}. It was found (rather
remarkably) in this work that the observed CP violation of the CKM form with the desired
Wolfenstein paramters can be realized within the pattern of mass-matrices of
Ref.~\cite{24}, while preserving its successes.

Turning to the case of real parameters for simplicity, thereby ignoring CP violation, the
parameters $(\sigma,\eta, \epsilon, \epsilon',\eta', {\cal M}_u^0, {\cal M}_D^0,\mbox{
and } y)$ can be determined by using, for example, $m_t^{\rm phys}=174$ GeV,
$m_c(m_c)=1.37$\,GeV, $m_S(1\,\mbox{GeV})=110$--116\,MeV, $m_u(1\,\mbox{GeV})=6$\,MeV,
the observed masses of $e$, $\mu$, and $\tau$ and $m(\nu_2)/m(\nu_3)\approx (1/6)$, as
inputs. One is thus led, {\it for this CP conserving case}, to the
 following fit for the parameters, and the
associated predictions.\cite{24} [In this fit, we leave the small quantities $x$ and $z$
in $M_R^\nu$ undetermined and proceed by assuming that they have the magnitudes suggested
by flavor symmetries (i.e., $x\sim (10^{-4}$-$10^{-5})$ and $z\sim 10^{-2}$(1 to 1/2)
(see remarks below Eq. (13))]:
\begin{eqnarray}
\label{eq:fit}
\begin{array}{c}
\sigma\approx 0.110, \quad \eta\approx 0.151, \quad \epsilon\approx -0.095,
 \quad |\eta'|\approx 4.4 \times 10^{-3},\\[3pt]
\begin{array}{cc}
\epsilon'\approx 2\times 10^{-4},& {\cal M}_u^0\approx m_t(M_U)\approx 120 \mbox{ GeV},\\[3pt]
{\cal M}^0_D\approx m_b(M_U)\approx 1.5 \mbox{ GeV}, & y\approx -(1/18).
\end{array}
\end{array}
\end{eqnarray}

The hierarchical entries in the Dirac and Majorana mass-matrices necessarily predict a
hierarchical pattern with {\it normal hierarchy} for the seesaw-generated masses of the
light neutrinos (i.e. $m(\nu_3)\gg m(\nu_2)> m(\nu_1)$). Taking this into account, the
mass-matrices given by Eqs.~(11) and (13) lead to the following predictions for
the  masses and mixings of the quarks and some of the light
neutrinos \cite{24}:
\beqa
& & 
m^\circ_b (M_U)\approx m^\circ_\tau (1-8\epsilon^2)\Rightarrow
m_b(m_b)\approx(4.7\mbox{-}4.9)\mbox{ GeV}, \CR
&&\sqrt{\Delta m^2_{\rm atm}}\equiv \sqrt{\Delta m^2_{31}}\approx
m(\nu_{3}) \approx (1/24\ \mathrm{eV})(1/2 - 2),\ \hbox{(see Sec.~6)}\CR
& &
V_{cb}\approx\left|\sqrt{\frac{m_s}{m_b}}\left|\frac{\eta+\epsilon}
{\eta-\epsilon}\right|^{1/2}- \sqrt{\frac{m_c}{m_t}}\left|\frac{\sigma
+\epsilon}{\sigma-\epsilon}\right|^{1/2}\right|\approx 0.044,\CR
&& \left\{ \begin{array}{l}
\theta^{\nu}_{23}\approx\left|\sqrt{\frac{m_\mu}{m_\tau}}
\left|\frac{\eta-3\epsilon}{\eta+3\epsilon}\right|^{1/2}+
\sqrt{\frac{m_{\nu_2}}{m_{\nu_3}}}\right|\approx |0.437+0.408|,\\
\noalign{\vspace*{8pt}} \mbox{Thus\ \cite{103}},\ \sin^2 2\theta^{\nu}_{23}\approx
0.994,\ \  (\mbox{for } m(\nu_{2})/m(\nu_{3}) \approx 1/6),\\
\end{array}\right.\CR
&&V_{us}\approx
\left|\sqrt{\frac{m_d}{m_s}}-\sqrt{\frac{m_u}{m_c}}\right|
\approx 0.20,\CR
& & \left|\frac{V_{ub}}{V_{cb}} \right|\approx
\sqrt{\frac{m_u}{m_c}}\approx
0.07,\CR
&& 
m_d(\mbox{1 GeV})\approx \mbox{8 MeV}.
\eeqa{eq:pred}
Before discussing the empirical successes of the predictions listed above, let me first
comment on some aspects of the neutrino-sector pertaining to the first family. As regards
$\nu_e-\nu_{\mu}$ and $\nu_e-\nu_{\tau}$ oscillations, the standard seesaw-contributions,
based on the hierarchical mass-matrices $M^D_\nu$ (Eq.~(8)) and $M^\nu_R$ (Eq.~(13)),
typically lead to rather small oscillation angles (${\lesssim}0.1$) \cite{24}. It has
however, been noted \cite{91} that non-seesaw contribution, e.g. to $\nu^e_{\rm
L}\nu^\mu_{\rm L}$ mixing mass-term $\sim (2-6)\times 10^{-3}$\,eV, can quite plausibly
arise through higher dimensional operators in accord with flavor symmetry (see
Ref.~\cite{91} and \cite{104}). Such a contribution, combined with the standard
seesaw contributions based an $M^D_\nu$ and $M^\nu_{\rm R}$, that yields a diagonal
($\nu^\mu_{\rm L}\nu^\mu_{\rm L}$) mass $\sim(3-10)\times^{-3}$\,eV, can plausibly lead
to $\theta^\nu_{12}$ as large as ${\sim}(30^\circ{-}35^\circ)$, in accord with the LMA
MSW solution of the solar neutrino problem (For a review of current experimental status,
see Refs.~\cite{105} and \cite{106}). Similarly, one can obtain through non-seesaw
contribution $\theta^\nu_{13}\approx (2^\circ{-}10^\circ)$, to be compared with the
observed value of ${\approx}9^\circ$ \cite{105,106}. Thus, including the seesaw (from
$M^D_\nu$ and $M^\nu_{\rm R}$) and non-seesaw contributions, together with a natural
value of $y\approx -1/18$ (which is expected to be of order 1/10 by flavor
symmetry\cite{100}), one gets:
\begin{eqnarray}
m(\nu_2)&\approx & (8.5)\times 10^{-3}\,{\rm eV}\
\hbox{(seesaw)}\nonumber\\[3pt]
m(\nu_1)&\approx & (1-\hbox{few})\times 10^{-3}\,{\rm eV}; \hbox{thus}\ \Delta
m^2_{21}\approx 7\times 10^{-5}\,{\rm eV}^2\nonumber\\[3pt]
\theta^\nu_{12} &\approx & (20^\circ{-}35^\circ)\ \hbox{(non-seesaw)}\nonumber\\[3pt]
\theta^\nu_{13}&\approx&(2^\circ{-}10^\circ)\ \hbox{(non-seesaw)}
\end{eqnarray}
 Here, $m(\nu_2)$ should be regarded as a reasonable expectation within the
model, within a factor of 2 or so either way, corresponding to a natural choice of
$|y|\approx 1/10-1/25$ (say). Unlike the results listed in Eq.~(15), which are compelling
predictions (postdictions) of the model, however, those on $\theta^\nu_{12}$ and
$\theta^\nu_{13}$ noted above should be considered only as plausible and consistent
possibilities within the   model.

The Majorana masses of the RH neutrinos ($N_i\equiv \nu^i_{\rm R}$) are given by (see
discussion in Sec.~6).
\begin{eqnarray}
\label{eq:MajM}
M_{3}& \approx & M_R\approx 10^{15}\mbox{ GeV (1/2-2)},\nonumber\\
M_{2}& \approx & |y^2|M_{3}\approx \mbox{($3\times 10^{12}$\,GeV)(1/2-1)},\\
M_{1}& \approx & |x-z^2|M_{3}\sim (1/4\mbox{-}2)10^{-5}M_{3}\sim \mbox{$10^{10}$
GeV(1/10-2)}.\nonumber
\end{eqnarray}
where $y\approx -1/18$ and $x\sim z^2\sim 10^{-4} (1/2-2)$ have been used in accord with
flavor symmetry.\cite{100} {\it Note that we necessarily have a hierarchical spectrum for
the light as well as the heavy neutrinos  with normal hierarchy $m_1<m_2\ll m_3$ and
$M_1\ll M_2\ll M_3$}. Leaving out the oscillation angles $\theta^\nu_{12}$ and
$\theta^\nu_{13}$, it seems remarkable that the first six predictions in Eq.~(15) agree with
observations, pretty well. Particularly intriguing is the (B-L)-dependent {\it
group-theoretic correlation} between the contribution from the first term in $V_{cb}$ and
that in $\theta^{\nu}_{23}$, which explains simultaneously why one is small ($V_{cb}$)
and the other is large ($\theta^{\nu}_{23}$).

Another interesting point of the hierarchical BPW model\cite{24} is that with $|y|$ being
hierarchical (of order 1/10 as opposed to being of order 1) and $m(\nu_2)/m(\nu_3)$ being
of order 1/5--1/10, {\it the mixing angle from the neutrino sector
$\sqrt{m(\nu_2)/m(\nu_3)}$ necessarily adds (rather than subtracts) to the contribution
from the charged lepton sector (see Eq.~(15))}, as shown in Ref.~\cite{24}. As a
result, in the BPW model, both charged lepton and neutrino-sectors give medium-large
contribution ($\approx 0.4$) which {\it add} to naturally yield a maximal
$\theta^{\nu}_{23}$. At the same time, analogous entries for $V_{\rm cb}$ subtract,
leading to its smallness (see Eq.~(15)). This thus becomes a simple and compelling
prediction of the model, based essentially on the group theory of the minimal Higgs
system in the context of SO(10) or G(224) and the hierarchical nature of the
mass-matrices.\footnote{The explanation of the largeness of $\theta^{\nu}_{23}$ together
with the smallness of $V_{\rm cb}$ outlined above, based on medium-large contributions
from the charged lepton and neutrino sectors, is quite distinct from alternative
explanations. In paricular, in the lop-sided Albright-Barr model\cite{93}, the largeness
of $\theta^{\nu}_{23}$ arises almost entirely from the lop-sidedness of the charged
lepton mass matrix. This distinction between the BPW and the AB models leads to markedly
different predictions for the rate of $\mu\to e\gamma$ decay in the two
models \cite{107}.}

The success of the model as regards especially the first six predictions in Eq.~(15)
provides some confidence in the ${\it gross\ pattern}$ of the Dirac and Majorana mass
matrices presented above. I now proceed to discuss primarily the implications of the
mass-scale of the atmospheric neutrino oscillation in the next section.


\section{Neutrino Masses Shedding Light on Unification and Our Origin}

Since the discoveries (confirmations) of the atmospheric\cite{11} and solar neutrino
oscillations,\cite{12} at SuperKamiokande and SNO respectively, the neutrinos have emerged as being the most effective probes into
the nature of higher unification. To add to these, a set of ingenius experiments have
come into play to study accelerator and reactor neutrinos involving disappearance and
appearance phenomena, including those at KamLAND, K2K, MINOS, OPERA, Double Chooz, T2K,
Daya Bay, RENO and Nova, giving new insights, sometimes with higher precision and/or
reconfirmations (For a review and notations, see e.g. Refs.~\cite{105} and
\cite{106} and references there in. These together have led to a measurement of the
two independent (mass)$^2$-differences of: $|\Delta m^2_{32}| \cong |\Delta
m^2_{31}|\equiv \Delta m^2_{\rm Atm} \approx (1/20\,{\rm eV})^2$, and $\Delta
m^2_{21}\equiv \Delta m^2_{\rm solar} \cong (1/115\,{\rm eV})^2$, and the three
oscillation angles of: $\theta_{23}\approx \pi/4$ (within a few degrees either way),
$\theta_{12} \approx \pi/5.4$, and $\theta_{13}\approx \pi/20$. Supplementing these measurements, 
cosmological studies by PLANCK now provide an upper limt on the sum of the three neutrino masses 
to be about 0.23 {\rm eV}~\cite{sum-nu}.                                         

In attempting to understand the mass-scales of the neutrino oscillations, one finds that,
although almost the feeblest of all entities of nature, simply by virtue of their tiny
masses, {\it the neutrinos seem to possess a subtle clue to some of the deepest laws of
nature pertaining to the unification-scale as well as the nature of the unification
symmetry}. In this sense, the neutrinos provide us with a rare window to view physics at
truly short distances which turn out to be as short as nearly 10$^{-30}$\,cm. Furthermore
it appears most likely that the origin of their tiny masses may be at the root of the
origin of matter-antimatter a symmetry in the early universe\cite{14} and thereby at the
root of our own origin.

Before discussing an understanding of the neutrino masses in more detail, let me explain
qualitatively why the neutrinos are able to provide us with the \hbox{window} to view
physics at truly short distances. The reason is two-fold. First, it is because of the
seesaw mechanism\cite{13} (I will confine to Type I seesaw only), that combines the heavy
(or superheavy ${\gg}1\,{\rm TeV}$) Majorana mass $M^{\nu}_R$ of the RH neutrino with the
familiar Dirac mass of the neutrino ($m^\nu_D \lesssim 100$\,GeV), to yield a light LH
neutrino with a mass $m(\nu_L)\cong (m^\nu_D)^2/M^\nu_R\ll m^\nu_D$. Thus, the seesaw
mechanism plays a crucial role by providing us with a natural reason for the neutrinos to
be light or even superlight, as specified in the following.

But, contrary to to the impression often found in the literature, {\it the seesaw
mechanism, by itself, is not sufficient to provide even an estimate having any degree of
certainty as regards {\it how light or how heavy can the neutrinos be}, while still
remaining much lighter than (say) 10--100\,GeV}. This is because, without the constraints
of an appropriate unification-symmetry and the unification-scale, {it has no clue to the
magnitude of either the Majorana mass} $M^\nu_R$ (which, in general, can vary from the
string or Planck scale ${\sim}10^{18}\,{\rm GeV}$ to (say) a few TeV), {or the Dirac
mass} $m^\nu_D$ (which, for any given species, unconstrained by symmetry, could lie
anywhere in the range of 100\,GeV to (say) 1\,MeV). Hence the uncertainty in the estimate
of the seesaw-mass which can exceed even twenty orders of magnitude.

But, as I will discuss below, if the seesaw mechanism arises in the context of an
appropriate unification symmetry possessing SU(4)-color, as in SO(10) or a
\hbox{string-unified} $G(2,2,4)$-symmetry, and is tied to the unification-scale ($M_U
\sim 2 \times 10^{16}$\,GeV), such that the latter is essentially the (B-L)-breaking
scale, both the Majorana mass $M^\nu_R$ and the Dirac mass $m^\nu_D$, suitably
generalized to \hbox{$3\times 3$-matrices} for the three-family case, get constrained. In
this case, as explained below, one quite naturally obtains the desired magnitudes for the
mass-scales of the \hbox{atmospheric} and solar neutrino oscillations. It is thus the
{\it combination} of the three ingredients: (a) the seesaw mechanism, (b) a suitable
unification-symmetry, and (c)~the unification-scale ${\sim}2\times 10^{16}$\,GeV that is
needed to provide an understanding of the neutrino mass-scales. This, in turn, is the
reason why the neutrinos turn out to be so revealing about: (i)~the nature of the
unification-symmetry, (ii)~the scale of unification (in particular that of
(B-L)-breaking),  and (iii)~the nature of the neutrino mass (being Majorana not Dirac),
all of which pertain to physics at truly high energies (${\sim}10^{16}$\,GeV) and thus
short distances ${\sim}10^{-30}$\,cm.

I now turn to discuss these aspects in more detail by exhibiting the steps that lead to
the mass-scale $(m(\nu_3))$ of the heaviest of the three light neutrinos. This, as we
will see, will represent the mass-scale of the atmospheric neutrino oscillation as well,
as quoted in Eq.~(12). I will follow the framework discussed in Sec.~5, which is based on
the symmetry SO(10) or a string-unified $G(2,2,4)$ symmetry, combined with an assumed
U(1)-flavor symmetry\cite{100} that serves to provide the desired inter-family
mass-hierarchy. For reasons discussed in Sec.~4, the low-dimensional Higgs system of
Eq.~(9) is used to break the symmetry SO(10) to the low-energy symmetry ${\rm SU}(3)^{\rm
c}\times {\rm U}(1)_{\rm em}$. (An analogous system applies for the symmetry $G(2,2,4)$).
The resulting hierarchical pattern of masses and mixings of all fermions including those
given by the Dirac and Majorana mass-matrices of the neutrinos are exhibited in Eqs.~(11)
and (13), which have their origins in the effective couplings of Eqs.~(10) and (12)
respectively.

To obtain an estimate of the mass-scale of the heaviest of the three light neutrinos, it
is useful to consider first only the third family, ignoring the first two families for a
moment, although the final answer will be obtained by considering the full three-family
system, including mixings among the families.\cite{24} Using the effective couplings
given by Eq.~(12), we expect the heaviest Majorana mass $M_3\approx M_{\rm R}$ of Eq.~(13)
to be given by:
\begin{eqnarray}
M_{\rm R}=\frac{f_{33}\langle \overline{16}_{\rm H}\rangle^2}{M}\approx (10^{15}\,{\rm
GeV}) (1/2-2)
\end{eqnarray}
where I have put $\langle \overline{16}_{\rm H}\rangle \approx M_U \approx 2\times
10^{16}\,{\rm GeV}, M \approx M_{\rm string} \approx 4\times 10^{17}\,{\rm
GeV}$\cite{76}, and the flavor-symmetry-allowed leading Majorana coupling $f_{33}\approx
1$, in line with the value of the leading Dirac Yukawa coupling $h_{33}\approx h_{\rm
top}$. An uncertainty factor of $(1/2-2)$ is allowed around a centrally expected value of
$M_{\rm R}$ of about $10^{15}$\,GeV. Using the seesaw formula, in the absence of the
first two families, one then obtains the light LH $\nu^\tau_{\rm L}$ with a mass
$m(\nu^\tau_{\rm L})\approx m(\nu^\tau)^2_{\rm Dirac/M_{R}}$. Now, allowing for mixings
among the three families as given in Eqs.~(11) and (13) (with
hierarchical entries, only
the 2--3 mixing turns out to be important for the purpose), the heaviest seesaw-generated
mass is given by\cite{24}{:}
\begin{equation}
m(\nu_{3}) \approx B \frac{m(\nu_{\mathrm{Dirac}}^{\tau})^{2}}{M_{R}} \label{seesaw}
\end{equation}

\noindent The quantity $B$ represents the effect of 2-3 family-mixing (which, as
mentioned above, is the dominant effect, representing mixing) and is given by \(B =
(\sigma + 3\epsilon)(\sigma + 3\epsilon - 2y)/y^{2}\) (see Eq.~(24) of
Ref.~\cite{24}). Recall that the same mixing successfully explains the near maximality
of $\sin^2 2\theta^\nu_{23}\approx 0.994$ (see Eq.~(15) and discussion in Sec.~5). Thus
$B$ is fully calculable within the model since the parameters $\sigma$, $\eta$,
$\epsilon$, and $y$ are determined in terms of inputs involving some quark and lepton
masses (as noted in Eq.~(14)). In this way, one obtains $B \approx (2.9 \pm 0.5)$.  The
Dirac mass of the tau-neutrino is obtained by using the $SU(4)$-color relation (see
Eq.~(6)): \(m(\nu_{\mathrm{Dirac}}^{\tau}) \approx m_{\mathrm{top}}(M_{U}) \approx 120\
\mathrm{GeV}\). The mass of the heaviest of the three light neutrinos, including mixing, is thus given by:
\begin{eqnarray}
m(\nu_{3}) & \approx & \frac{(2.9)(120 \mbox{\ GeV})^{2}}{10^{15} \mbox{\ GeV}}
(1/2\mbox{--}2)\nonumber \\
& \approx & (1/24 \mbox{\ eV})(1/2\mbox{--}2) \label{seesaw2}
\end{eqnarray}
Noting that for hierarchical entries --- i.e. for ($\sigma$, $\epsilon$, and $y$) \(\sim
1/10\) --- one naturally obtains a hierarchical spectrum of neutrino-masses: \(m(\nu_{1})
\lesssim m(\nu_{2}) \sim (1/10)m(\nu_{3})\), we thus get:
\begin{equation}
\sqrt{\Delta m_{\rm atm}^{2}}\equiv \sqrt{|\Delta m^2_{31}|}\approx m(\nu_{3}) \approx
(1/24 \mbox{\ eV})(1/2\mbox{--}2) \label{delta}
\end{equation}
This agrees remarkably well with the SuperK value of \((\sqrt{\Delta m_{\rm
Atm}^{2}})_{\mathrm{SK}} (\approx 1/20 \mbox{ eV})\).  As mentioned in the introduction,
the success of this prediction provides clear support for (i) the existence of $\nu_{R}$,
(ii) the notion of $SU(4)$-color symmetry that gives $\nu_{\rm R}$, (B-L) and
$m(\nu_{\mathrm{Dirac}}^{\tau})$, (iii) the SUSY unification-scale, with the assumption
of a single-step breaking of SO(10) to the SM, that gives $M_{{\rm B\hbox{-}L}}\approx
M_U$ and there by $M_{R}$, \emph{and}, of course, (iv) the seesaw mechanism.

We note that alternative symmetries such as $SU(5)$ would have no compelling reason to
introduce the $\nu_{R}$'s.  Even if one did introduce $\nu_{R}^{i}$ by hand, there would
be no symmetry to relate the Dirac mass of $\nu_{\tau}$ to the top quark mass.  Thus
$m(\nu_{\mathrm{Dirac}}^{\tau})$ would be an arbitrary parameter in $SU(5)$, which, could
well vary from say 1~GeV to 100~GeV. Furthermore, without B-L as a local symmetry, the
Majorana masses of the RH neutrinos, which are singlets of $SU(5)$, can well be as high
as the string scale \({\sim}4 \times 10^{17}\ \mathrm{GeV}\) (say), and as low as say
1~TeV.  Thus, as mentioned above, within $SU(5)$, the absolute scale of the mass of
$\nu_3$, obtained via the familiar seesaw mechanism\cite{13}, would be uncertain by some
twenty orders of magnitude.

Other effective symmetries such as $[SU(3)]^{3}$\cite{15} and \(SU(2)_{L} \times
SU(2)_{R}\times U(1)_{B-L}\times SU(3)^{C}\) (see Refs. \cite{eight}, \cite{55}) would
give $\nu_{R}$ and B-L as a local symmetry, but not the desired $SU(4)$-color
mass-relations: \(m(\nu_{\mathrm{Dirac}}^{\tau}) \approx m_{t}(M_{X})\) and
\(m_{b}(M_{X}) \approx m_{\tau}\) (see Eqs.~ (5) and (6)).Flipped \(SU(5) \times U(1)\)\cite{108} on the other
hand would yield the desired features for the neutrino-system, but not the empirically
favored $b$-$\tau$ mass relation (Eq.~(4a)). Thus, combined with the observed $b/\tau$
mass-ratio, the SuperK data on atmospheric neutrino oscillation seems to clearly select
out the effective symmetry in 4D being either string-unified $G(2,2,4)$ or $SO(10)$, as
opposed to the other alternatives mentioned above. \emph{It is in this sense that the
neutrinos, by virtue of their tiny masses, provide crucial information on the
unification-scale as well as on the nature of the unification-symmetry in 4D, as alluded
to in the   introduction}.

It is not just the mass-scale of the atmospheric neutrino oscillation that receives an
explanation within the $G(2,2,4)/ SO(10)$-framework of Sec.~5. Within the same framework, and with a
natural choice of the parameter $|y|\approx 1/10-1/20$, motivated by the flavor symmetry
(see Eq.~(10) and discussion in Sec.~5), one also obtains the desired magnitude for the
mass-scale of solar neutrino oscillation~--- i.e.
\begin{equation}
\sqrt{\Delta m^2_{\rm solar}} =\sqrt{\Delta m^2_{21}}\approx m(\nu_{2}) \sim (1/10)\sqrt{ |\Delta m^2_{\rm atm}|}
\end{equation}
, as observed \cite{105,106}. In short, it
seems quite a feat that the four ingredients (i)--(iv) listed below Eq.(21), acting together,
combining the interplay of physics at two vastly different scales~--- $M_U \approx 2
\times 10^{16}$\,GeV and $M_{\rm EW}\sim 100$\,GeV~--- end up in yielding both the
atmospheric and the solar neutrino oscillation mass-scales with the right magnitudes
(within factors of 2--10)!


\subsection{Neutrinos at the root of our Origin?}

I will now discuss briefly that the tiny neutrino masses not only shed light on the
nature of the unification symmetry and the unification scale, but they may well be at the
root of our own origin. This comes about because of a combination of two
reasons \cite{14}: (a) The heavy Majorana masses of the RH neutrinos (Eq.~(17)), which
enter into an understanding of the tiny neutrino masses via the seesaw mechanism,
necessarily violate lepton number and $B-L$ by two units \hbox{($|\Delta
L|=|\Delta(B-L)|=2$),} and (b) the discovery of non-perturbative electroweak sphaleron
effects \cite{14a}, which violate $B+L$ but conserve $B-L$, and remain in thermal
equilibrium in the temperature range of about $10^{12}$\,GeV to 200\,GeV.

As a result, they efficiently erase any pre-existing baryon/lepton asymmetry that
satisfies $\Delta (\mathrm{B+L}) \neq 0$, but $\Delta (\mathrm{B-L}) = 0$. This is one
reason why standard GUT-baryogenesis satisfying $\Delta (\mathrm{B-L}) = 0$ (as in
minimal SU(5)) (however small) becomes irrelevant to the observed baryon asymmetry of the
universe. On the other hand, purely electroweak baryogenesis based on the sphaleron
effects~--- although a priori an interesting possibility~--- appears to be excluded for
the case of the standard model without supersymmetry, and highly constrained as regards
the available parameter space for the case of the supersymmetric standard model, owing to
the mass of the Higgs boson being rather high (${\approx}$125\,GeV). As a result, in the
presence of electroweak sphalerons \cite{14a}, the mechanism of baryogenesis via
leptogenesis\cite{14} has emerged as perhaps the most attractive and promising mechanism
to generate the observed baryon asymmetry of the universe.

Intriguingly, this mechanism  directly relates our
understanding of the light neutrino masses to our own origin. The question of whether
this mechanism can quantitatively explain the magnitude of the observed baryon-asymmetry
depends however crucially on the Dirac as well as the Majorana mass-matrices of the
neutrinos, including the phases and the eigenvalues of the latter-i.e. $M_1$, $M_2$ and
$M_3$ (see Eq.~(14)).

This issue has been discussed in many excellent reviews \cite{109}. Here, to be concrete,
I will present briefly the work of Ref.~\cite{110}, which is carried out in the
context of a realistic and predictive framework for fermion masses and neutrino
oscillations \cite{24}, based on the symmetry $G(2,2,4)$ or $SO(10)$, as discussed in
Sec.~5, with CP violation treated as in Ref.~\cite{102}. I will primarily quote the
results and refer the reader to Ref.~\cite{110} for more details, especially for
the discussion on inflation and relevant references.

The basic picture is this. Following inflation, the lightest RH neutrinos ($N_1$'s) with
a mass ${\approx}10^{10}$\,GeV ($1/4\ -\ 3$) (see Eq.~(14)) are produced either from the
thermal bath following reheating ($T_{RH}\approx \mbox{ few} \times 10^9$ GeV), or
non-thermally directly from the decay of the inflaton\footnote{In this case the inflaton
can naturally be composed of the Higgs-like objects having the quantum numbers of the RH
sneutrinos ($\tilde{\nu}_{RH}$ and $\bar{\tilde{\nu}}_{RH}$) lying in $(1,\ 2,\ 4)_H$ and
$(1,\ 2,\ \bar{4})_H$ for $G(2,2,4)$ (or $16_H$ and $\bar{16}_H$ for $SO(10)$), whose
VEV's break B-L
 and give Majorana masses to the RH neutrinos via the coupling shown in
Eq.~(12).} (with $T_{RH}$ in this case being about $10^7$\,GeV). In either case, the RH
neutrinos having Majorana masses decay (out of equilibrium) by utilizing their Dirac
Yukawa couplings into both $l+H$ and $\bar{l}+\bar{H}$ (and corresponding SUSY modes),
thus violating B-L. In the presence of C and CP violating phases, these decays produce a
net lepton-asymmetry $Y_L=(n_L-n_{\bar{L}})/s$ which is converted to a baryon-asymmetry
$Y_B=(n_B-n_{\bar{B}})/s=C Y_L$ ($C\approx -1/3$ for MSSM) by the EW sphaleron effects.
Using the Dirac and the Majorana mass-matrices of Sec.~5, with the introduction of
CP-violating phases as in Ref.~\cite{102}, the lepton-asymmetry produced per $N_1$ (or
($\tilde{N}_1+\bar{\tilde{N}}_1)$-pair) decay is found to be\cite{110}{:}
\begin{eqnarray}
\epsilon_{1} & \approx & \frac{1}{8\pi}\left(\frac{\mathcal{M}_u^0}{v}\right)^2
|(\sigma+3\epsilon)-y|^2\sin(2\phi_{21})\times (-3)\left(\frac{M_1}{M_2}\right)\nonumber\\
& \approx & -(2.0\times 10^{-6}) \sin\left(2\phi_{21}\right) \label{eq:28}
\end{eqnarray}
\looseness-1
Here $\phi_{21}$ denotes an effective phase depending upon phases in the
Dirac as well as Majorana mass-matrices (see Ref.~\cite{110}). Note that the
parameters $\sigma$, $\epsilon$, $y$ and $(\mathcal{M}_u^0/v)$ are already determined
within our framework (to within 10\%) from considerations of fermion masses and neutrino
oscillations\cite{24} (see Sec.~5), including CP violation \cite{102}. Furthermore, for
concreteness (for the present case of thermal leptogenesis), we have put $M_1\approx
4\times 10^9$\,GeV and $M_2\approx 2\times 10^{12}$\,GeV, in accord with Eq.~(14). In
short, leaving aside the phase factor $\phi_{21}$, the RHS of Eq.~(19) is pretty well
determined within our framework (to within about a factor of 5--10), as opposed to being
uncertain by orders of magnitude either way. \emph{This is the advantage of our obtaining
the lepton-asymmetry in conjunction with a predictive framework for fermion masses and
neutrino oscillations.} Now the phase angle $\phi_{21}$ is uncertain because we do not
have any constraint yet on the phases in the Majorana sector $(M^\nu_R)$. At the same
time, since the phases in the Dirac sector are relatively large (see Sec.~5 and
Ref.~\cite{102}), barring unnatural cancellation between the Dirac and Majorana
phases, we would naturally expect $\sin(2\phi_{21})$ to be sizable-i.e. of order $1/10$
to $1$ (say).

The lepton-asymmetry is given by $Y_L= \kappa (\epsilon_1/g^*)$, where $\kappa$ denotes
an efficiency factor representing wash-out effects and $g^*$ denotes the light degrees of
freedom ($g^*\approx 228$ for MSSM). For the model being considered \cite{24}, using
discussions on $\kappa$ from Ref.~\cite{111}, we obtain: $\kappa\approx(1/18 - 1/60)$,
for the thermal case, depending upon the $''31''$ entries in the neutrino-Dirac and
Majorana mass-matrices (see Refs.~\cite{24}, \cite{102}). Thus, for the thermal
case, we obtain:
\begin{equation}
(Y_B)_{thermal}/\sin(2\phi_{21})\approx (10 - 30)\times 10^{-11} \label{eq:29}
\end{equation}
In this case, for $M_1\approx 4\times 10^9$ GeV, the reheat temperature would have to be
about few $\times 10^9$ GeV so that $N_1$'s can be produced thermally. We see that the
derived values of $Y_B$ can in fact account for the observed value, based on the latest
PLANCK measurements, of $Y_B\approx (8.65 \pm 0.10)\times 10^{-11}$ \cite{112}, for a
natural value of the phase angle $\sin(2\phi_{21})\approx (1/3- 1)$. This case seems,
however, to be in conflict with the familiar gravitino-constraint, with the gravitinos being unstable
against decay to lighter MSSM-LSP particles \cite{111,113,115}. The constraint may, however, be avoided if the gravitino is
somehow heavier than about 10 TeV, or if it is the LSP being as light as about 1~keV~\cite{Murayama}.
An alternative and attractive possibility is the the case of non-thermal
leptogenesis discussed below. This case typically needs a
significantly lower reheat temperature, in accord with the gravitino constraint (noted above), and it can allow 
lower phase angles as well, compared to the thermal case.

For the non-thermal case, to be specific, I will assume an effective
superpotential suggested in \cite{116}: $W^{infl}_{eff}=\lambda \hat{S}
(\bar{\Phi}\Phi-M^2)\,{+}\,$(non-ren.terms) so as to implement hybrid inflation\footnote{For alternative attempts in inflationary leptogenesis, based on subcritical hybrid inflation, 
within the $G(2,2,4)$ symmetry, see Ref.\cite{bryant}, and for models of so-called ``semi-shifted '' hybrid inflation, also based on the symmetry $G(2,2,4)$, 
which possess certain desirable features, see\cite{ss-hybrid}.} Here
$\hat{S}$ is a singlet field\cite{117} and $\Phi$ and $\bar{\Phi}$ are Higgs fields
transforming as $(1, 2, 4)$ and $(1, 2, \bar{4})$ of $G(2,2,4)$ which break $G(2,2,4)$ to
the SM and thereby B-L at the GUT scale and give Majorana masses to the RH neutrinos.
Following the discussion in,\cite{110,116} one obtains: $m_{infl}=\sqrt 2 \lambda M$,
where $M=<(1, 2, 4)_H>\approx 2\times 10^{16}$ GeV;
$T_{RH}\approx(1/7)(\Gamma_{infl}M_{Pl})^{1/2}\approx(1/7)(M_1/M)
(m_{infl}M_{Pl}/8\pi)^{1/2}$ and $Y_B\approx-(1/2)(T_{RH}/m_{infl}) \varepsilon_1$.
Taking the coupling $\lambda$ in a plausible range $(10^{-5} - 10^{-6})$, we get
$m_{infl}\approx 3\times(10^{10}-10^{11}$\,Gev), and a desired, reheat temperature (see
below). Taking $M_1\approx (4/3-2/3)\times10^{10}$\,GeV and $M_2\approx 2\times
10^{12}$\,GeV, in accord with Eq.~(14), the inflaton would decay into a pair of N$_1$'s
utilizing the coupling of Eq.~(12), which in turn decay both into $l+H$ and $\bar{l}+\bar{H}$,
as noted above, causing lepton asymmetry. Taking the asymmetry parameter as in Eq.~(23),
 one quite plausibly obtains
\begin{eqnarray}
(Y_B)_{Non-Thermal}\approx(8 - 9)\times10^{-11}
\end{eqnarray}
in full accord with the PLANCK data, for natural values of the phase angle
$\sin(2\phi_{21})\approx(1/3 - 1/6)$, and with $T_{RH}$ being as low as $10^7$\,GeV
$(2-1/2)$. Such low values of the reheat temperature are consistent with the
gravitino-constraint for $m_{3/2}\approx $ 400 GeV-1~TeV (say), even if one
allows for possible decays of the gravitinos for example via
$\gamma\tilde{\gamma}$-modes.

\begin{figure}[p]
\centering
\includegraphics[width=0.5\hsize]{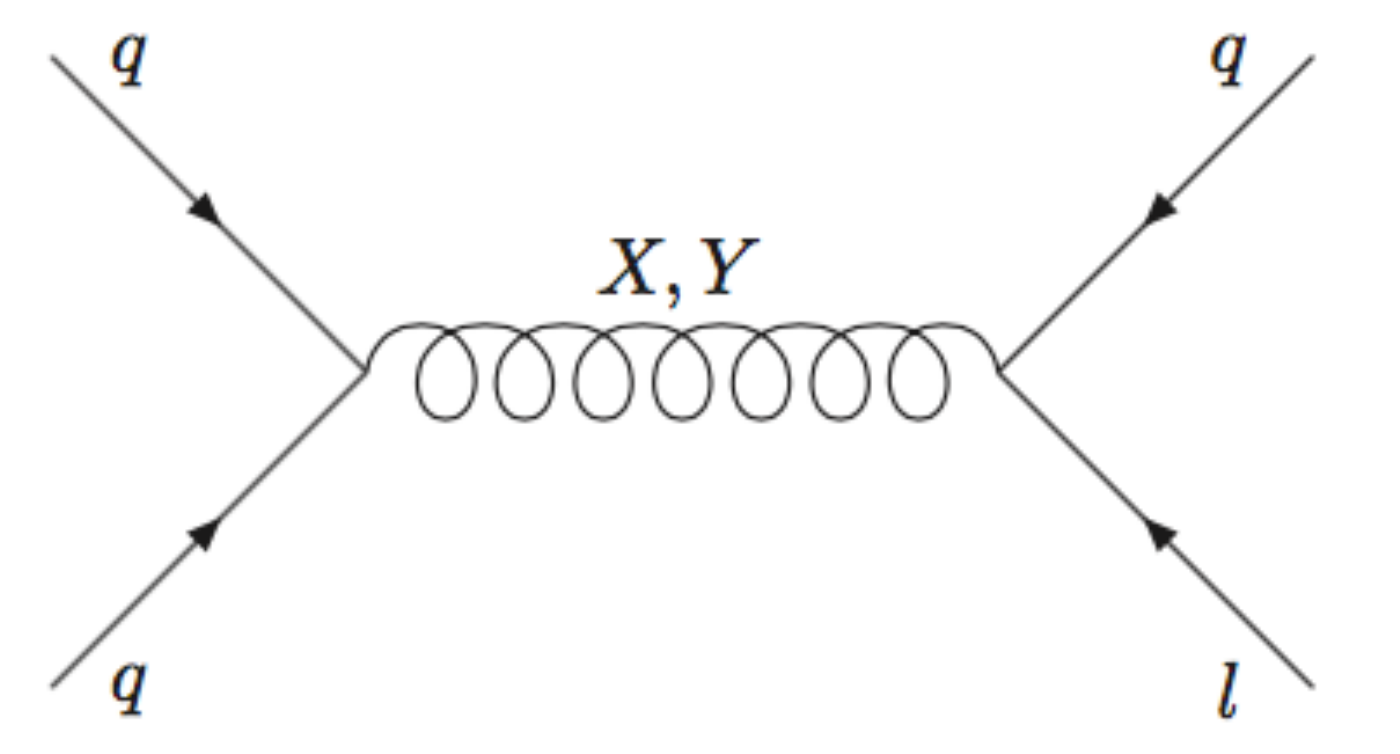}
\caption{$d=6$ proton decay operator
\label{4fig2}}%
\includegraphics[width=0.5\hsize]{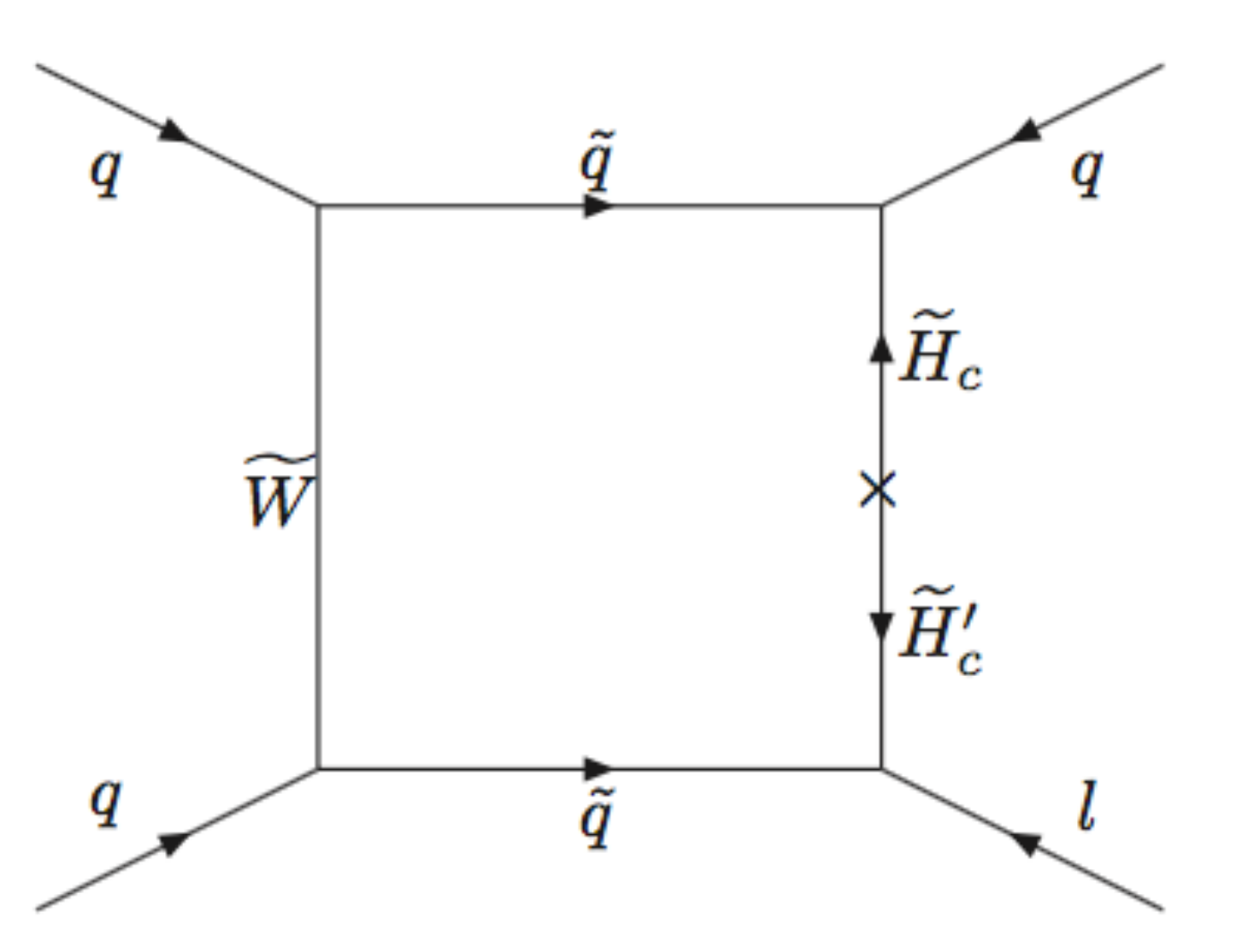}
\caption{The standard $d=5$ proton decay operator. The $\widetilde{H}_c$
($\widetilde{H}^{\prime}_c$) are color triplet(anti-triplet) Higgsinos belonging to
$5_H$($\overline{5}_H$) of $SU(5)$ or $10_H$ of  $SO(10)$.
\label{4fig3}}
\includegraphics[width=0.5\hsize]{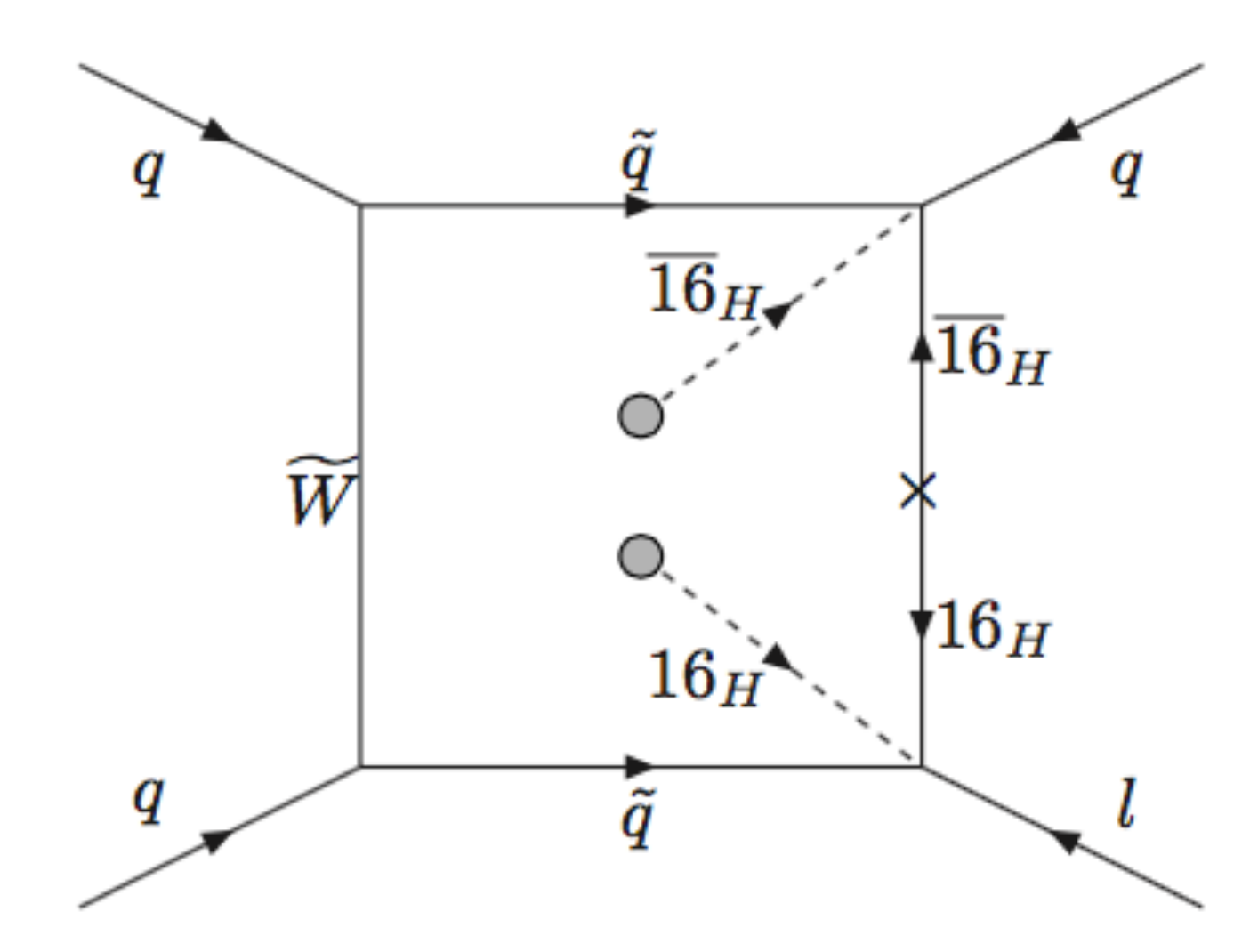}
\caption{The ``new'' $d=5$ operators related to the Majorana masses of the
RH neutrinos. Note that the vertex at the upper right utilizes the coupling in Eq.~(9)
which assigns Majorana masses to $\nu_R$'s, while the lower right vertex utilizes the
$g_{ij}$ couplings in Eq.~(7) which are needed to generate CKM mixings.
\label{4fig4}}
\end{figure}

In summary, I have presented two alternative scenarios (thermal as well as non-thermal)
for inflation and leptogenesis. We see that the $G(2,2,4)/SO(10)$-framework provides {\it
a simple and unified description} of not only fermion masses, and neutrino oscillations,
but also of baryogenesis via leptogenesis, the latter being in accord with all
constraints for the non-thermal case. Each of the following features: (a)~the existence
of the RH neutrinos, (b)~B-L local symmetry, (c)~$SU(4)$-color, (d)~the SUSY unification
scale, (e)~the seesaw mechanism, and (f)~the pattern of $G(2,2,4)/SO(10)$ mass-matrices
based on the minimal low-dimensional Higgs system (see Sec.~4), have played crucial roles
in realizing this \emph{unified and successful description}. I now turn to discuss the
most intriguing consequence of grand unification.


\section{Proton Decay: The Hallmark of Grand Unification}

\subsection{Preliminaries}

Perhaps the most dramatic prediction of grand unification is proton decay. This topic has
been discussed in the context of the SUSY SO(10)/G$(2,2,4)$-framework, presented in
Secs~4 and 5, in some detail in the review articles of Refs.~\cite{91} and
\cite{122} which are updates of the results obtained in Ref.~\cite{24}. For a concise review of works by several authors on proton decay in SUSY GUTs,
 see e.g. Ref.~\cite{64}. Here I will
recall the older works and present briefly the more recent works\cite{25,25a} which
provide a sharpening of the theoretical expectations for proton-decay lifetimes within a
well-motivated class of supersymmetric SO(10) models by doing two things: (i) realizing a
natural and stable doublet-triplet splitting (to be explained below); and (ii) including
the GUT-scale threshold corrections to the running of the gauge couplings. It turns out
that these two steps, carried out in the context of a minimal set of low-dimensional
Higgs multiplets, lead to an intriguing {\it correlation equation} that inversely relates
the $d=6(p\to e^{+}\pi^\circ)$ and $d=5(p\to \overline{\nu} K^{+})$ decay
amplitudes \cite{25}. Together with the empirical lower limits on the inverse rates for
these two decay modes, the correlation equation allows one to derive constrained upper
limits for the same, for any given choice of the SUSY spectrum, which is updated in light
of the LHC searches \cite{25a}. The discussion to follow will also include comments on the
importance of the contributions to the $d=5(p\to \overline{\nu}K^{+})$ decay amplitude
from {\it a new class of diagrams}\cite{23} directly related to an understanding of the
tiny neutrino masses, which are invariably ignored in the literature.

To provide a background for this discussion, in SUSY grand unification, there exist three
distinct mechanisms for proton decay, exhibited in Figs.~2--4.

\begin{enumerate}[label=(\roman*)]
\item {\bf The familiar d=6 operators} mediated by $X$ and $Y$ gauge bosons of SO(10)
(Fig.~2) (similarly for SU(5) as well) which yield $p\to e^{+}\pi^{\circ}$ as a dominant
decay mode, with comparable $p\to\overline{\nu}\pi^+$ mode. Generalizing the result for
the minimal SUSY SU(5)-case derived in Ref.~\cite{118}, to the case of SO(10), one
obtains:
\begin{eqnarray}
&&\Gamma^{-1}_{d=6} (p\to e^{+}\pi^{\circ}) \simeq(1.30\times 10^{35}\,{\rm
yrs})\left(\frac{0.012\,{\rm GeV}^3}{|\alpha_H|}\right)^2\left(\frac{2.5}{A_R}\right)^2\nonumber\\[3pt]
&&\quad \times\left(\frac{5.12}{f(p)}\right)\left(\frac{1/25}{\alpha_X}\right)^2
(M_X/10^{16}\,{\rm GeV})^4\nonumber\\[3pt]
&&\Gamma(p-\overline{\nu}\pi^+)/\Gamma (p\to e^+\pi^\circ)\approx 2[(f(p)-4)/f(p)]
\end{eqnarray}
Here $M_X$ is the mass of the $X$, $Y$ gauge bosons which mediate proton decay; $\alpha_X
=g^2_X/4\pi$ denotes the $(X, Y)$-boson coupling at $M_X; |\alpha_H|\simeq 0.012\,{\rm
GeV}^3$ is the relevant proton decay matrix element; $A_R\simeq 2.5$ is the net
renormalization of the $d=6$ proton decay operator; the function $f(p)=4+(1+1/(1+p^2))^2$
varies between 8 and 5 as the parameter $p\equiv 2\,{<}16_{\rm H}{>}/{<}45_{\rm H}{>}$
varies from 0 to $\infty$, where $f(p)=5$, obtained in the limit $p\to \infty$,
corresponds to the SU(5)-case. The result quoted in Eq.~(26) has assumed a value for the
relevant chiral lagrangian parameter $D+F\simeq 1.27$. Varying $p$, one obtains:
$\Gamma(p\to e^+\pi^\circ)/\Gamma (p\to\overline{\nu}\pi^+)\simeq (1,1.4,2.5)$ for
$(p\lesssim 1/3, p\approx 1, p\gg 1)$. Thus, if this branching ratio is found to be
significantly lower than 2.5, that would be strongly suggestive of SO(10) (as opposed to
SU(5)).

\quad While the $d=6$ inverse decay rate quoted above is largely independent of the
details of the Yukawa couplings and the SUSY spectrum, it depends sensitively on the
value of $M_X$. Often in the literature, a value of $M_X=M_U\approx 2\times 10^{16}\,{\rm
GeV}$ is used to obtain an estimate of this inverse decay rate. Although, we expect $M_X$
to be of order $M_U$, there is no reason to expect $M_X=M_U$. If one allows an
uncertainty in $M_X$ by a factor of 3 (say) around $M_X=M_U$ either way, one would
obtain: $\Gamma^{-1}(p\to e^+\pi^\circ)^{\rm estimated}\sim (10^{33}- 10^{37})\,{\rm
yrs}$, having a large uncertainty by four orders of magnitude. Naively, however, we would
expect $M_X$ and the masses of other GUT-scale split multiplets to be somewhat below
$M_U$, so that they can be neglected in the running of the gauge couplings to achieve
unification at $M_U$. We will see that our considerations of GUT-scale threshold
corrections would lead to an upper limit on $M_X$, and thereby on $\Gamma^{-1}(p\to
e^+\pi^\circ)$ \cite{25}, in accord with the naive expectations.

\item {\bf The ``Standard'' d=5 operators}\cite{119} (Fig.~\ref{4fig3}) of the form
$Q_iQ_jQ_kQ_l/M$ in the superpotential, which arise through exchanges of color triplet
Higgsinos, which are the GUT-partners of the standard Higgs(ino) doublets in the
$5+\overline{5}$ of SU(5) or 10 of SO(10). Thus in SUSY grand unification based on
symmetries like SU(5) or SO(10), it is crucial, for consistency with the empirical lower
limit on proton life time, that a {\it suitable doublet-triplet splitting mechanism}
should exist that assigns GUT-scale masses to the color triplets in the 10$_H$ of SO(10),
or in the 5$_{H}+\overline{5}_{H}$ of {\it SU}(5), while keeping their electroweak
doublet partners light.

\quad Now for minimal SUSY SU(5), without large-dimensional Higgs multiplets, such a
splitting can only be achieved only by extreme fine-tuning. As alluded to before in
Sec.~4, for SUSY SO(10), on the other hand, there exists a natural mechanism of
group-theoretic origin\cite{84,85} involving only low-dimensional Higgs multiplets that
achieves such a splitting without any fine-tuning. The mechanism involves the
introduction of a 10$'$, in addition to the minimal set given by Eq.~(9), where
10$^{\prime}$ is assumed to have an effective coupling of the form $10_{H}\cdot
45_{H}\cdot 10^{\prime}$, and $45_{H}$ aquires a VEV (consistent with minimization of the
potential) along the  B-L direction of the form: $\langle 45_{H}\rangle =i, \sigma_2
\otimes {\rm Diag}\ (a, a, a, 0, 0)$. The 10$'$ does not have a VEV and does not couple
to the matter multiplets 16$_i$. It has a mass that is suppressed by 3 to 4 orders of
magnitude compared to the GUT-scale owing to flavor symmetries (the same symmetries serve
to stabilize the doublet-triplet (DT) splitting against {\it all} higher order operators,
see Ref.~\cite{25} for details of this discussion). It~can~be seen that with the
coupling of 10$'$ and the VEV of $\langle 45_{H}\rangle$ as above, the color triplets in
the 10$_{H}$ acquire GUT-scale masses, while the EW doublets remain massless in the SUSY
limit. In short, the DT splitting is realized in this case of SUSY SO(10) without any
fine-tuning.

\quad Now, owing to (a) Bose symmetry of the superfields in $QQQL/M$, (b) color
antisymmetry, and especially (c) the hierarchical Yukawa couplings of the Higgs doublets,
it turns out that these standard $d=5$ operators lead to dominant $\overline{\nu}K^{+}$
and comparable $\overline{\nu}\pi^{+}$ modes, but in all cases to highly suppressed
$e^{+}\pi^{0}$, $e^{+}K^{0}$ and even $\mu^{+}K^{0}$ modes. For instance, for minimal
SUSY SU(5), one obtains (with $\tan\beta\leq20$, say):
\begin{eqnarray}
[\,\Gamma(\mu^{+}K^{0})/\Gamma(\overline{\nu}K^{+})\,]^{SU(5)}_{std}\,\sim\,[m_{u}/m_{c}\,\sin^{2}\theta]\,R\,\approx\,10^{-3}\,,
\label{e34}
\end{eqnarray}
where $R\approx 0.1$ is the ratio of the relevant $|$matrix element$|^{2}\times$(phase
space), for the two modes.

\quad It is clear from Fig.~\ref{4fig3} that, following loop-integration, the $d=5$
proton-decay amplitude will be characterized by (for
$m_{\widetilde{W}}\,{\ll}\,m_{\tilde{q}})$:
\begin{eqnarray}
A_{d=5} (p\to \overline{\nu} K^+)_{\rm std}\,\infty\,\alpha_2
(hh'/M)(m_{\widetilde{W}}/m^2_{\tilde{q}})
\end{eqnarray}
where $h$ and $h'$ are the Yukawa couplings that enter at the top and bottom corners on
the right side of the loop in Fig.~\ref{4fig3}.

\quad It can be seen that for minimal SUSY SU(5) (following DT splitting through
fine-tuning), the $(d=5)$-amplitude is scaled by $M=M_{H_{c}}\sim M_{GUT}$, where
$M_{H_{c}}$ denotes the physical mass of the color triplets. It turns out that, in this
case, gauge coupling unification requires that the color triplets be lighter than $M_{\rm
GUT}$. Thus, $M$ (minimal SUSY SU(5)) $=$ $M_{H_{c}} \lesssim M_U$. Despite the smallness
of the relevant Yukawa couplings (including CKM mixings) entering into Fig.~3, with the
amplitude being suppressed by only one power of a super heavy mass $M \lesssim M_U$
(Eq.~(24)), the minimal SUSY SU(5) model seems to be in conflict\cite{120} with the
current SuperK limit on $\Gamma^{-1}(p\to \overline{\nu} K^+)$, at least with the
superparticle masses being lighter than about 3--4\,TeV. (see, however, comments   in
Ref.~\cite{121}).

\quad For SUSY SO(10), with DT splitting achieved naturally through the coupling
$10_{H}\cdot 45_{H}\cdot 10'$, the situation is different. Here, one would need the
insertion of the mass of 10$'$ in the right leg of Fig.~\ref{4fig3}. Thus M is given by
an effective mass (see Refs.~\cite{24} and \cite{25}):
\begin{eqnarray}
M(SO(10))= M_{\rm eff}\approx M^2_{\rm GUT}/M_{10'}
\end{eqnarray}
Since the mass of $10'$ is suppressed compared to the GUT-scale owing to flavor
symmetries\cite{24,25}, we expect $M=M_{\rm eff}$ to be as high as
${\sim}10^{19}-10^{20}\,{\rm GeV}$. This in turn would provide a significant suppression
for the $d=5 (p\to \overline{\nu}K^+)$-amplitude in SUSY SO(10). Together with such a
suppression, SO(10), however, possesses an enhancement of the same amplitude (relative to
SUSY SU(5)) owing in part to constraint from the nature of Yukawa coupling in SO(10) (see
discussion in the Appendix of Ref.~\cite{24}). It turns out that, combining the
suppression with the enhancement, SUSY SO(10) predicts $d=5 (p\to \overline{\nu}K^+)$
decay with inverse rates that are fully consistent with the current superK limits, but
lie in an interesting range which can be probed by experiments in the near future. I will turn
to this in light of recent work shortly. First, I will discuss a third mechanism, having
some special features, which arises naturally within the SUSY SO(10)/$G(2,2,4)$-framework
and can induce $d=5$ proton decay.

\item \textbf{The so called ``new''
   d=5
operators}\cite{23,24,122} (see Fig.~4) which can generically arise through the exchange of color-triplet
Higgsinos in the  Higgs multiplets like $(16_H +\bar{16}_H)$ of $SO(10)$, which have been
used in an essential manner to give masses and mixings to the fermions including the RH
neutrinos, and to break SO(10) (see below). Such exchanges are possible by utilizing the
joint effects of (a)~the couplings given in Eq.~(12) which assign superheavy Majorana
masses to the RH neutrinos through the VEV of $\overline{16}_H$, (b)~the coupling of the
form $g_{ij} 16_i 16_j 16_{H}16_{H}/M$ (see Eq.~(10)) which are needed, at least for the
minimal Higgs-system, to generate CKM-mixings, and (c) the mass-term $M_{16}16_{\rm
H}\cdot \overline{16}_{\rm H}$ \cite{123}. These operators also lead to
$\overline{\nu}K^+$ and $\overline{\nu}\pi^+$ as being among the dominant modes,
together, quite possibly, with the $\mu+{\rm K}^{\circ}$ mode (see remarks below), and
they can plausibly lead to lifetimes in the range of $10^{32}-10^{35}$ yrs [see below].
These operators, though most natural in a theory with Majorana masses for the RH
neutrinos, especially in the context of a low-dimensional Higgs system ( see Eq.~(9)), have, however, been invariably omitted in the literature.
\end{enumerate}

One distinguishing feature of the new $d=5$ operator is that they directly link proton
decay to neutrino masses via the mechanism for generating Majorana masses of the RH
neutrinos. \emph{The other, and perhaps most important, is that these new $d=5$ operators
can induce proton decay even when the $d=6$ and standard $d=5$ operators mentioned above
are absent.} This is what would happen if the string theory or a higher dimensional
GUT-theory would lead to an effective $G(2,2,4)$-symmetry in 4D
(along the lines discussed in Sec.~3), which would be devoid of both $X$ and $Y$ gauge
bosons and the dangerous color-triplets in the $10_H$ of $SO(10)$. \emph{By the same
token, for an effective $G(2,2,4)$-theory, these new $d=5$ operators become the sole and
viable source of proton decay leading to lifetimes in an interesting range (see below).}
And this happens primarily because the RH neutrinos have a Majorana mass!

In evaluating the contributions of the new $d=5$ operators to proton decay, allowance
needs to be made for the fact that for the $f_{ij}$ couplings (see Eq.~(12)), there are
two possible $SO(10)$-contractions (leading to a $\mathbf{45}$ or a $\mathbf{1}$) of the
pair \(\mathbf{16}_{i} \overline{\mathbf{16}}_{H}\), both of which contribute to the
Majorana masses of the $\nu_{R}$s, but only the contraction via the $\mathbf{45}$
contributes to proton decay. In the presence of non-perturbative quantum gravity one
would in general expect both contractions to be present having comparable strengths.  For
example, the couplings of the $\mathbf{45}$s lying in the string-tower or possibly below
the string scale, and likewise of the singlets to the \(\mathbf{16}_{i}
\overline{\mathbf{16}}_{H}\) pair would respectively generate the two contractions.
Allowing for a difference between the relevant projection factors for $\nu_{R}$-masses
versus proton decay operator, we set \((f_{ij})_{p} \equiv (f_{ij})_{\nu}K\), where
$(f_{ij})_{\nu}$ defined in Sec.~5 directly yields \hbox{$\nu_{R}$-masses} and $K$ is a
relative factor of order unity \cite{124}. As a plausible range, we take $K \approx 1/5 -
2$ (say), where $K=1/5$ seems to be a conservative value on the low side that would
correspond to proton lifetimes near the   upper end.


The results of Ref.~\cite{24} (see Eqs.~(41) and (45) of this reference) giving the
contributions of the new $d=5$ operators to the $p\to \overline{\nu}K^+$ decay needs to
be updated in two respects: (i) by including the projection factor $K$ mentioned above,
and (ii) by taking the constraints of the LHC searches on SUSY particles\cite{125,126}
into account. One possible scenario incorporating these constraints, while preserving
reasonable degree of SUSY naturalness, will be considered shortly in the course of
discussing an update of Ref.~\cite{25}.

For concreteness, it assumes an inverted sfermion mass-hierarchy, along the lines
considered within the GUT-framework in Ref.~\cite{127}, with a light stop
(${\sim}(500{-}1000)$GeV) \cite{128,129}, lighter neutalino (possibly close to stop-mass),
heavy first two generations ($\sim$(15--20) TeV), $m_{\widetilde{W}}\sim
(800{-}1200)$\,GeV and $m_{\widetilde{g}}\sim (2.5{-}3.5)$\,TeV. Using Eqs.~(41) and (45)
of Ref.~\cite{24}, we find that for a SUSY-spectrum as above, and in the absence of
the standard $d=5$ operators discussed above (so that there is no interference between
them), the new $d=5$ operators by themselves lead to:
\begin{eqnarray}
\Gamma^{-1} (p\to \overline{\nu}_{\tau} K^+)_{{\rm new}\, d=5} \approx[(5\times
10^{31})-10^{35}]\,{\rm Yrs}
\end{eqnarray}
Here, $K=(1-1/5)$ has been used. Such an inverse rate is in fact quite comparable to the
kind of lifetimes that would be expected for the $p\to \overline{\nu}K^+$ decay modes
from the standard $d=5$ operators (Fig.~3), with the same or similar SUSY spectrum as
above (see discussion in the next sub-section).

There are three special features of the new $d=5$ operators that
are   worth noting:

(1) Unlike the standard $d=5$ operators (Fig.~3), {\it the new $d=5$ operators
$($Fig.~$4)$ are free from the doublet-triplet splitting problem}, even if the effective
symmetry is SUSY SO(10) rather than a string-unified $G(2,2,4)$-symmetry.

(2) By the same token, the new $d=5$ operators are independent of $M_{\rm eff}$ (see
Eq.~(29)).

(3) Together with the $\overline{\nu} K^+$ and $\overline{\nu}\pi^+$ modes, as a
distinguishing feature, the new $d=5$ operators lead to the $\mu^+ K^\circ$-decay mode
that is typically quite prominent, more so than what one would expect from the standard
$d=5$ operators, even in the case of SUSY SO(10) (see Ref.~\cite{24}). Specifically,
one would expect:
\begin{eqnarray}
[\Gamma (p\to \mu^+ K^\circ)/\Gamma (p\to \overline{\nu}K^+)]_{{\rm new}\, d=5}\approx
(5-50)\%
\end{eqnarray}
This is to be compared with an expected branching ratio of about (5--10)\% for the case
of standard $d=5$ operators for SUSY SO(10), and about $10^{-3}$ for SUSY SU(5) (see
Eq.~(27)). {\it Thus, the $(\mu^+K^\circ)$-mode can serve as a signature for the new
$d=5$ operators}. Observation of a large branching ratio of the $(\mu^+K^\circ)$-mode
(compared to the $(\overline{\nu}K^+)$-mode) of (30--50)\% (say) would be a clear signal
for the relevance of the neutrino-mass related new $d=5$ operators for proton decay, in the context of a SUSY $SO(10)$ or string-$G(2,2,4)$ model in 4D. That
would be a valuable piece of information.

Before considering a sharpening of the proton decay lifetimes based on recent
works \cite{25}, a general comment about the gauge-boson-mediated $d=6$ operator that
yields the ($e^+\pi^\circ$)-mode as the dominant one, is worth making. While, as
mentioned before, naively we expect $M_X$ to lie below the unification scale $M_U\simeq
2\times 10^{16}$\,GeV, in case $M_X$ is as high as about $(1.5{-}1.7)\times
10^{16}$\,GeV, not quite in accord with the naive expectations, $\Gamma^{-1}(p\to
e^+\pi^\circ)$ may well be as high as ${\approx}(5{-}10)\times 10^{35}$\,yrs (see
Eq.~(23)). In this case, the $d=5$ ($p\to \overline{\nu}K^+$) decay mode (depending on
the SUSY spectrum) may well be the dominant mode with lifetime ${\approx}({\rm few\
to}\,10)\times 10^{34}$\,yrs.

It should be stressed, however, that the $e^{+}\pi^{0}$-mode is the \emph{common
denominator} of all GUT models ($SU(5)$, $SO(10)$, etc.) which unify quarks and leptons
and the three gauge forces.  Its rate is determined essentially by the matrix element
$\alpha_{H}$ and the mass of the ($X, Y$) gauge bosons related to the SUSY unification
scale, without the uncertainty of the SUSY spectrum. I should also mention that the
$e^{+}\pi^{0}$-mode is predicted to be the dominant mode in the flipped \(SU(5) \times
U(1)\)-model \cite{108}, and also as it turns out in certain higher- dimensional orbifold
GUT-models, discussed in Sec.~3.1 (see Refs.~\cite{Kawamura,Asaka}), as well as in a model of compactification of M-theory on a manifold
of $G_2$ holonomy \cite{131}. For these reasons, intensifying the search for the
$e^{+}\pi^{0}$-mode to the level of sensitivity of about (a few) $\times\ 10^{35}$ years
in a next-generation proton decay detector and, if need be, to that of 10$^{36}$ yrs in a
next-to-next generation detector, should be well worth the   effort.

I will now discuss recent works, which yield expected upper limits on the lifetimes for
the $(e^+\pi^\circ)$ and $(\overline{\nu}K^+)$ decay modes within a class of
well-motivated supersymmetric SO(10)-models.

\subsection{Constraining Proton Lifetime in SUSY SO(10) with
  Stabilized Doublet-Triplet
Splitting}

Following the preliminary discussion on the three mechanisms for proton decay noted in
the preceding sub-section, I would now present the gist of a more recent work \cite{25} and its
update\cite{25a} on sharpening the inverse rates of proton decay induced by the $d=6$ and
$d=5$ operators exhibited in Figs.~2 and 3 respectively. The former leads to
$e^+\pi^{\circ}$ and comparable $\overline{\nu}\pi^+$, as the dominant decay modes, while
the latter leads to $\overline{\nu}K^+$ as the dominant mode. The works of
the two references cited above, incorporating the LHC searches for SUSY, allow
one to set conservative upper limits for the inverse rates of both the $p\to
e^+\pi^\circ$ and $p\to \overline{\nu}K^+$ decay modes within a well-motivated class of
SUSY SO(10)-models, based on a minimal set of low-dimensional Higgs multiplets of the
type presented in Eq.~(9). This comes about through the following set of steps:

(1) First, recognizing that the doublet-triplet (D-T) splitting, requiring a large
hierarchy of some 13 orders of magnitude between the masses of the doublet and the
triplet, mentioned in the preceding sub-section, poses a major issue for all SUSY GUT
models (SU(5), SO(10), E$_6$ etc.), Babu, Tavartkiladze and I attempted to ensure: (a)
that the D-T splitting which is naturally induced by the missing VEV mechanism of
Refs.~\cite{84,85}, with 45$_{\rm H}$ having a ${\rm VEV} =i\sigma_2 \otimes {\rm
Diag}\ (a,a,a,0,0)$ along the (B-L)-direction, is stable to a very high accuracy in the
presence of all allowed higher dimensional operators; (b) that there does not exist any
undersirable pseudo-Goldstone bosons; and (c) that there are no flat directions which
would lead to VEVs of fields undetermined. Furthermore, (d) one must also examine by
including all GUT-scale threshold corrections to the gauge couplings, the implications of
D-T splitting on coupling unification and on proton decay. While some of these issues had
been partially addressed in the literature, and a major progress was made in
Ref.~\cite{132}, simultaneous resolution of all four issues had remained a challenge
before the   work of Ref.~\cite{25}.

(2) A predictive class of SUSY SO(10) models based on a minimal low-dimensional Higgs system
(that includes the multiplets of Eq.~(9), together with an additional pair of
$16'+16'$-bar\cite{132} and two SO(10)-singlets) was introduced in Ref.~\cite{25}, in
which all the issues of D-T splitting mentioned above are resolved, and the threshold
corrections to the gauge couplings and their implications for proton decay are properly
studied as well. A postulated anomalous U(1)$_{\rm A}$ gauge symmetry, together with a
discrete symmetry $Z_2$, both of which may have a string-origin, plays a crucial role in
achieving the desired results mentioned above.

(3) The minimal low-dimensional Higgs system lead to smaller threshold corrections unlike
in the case of higher dimensional multiplets (like $126_{\rm H}, \overline{126}_{\rm H},
210_{\rm H}$, possible $54_{\rm H}$). It turns out that within such a low-dimensional
Higgs system, subject to the symmetry as mentioned above, there are a large set of
cancellations between different GUT-scale threshold contributions (see Ref.~\cite{25} for explanation). As
a result, somewhat \hbox{surprisingly,} the GUT-scale threshold corrections to
$\alpha_3(m_z)$ are determined in terms of a very few parameters. This makes the model
rather predictive for proton decay.

As a novel feature, by incorporating D-T splitting as indicated above and GUT-scale
threshold corrections, we find an interesting {\it inverse correlation} between the
mass-scales $M_X$ and $M_{\rm eff}$, which respectively control the
$d\,{=}\,6\ (p\to
e^+\pi^\circ)$ and $d\,{=}\,5\ (p\to\overline{\nu}K^+)$ decay amplitudes (see Eqs.~(22) and
(24)), of the following   form \cite{25}:
\begin{eqnarray}
M_{\rm eff} \propto K_{\rm SUSY} [10^{16}\,{\rm GeV}/M_X]^3
\end{eqnarray}
 Here $K_{\rm SUSY}$ depends (rather mildly) on the SUSY spectrum at the EW
scale, and also on a ratio of two GUT-scale masses. The ratio is varied within a wide
range (within reason \cite{25}) so as to get conservative upper limits on the lifefitmes
(see below).

Now the empirical lower limit on $\Gamma^{-1}(p\to \overline{\nu}K^+)$ sets a lower limit
on $M_{\rm eff}$, corresponding to any given SUSY spectrum. That in turn yields, via the
inverse correlation given in Eq.~(32), a {\it theoretical upper limit} on $M_X$ and
thereby on $\Gamma^{-1}(p\to e^+\pi^{0})$. Likewise, the empirical lower limit on
$\Gamma^{-1}(p\to e^+\pi^{0})$ yields, via the correlation Eq.~(32), a theoretical
upper limit on $\Gamma^{-1}(p\to \overline{\nu}K^+)$. This chain of arguments thus allows
the unusual result leading to predicted upper limits (corresponding to any given SUSY
spectrum) on the inverse rates for proton decaying via both the
$e^+\pi^{0}$ and the
$\overline{\nu}K^+$ modes. Interestingly, as discussed below, these upper limits turn out
to be at striking distance from the current empirical lower limits suggesting that proton
decay ought to be discovered in the next-generation experiments.



To be quantitative on the predictions mentioned above, we first need the empirical lower
limits on proton decay lifetimes. Based on the currently most sensitive searches at
SuperKamiokande, the limits on the inverse rates of the two dacay modes are given
by\cite{65,65a}{:}
\begin{eqnarray}
\Gamma^{-1}(p\to e^+\pi^{0})_{expt}&>&1.6\times 10^{34}\,{\rm yrs}\\[6pt]
\Gamma^{-1}(p\to \overline{\nu}+K^+)_{expt}&>&6.6\times 10^{33}\,{\rm yrs}
\end{eqnarray}
On the theoretical side, to derive a lower limit on $M_{\rm eff}$ by using the empirical
lower limit on $\Gamma^{-1}(p\to \overline{\nu}K^+)$, we need two things: (i) the
relevant Yukawa couplings, and (ii) the SUSY spectrum, both of which enter into Fig.~3.
Now the Yukawa couplings (including their phases) get determined by relating the
predictive SO(10)-framework to the masses and mixings of all fermions including
neutrinos, and to the observed CP violation\cite{25}.

The major unknown at present is the SUSY spectrum. At the same time, as noted in Sec.~3,
the motivations for low-energy supersymmetry in some form seem to be compelling. These
include in particular: (i) the need to understand the smallness of the Higgs mass
compared to the GUT or Planck scale (or equivalently that of the gauge-hierarchy ratio
$(m_W/M_U)\sim 10^{-14}$), without introducing unnatural extreme fine-tuning; and (ii)
preserving gauge coupling unification with a successful prediction of the weak angle,
which calls for both supersymmetry and grand unification, as discussed in Sec.~3.

That said, consistent with the LHC-1 and LHC-2 searches, the Higgs boson mass, and flavor
and CP-violating processes, there are, however, different possibilities for the SUSY
spectrum that would be compatible with reasonable ``SUSY naturalness''. The latter
corresponds to avoiding unnatural fine-tuning in the Higgs mass.  For a good discussion
of these possibilities, see e.g. Ref.~\cite{133}, and references there in..

For our purposes, we adopt two guidelines: (a) simple-minded reasonable SUSY naturalness
that suggests a light stop with a mass ${\sim}$(500--1000) GeV, say, together with a
lighter higgsino that may need to be close to the stop-mass for consistency with LHC
searches (see e.g. Ref.~\cite{128}), and (b) a simple solution to the supersymmetric
FCNC and CP problems that suggests heavy sfermions (\hbox{${\sim}$(15--20)} Tev, say) of
the first two generations. Such an inverted hierarchy spectrum for the sfermions together
with a light stop has been considered by many authors, see e.g. Refs.~\cite{127} and
\cite{134}. As mentioned before, for concreteness, we essentially follow the work of
Badziak, Dudas, Olechowski and Pokorski\cite{127}, which is cast within the
GUT-framework. An inverted hierarchy of the type mentioned above can be obtained at the
electroweak scale by using partial universality in the soft parameters at the GUT-scale
as follows:
\begin{eqnarray}
m_{\rm o}(1,2)\sim(15-25)\,{\rm TeV}&\gg&m_{\rm o}(3)\sim(3-3.8)\,{\rm TeV}\gg
m_{1/2}\sim(1.2-2)\,{\rm TeV},\nonumber\\[3pt]
{\rm with}\quad m_{\rm o}(Hu)&=&m_{\rm o}(3)\neq m_{\rm o}(Hd);\ A_{\rm o}=0\ {\rm to}\
-2\,{\rm TeV}
\end{eqnarray}
The $\mu$-parammeter (at EW scale) is determined by radiative electroweak symmetry-breaking condition to be ${\sim}$(500--800)\,GeV, with
$\tan\beta=10$, as input. The inverted hierarchy in the soft masses as shown above can
be realized consistently through the use of flavor symmetries (like the Q$_4$-symmetry in
our case\cite{25}, and an Abelian $U(1)$-symmetry in\cite{127}), which distinguish
between the third family and the first two). Input parameters as in Eq.~(35) lead to a
mass-pattern at the EW scale as follows (only a few relevant masses are listed):
\beqa
m_h&\approx & 125\,{\rm GeV},\quad m_{\tilde{t}_1}\sim(500-1000)\,{\rm GeV},\CR
m_{\tilde{t}_2}&\sim&(1.5- 2.2)\,{\rm TeV},\quad m_{\tilde{q}_{1,2}}\sim(18-21)\,{\rm
TeV},\CR
m_{\tilde{b}_{1,2}}&\sim&(1.8-3)\,{\rm TeV},\quad m_{\tilde{l}_{1,2}}\sim(18-21)\,{\rm
TeV},\CR
m_{\tilde{l}_{3}}&\sim&(3.1-3.5)\,{\rm TeV},\quad m_{\tilde{\chi}_{i}}\approx
m_{\tilde{B}} \sim(460-800)\,{\rm
GeV},\CR
m_{\tilde{W}}&\sim&(830-1200)\,{\rm GeV},\quad m_{\tilde{g}}\sim(2.5-3.5)\,{\rm TeV},
\eeqan
Note the possibility of a light stop (${\sim}$(500 - 1000) GeV) with a higgsino/bino being
lighter but close to it within about (20--100)\,GeV, which seems to be consistent with the
LHC-13 searches\cite{125,126,128}. Spectra of the type presented above have been shown\cite{127}
to be consistent with BR(b$\to$s$\gamma$), with either the lightest higgsino or an almost pure bino being the LSP dark matter,
 and BR($B_S\to \mu^+\mu^-$).

The fine-tuning parameter may be defined as in Ref.~\cite{127} by $\Delta \equiv
\max\{\Delta_a\}$, where $\Delta_a\equiv \left|\frac{\partial (ln\, m_h)}{\partial (ln\,
a)}\right|$; here ``a'' represents any soft term or $\mu$. Spectra of the type given
above correspond to $\Delta \approx 150-250$. We regard this as reasonable naturalness
(in contrast to extreme fine-tuning).

Using the SUSY spectrum of the type given in Eq.~(36), and the empirical lower limits on
proton decay lifetimes given in Eqs.~(33) and (34), one can now utilize the correlation
Eq.~(32) to calculate the upper limits on the inverse rates for the proton decaying via
both the $e^+\pi^{0}$ and the $\overline{\nu}K^+$ modes. To arrive at conservative
values for these upper limits, we allow for a wide variation in the ratio of the
GUT-scale masses that enters into $K_{\rm SUSY}$ (of Eq.~(32))\cite{25}. We also consider
some variation in the SUSY spectrum relative to the type exhibited in Eq.~(36), including
the possibility that the sfermions of the first two families have medium-heavy
(${\sim}$4\,TeV) rather than very heavy (${\sim}$20\,TeV) masses, always requiring
reasonlable SUSY naturalness and $m_h\approx 125$\,GeV. In addition we allow
uncertainties in the lattice-value of the matrix elements and $\alpha_3(m_z)$. Including
these uncertainties, the correlation equation and the empirical lower limits on the
proton decay lifetimes yield the following theoretical upper limits for the same:
\beqa
\Gamma^{-1}(p\to e^+\pi^{0})_{\rm Theory}&\lesssim&(2-10)\times 10^{34}\,{\rm
Yrs}\CR
\Gamma^{-1}(p\to \overline{\nu}K^+)_{\rm Theory}&\lesssim&(1-8)\times 10^{34}\,{\rm
Yrs}
\eeqan
These should be regarded as conservative upper limits because the uncertainties of the
type mentioned above are all stretched together so as to prolong the proton lifetimes.
The actual lifetimes can be quite a bit shorter than the upper limits exhibited above.


As we see, the predicted upper limits (Eq.~(37)) are within factors of five to ten above
the current SuperKamiokande limits (Eqs.~(33), (34)). I should add that supersymmetric
grand unified theories that are in accord with the observed masses and mixings of all
fermions, including neutrinos, typically yield estimated proton lifetimes in the range as
mentioned above (see Ref.~\cite{64} for an overview). Thus, the prospects for
discovery of proton decay in the next-generation deep underground detectors~--- including
especially the 560\,kt water Cherenkov detector at \hbox{HyperKamiokande} and the
(20--70)\,kt Liquid Argon detector at LBNF-DUNE~--- would be high.

\subsection{Proton Decay as a Unique Probe to Physics at Ultrashort Distances}

Proton decay, if discovered, would provide a {\it unique window} to view physics at truly
high energies ${\sim}10^{16}$\,GeV, or equivalently at truly short distances
${\sim}10^{-30}$\,cm. This cannot be realized through accelerators/colliders in the
conceivable future. To be specific, some of the valuable insights which one may gain
through the discovery and subsequent study of proton decay include the following:
\begin{enumerate}[label=(\roman*)]
\item If $p\to \overline{\nu}K^+$ decay mode is seen with an
inverse decay rate ${\sim}(10^{34}$ to (a~few)$\,{\times}\,10^{35})\,{\rm yrs}$, say, in the context of higher unification in 4D, it
would imply that either the standard $d=5$ (Fig.~3) or the new $d=5$ operator (Fig.~4),
or both, are relevant, involving physics at the unification-scale
${\sim}2\times^{16}$\,GeV. Importantly, it would mean that supersymmetry in some form
should exist at low energies. (The latter, hopefully, may be discovered at the LHC in the
meantime).

\item  If $p\to\mu^+K^\circ$ decay is seen with a decent branching ratio
(${\gtrsim}$30\%), in the context of a gauge-unification of forces in 4D, it would mean (as discussed in Sec.~7.1) that neither the $d=6$
(Fig.~2) nor the standard $d=5$ operator (Fig.~3) can account for such an observation.
And, the neutrino-mass related new $d=5$ operators\cite{23,24} must be playing a role in
proton decay; {\it that would mean proton decay knows about the origin of neutrino
masses and vice versa}! As noted in the later part of Sec.~3, the $p\to\mu^+K^\circ$ decay mode can be prominent or dominant through d=6 operators within a
 certain higher dimensional orbifold GUT models as well ( see the last four papers in Ref.~\cite{Kawamura}).

\item  If (B-L)-violating decay modes of the
nucleon\cite{135} such as $n\to e^{-}\pi^+$ or $p\to e^{-}\pi^+\pi^+$ (satisfying
$\Delta$ (B-L) $=-2$) are seen at all (a feature which I have not discussed), it would
mean that fundamental physics at intermediate scales $\ll M_U\approx 2\times
10^{16}$\,GeV is necessarily present. This would, of course, be incompatible with the
striking successes of the ``conventional'' picture of one-step breaking of SUSY-GUT like
SUSY-SO(10) to the SM at $M_U$, including those of the prediction of the weak angle
(Sec.~3) and an understanding of the mass-scales of neutrino-oscillations (Sec.~6). In
this sense, observation of (B-L)-violating decay modes of the proton would imply that the
successes as above are, somehow, accidents. Yet, experiment is the final arbiter. Thus,
one must keep an open mind and search for such decay modes as sensitively as possible.
Non-observation of such decay modes would, of course, serve to strengthen the conventional picture.

\item If $p\to e^+\pi^{\circ}$ decay is seen with a decent inverse rate
(${\lesssim}10^{36}$ yrs, say), then in the context of gauge-unification in 4D, it would imply that
the gauge-mediated $d=6$ operator (Fig.~2) is very likely relevant for the decay. That
would mean that not only $q\leftrightarrow l$ but also $q\leftrightarrow\bar{q}$ and
$q\leftrightarrow\bar{l}$ unifications are relevant. In the context of unification
through symmetries like SO(10), SU(5) or a string-unified $G(2,2,4)$-symmetry in 4D, it
would in turn imply that an intact GUT-symmetry like SO(10) or SU(5), rather than a
string-unified $G(2,2,4)$-symmetry (which is on par with SO(10) in explaining observed
neutrino oscillations\cite{24})\footnote{Within an effective $G(2,2,4)$-symmetry, the exchange of
a scalar $(1,1,6)$-multiplet having a GUT-scale mass can induce the $p\to e^+\pi^\circ$
decay, but such a decay mode will not be a compelling feature of the symmetry.} is very
likely operative in 4D. SUSY SU(5) would yield $p\to e^+\pi^{\circ}$-decay with an
inverse rate as above; it however seems to be disfavored on other grounds, especially by
observed neutrino oscillations (see Sec.~6). Thus, at least in the context mentioned
above, observation of the $p\to e^+\pi^\circ$ decay mode would clearly suggest that an intact SUSY
SO(10) is operative in 4D. As mentioned at the end of Sec.~7.1, such a decay mode
could, of course, also arise in other context such as, flipped $SU(5)\times
U(1)$ \cite{108}, or certain higher-dimensional orbifold GUT-models as in Ref.~ \cite{Kawamura,Asaka}, or string-theory models as
in Ref.~\cite{131}.

\item  On the other hand, if, $p\to e^+\pi^\circ$ decay mode is {\it not} seen with an inverse decay
rate as high as, (say) (10$^{36}$-even 10$^{37}$)\,yrs, but $p\to \overline{\nu}K^+$
decay mode is seen with an inverse rate ${\lesssim}10^{35}$\,yrs, it would first of all
mean, at least in the context of a gauge-unification of forces in 4D, that an intact GUT-symmetry like SO(10) is very likely {\it not} operative in 4D; instead
an effective symmetry like $G(2,2,4)$, or its close relative $G(2,1,4)$, very likely
having a string-origin, is operative in 4D with low-energy supersymmetry; {\it and it is
the neutrino-mass related $d=5$ operator (see Sec.~7.1) that induces such a decay}.
\end{enumerate}

In short, proton decay, if seen, will bring a wealth of knowledge of a fundamental nature
that can not be gained by any other means.

As we have seen, the potential for discovery of proton decay, within a well-motivated
class of grand unification models, is high. This is why an improved search for proton
decay is now most pressing. This can only be done with a large detector built deep
underground. Most desirably we would need both Water Cherenkov (as in HyperKamiokande)
and Liquid Argon (as in LBNF-DUNE) detectors, because the former is specially sensitive
to the ($e^+ \pi^\circ$)-mode, and the latter to the ($\overline{\nu}K^+$)-mode. Such a
detector, coupled to a long-baseline intense neutrino beam, can simultaneously
sensitively study neutrino oscillations so as to shed light on neutrino mixing
parameters, mass-ordering, and most importantly CP violation in the neutrino system. And
it can help efficiently study supernova neutrinos. In short such a detector (or rather
two such detectors with the HyperKamiokande being specially sensitive to the $e^+\pi^0$ mode of proton decay and the DUNE/LBNF to the $\bar\nu K^+$ mode) would have a unique multi-purpose value
with high discovery potential in all three areas.



\section{ Concluding Remarks}

Neutrinos seem to be as elusive as revealing. Simply by virtue of their tiny masses, they
provide some crucial information on the nature of the unification symmetry and on the
unification scale, more precisely on the (B-L)-breaking scale. In particular, as argued
in Sec.~6, combined with the $b/\tau$ mass-ratio, the mass-scale $(\sqrt{\Delta m^2_{\rm
Atm}}\approx \sqrt{\Delta m^2_{31}}\approx 1/20\,{\rm eV})$ of the atmospheric neutrino
oscillation provides a clear support for the following three features:
\begin{enumerate}[label=(\roman*)]
\item the existence of the SU(4)-color symmetry in 4D at and above the GUT-scale which
provides not only the RH neutrinos but also B-L as a local symmetry and a value for
$m(\nu^{\tau}_{\rm Dirac})$ (Eq.~(6));

\item the (B-L)-breaking scale being close to the familiar SUSY unification-scale $M_U\approx 2\times
10^{16}$\,GeV, which provides the mass-scale $M_R$ for the Majorana mass of the heaviest
RH neutrino (see Eqs.~(12) and (17)); and importantly

\item the seesaw mechanism that successfully explains the atmospheric neutrino
oscillation mass-scale by utilizing both (i) and (ii) (see Eqs.~(18)--(21)).
\end{enumerate}
{\it In turn this chain of argument selects out the effective symmetry in 4D being either a
string-derived $G(2,2,4)$ or $SO(10)$-symmetry, as opposed to other alternatives like
$SU(5)$, $[SU(3)]^3$ or even flipped $SU(5)\times U(1)$}. As a corollary, this supports
the idea that Nature intrinsically is left-right symmetric (parity-conserving)\cite{55}.

Furthermore, the success of the $G(2,2,4)/SO(10)$-based seesaw mechanism in accounting
for the neutrino oscillation mass-scales (both atmospheric and solar, see Sec.~6) implies
that the masses of both the heavy RH and the light LH neutrinos are Majorana, not Dirac,
which violate lepton number by two units. That in turn provides, by utilizing the
out-of-equilibrium decays of the RH neutrinos and the electroweak sphaleron process, the
promising mechanism of baryogenesis via leptogenesis, which can naturally yield
$Y_B\approx 10^{-10}$ (see Sec.~6). {\it In short, the neutrinos may well be at the root of
our own origin.} As a by-product, the needed Majorana nature of the neutrino masses clearly call
for most sensitive searches for neutrinoless double beta decay.

\looseness-1 Including the insight gained from the neutrinos as above, we are now in
possession of a set of facts which may be viewed as the {\it matching pieces of a puzzle}
in that all of them can be understood simply by just one idea~--- that of supersymmetric
grand unification. These include: (i) the quantum numbers of all the members in a family
including the RH neutrino; (ii) the quantization of electric charge, with $Q_{e-}=-Q_p$;
(iii) the dramatic meeting of the three gauge couplings (Fig.~1, right panel) or
equivalently the success of the associated prediction of the weak angle; (iv)
$m^\circ_b\approx m^\circ_{\tau}$; (v) $\sqrt{\Delta m^2_{\rm atm}}\approx 1/20$\,eV;
(vi) a nearly maximal $\theta^\nu_{23}\approx \pi/4$ with a minimal $V_{\rm cb}\approx
0.04$; and (vii) baryogenesis via leptogenesis leading to $Y_B\approx 10^{-10}$ (see
Sec.~6).

All of these features and more hang together within a {\it single unified framework}
based on a presumed string-derived $G(2,2,4)$ or SO(10) symmetry, with low-energy
supersymmetry. Unless this neat fitting of all the pieces within a single simple picture
is just a mere coincidence, it is hard to think that that can be the case, it is pressing
that dedicated searches be made to find the two missing pieces of this picture~--- that
is: proton decay and supersymmetry. 
I should add (as mentioned before) that {\it low-energy supersymmetry
is motivated independently of gauge coupling unification in that it provides a natural resolution of the 
gauge-hierarchy problem as well as a viable candidate for cold dark matter}.
The search for supersymmetry, which is now in progress at the LHC, thus needs to be continued
as intensely as possible at the LHC and beyond to cover the multi-TeV region, if need be,
at future accelerators and linear colliders. That for proton decay, as noted in the
previous section, needs megaton-size deep-underground detectors, like HyperKamiokande and
LBNF-DUNE and their successors (if need be), not only to search for this process as
sensitively as possible, but also to study the branching ratios of different decay modes,
should proton decay be discovered. As discussed in the previous section, the prospects
for discovery of proton decay with improvements of current SuperKamiokande limits by
factors of 5 to 10 are high.

The discovery of proton decay will no doubt constitute a landmark in the history of
physics. That of supersymmetry will do the same. The discovery of these two features~---
supersymmetry and proton decay~--- will fill the two missing pieces of a pretty
picture~--- a gauge unification of matter and of its forces.

On the theoretical side, it is but natural to dream that a deeper fundamental theory
should provide a unity of all the forces of nature including gravity, together with a
good quantum theory of gravity, while providing a predictive and realistic description of
the physical world. The string/M-theory,  with
its majestic beauty (in the words of Edward Witten), is undoubtedly the best candidate we
now have for such a deeper fundamental theory.

Notwithstanding the limited understanding we presently have of this theory
(especially in its non-perturbative aspects), because of the hanging-together of
several  pieces within one and the same picture as mentioned above, it stands to
reason to ask: can a ``preferred solution'' of string/M-theory (if it could exist) lead
to a grand-unified picture as above based on an effective $G(2,2,4)$   or SO(10)-like
symmetry in 4D, with the necessary ingredients to go with reality? If such a solution
does emerge, it would no doubt provide a {\it very useful and desirable bridge} between
string theory and the low-energy world described by the standard model. This is because,
as noted above, such a bridge seems to work in explaining a set of phenomena in a
non-trivial manner. Despite its successes, however, the answers to some fundamental issues~--- such as the origin of the
three families with their hierarchical masses and mixings and an understanding of the observed value of the dark energy ( cosmological constant)~--- are outside of its
reach. May be a string-derived grand-unified theory as above could provide a resolution of at least some of these major issues and provide a rationale for the choice of such an effective theory in 4D. Could some
clear glimpses of such an utopian picture, based on developments in experiments and
theory, with both proton  decay and supersymmetry in hand, and with a better
understanding of string theory describing reality in our possession, emerge by the 120th
birthday of Abdus Salam or even by the 150th? A fulfillment of some  or all of these wishes, no matter how compelling they may appear to be,
depends of course on what {\it Nature} has in store for us?

\section*{Acknowledgement}

I would like to thank especially Kaladi Babu, Alon Faraggi, Rabindra Mohapatra, Zurab
Tavartkiladze and Frank Wilczek for collaborative discussions over the years which have
shaped the content of this article. I have greatly benefitted from discussions with Marcin Badziak,
Pasquale Di Bari, Michael Peskin, Pran Nath, Stuart Raby, Goran Senjanovic, Qaisar Shafi and Edward
Witten on various aspects of the physics in the paper. Communications from Edward Kearns
and M. Shiozawa on the latest SuperKamiokande results on searches for proton decay have
been helpful.

I am deeply grateful to late Abdus Salam for the joy of a fruitful collaboration and for a warm
friendship  that lasted between us till he parted from this world. With gratitude and
respect I dedicate this talk to his memory.

I am thankful to Lars Brink, Mike Duff and Kok Khoo Phua for their hospitality, to
Chee-Hok Lim and the members of the Stallion Press, Chennai, India for their help in the
processing of this paper, and especially to Nikita Blinov, Michael Peskin and Brian Shuve as regards the latter. 

\end{document}

%% file: mymacros.tex
\def\beq{\begin{equation}}
\def\eeq#1{\label{#1}\end{equation}}
\def\eeqn{\end{equation}}

\newenvironment{Eqnarray}%
   {\arraycolsep 0.14em\begin{eqnarray}}{\end{eqnarray}}
\def\beqa{\begin{Eqnarray}}
\def\eeqa#1{\label{#1}\end{Eqnarray}}
\def\eeqan{\end{Eqnarray}}
\def\CR{\nonumber \\ }

\let\bar=\overbar

\def\lsim{\mathrel{\raise.3ex\hbox{$<$\kern-.75em\lower1ex\hbox{$\sim$}}}}
\def\gsim{\mathrel{\raise.3ex\hbox{$>$\kern-.75em\lower1ex\hbox{$\sim$}}}}

\def\del{\partial}
\def\Dslash{\not{\hbox{\kern-4pt $D$}}}
\def\dslash{\not{\hbox{\kern-2pt $\del$}}}

\def\msb{{\bar{\scriptsize M \kern -1pt S}}}

\def\drb{{\bar{\scriptsize D \kern -1pt R}}}

\makeatletter
\def\section{\@startsection{section}{0}{\z@}{5.5ex plus .5ex minus
 1.5ex}{2.3ex plus .2ex}{\large\bf}}
\def\subsection{\@startsection{subsection}{1}{\z@}{3.5ex plus .5ex minus
 1.5ex}{1.3ex plus .2ex}{\normalsize\bf}}
\def\subsubsection{\@startsection{subsubsection}{2}{\z@}{-3.5ex plus
-1ex minus  -.2ex}{2.3ex plus .2ex}{\normalsize\sl}}

\renewcommand{\@makecaption}[2]{%
   \vskip 10pt
   \setbox\@tempboxa\hbox{\small #1: #2}
   \ifdim \wd\@tempboxa >\hsize     
       \small #1: #2\par          
     \else                        
       \hbox to\hsize{\hfil\box\@tempboxa\hfil}
   \fi}

 \def\citenum#1{{\def\@cite##1##2{##1}\cite{#1}}}
 
\newcount\@tempcntc
\def\@citex[#1]#2{\if@filesw\immediate\write\@auxout{\string\citation{#2}}\fi
  \@tempcnta\z@\@tempcntb\m@ne\def\@citea{}\@cite{\@for\@citeb:=#2\do
    {\@ifundefined
       {b@\@citeb}{\@citeo\@tempcntb\m@ne\@citea\def\@citea{,}{\bf ?}\@warning
       {Citation `\@citeb' on page \thepage \space undefined}}%
    {\setbox\z@\hbox{\global\@tempcntc0\csname b@\@citeb\endcsname\relax}%
     \ifnum\@tempcntc=\z@ \@citeo\@tempcntb\m@ne
       \@citea\def\@citea{,}\hbox{\csname b@\@citeb\endcsname}%
     \else
      \advance\@tempcntb\@ne
      \ifnum\@tempcntb=\@tempcntc
      \else\advance\@tempcntb\m@ne\@citeo
      \@tempcnta\@tempcntc\@tempcntb\@tempcntc\fi\fi}}\@citeo}{#1}}
\def\@citeo{\ifnum\@tempcnta>\@tempcntb\else\@citea\def\@citea{,}%
  \ifnum\@tempcnta=\@tempcntb\the\@tempcnta\else
  {\advance\@tempcnta\@ne\ifnum\@tempcnta=\@tempcntb \else\def\@citea{--}\fi
    \advance\@tempcnta\m@ne\the\@tempcnta\@citea\the\@tempcntb}\fi\fi}
\makeatother

%% file: Pati-Memoir2.bbl
\begin{thebibliography}{100}
\bibitem{38}
By the time SU(5) was proposed, about a year later\cite{9}, the situation had changed.
Asymptotic freedom had been discovered\cite{39}, as also the weak neutral current
interaction\cite{40}, both of which gave strong support to the standard model.
\bibitem{103}
The expressions $V_{\rm cb}$ and $\theta^\nu_{23}$ are given in Eq.~(12) in terms of the
simple square-root-formulas only to exhibit the qualitative reason based on symmetry
(explained in the text) for the enhancement of $\theta^\nu_{23}$ and suppression of
$V_{\rm cb}$. The simple formulas do not really hold, however, when each square-root-term
is rather large, as in $\theta^\nu_{23}$. The values quoted in Eq.~(12) are based on
exact numerical   calculations.\cite{24}

\enlargethispage{12pt}
\bibitem{117}
Incorporating such an effective super potential in accord with the assignment of flavor
charges suggested in Refs.~\cite{100} and \cite{104} (for this purpose, $\Phi$ and
$\overline{\Phi}$ are quivalent to $\overline{16}_{\rm H}$ and 16$_{\rm H}$ respectively)
would involve two additional signlets with appropriate flavor charges. One or both of
those may have a VEV $\sim$  M.
\bibitem{124}
Thus the new set of proton decay operators become proportional to \((f_{ij})_{\nu}
g_{kl}K \langle \overline{\mathbf{16}}_{H} \rangle \langle \mathbf{16}_{H} \rangle /
(M^{2})(M_{16})\) where \(M \approx M_{\mathrm{st}}\) and
\(M_{16}({\sim}M_{\mathrm{GUT}})\) is the mass of the $\mathbf{16}_{H}$ (see
Ref.~\cite{122} for a discussion limiting the strength of this operator).
\end{thebibliography}
